\documentclass[11pt]{article}
\pdfoutput=1
\usepackage{jcapmod,parskip}
\usepackage{enumerate}
\usepackage{empheq}
\usepackage[english]{babel}

\allowdisplaybreaks[1]

\usepackage{bm}

\usepackage{color,xcolor}
\newcommand{\de}{\partial}
\newcommand{\be}{\begin{equation}}
\newcommand{\ee}{\end{equation}}

\makeatletter    

\begin{document}

\begin{titlepage}

\setcounter{page}{1} \baselineskip=15.5pt \thispagestyle{empty}

\bigskip\

\vspace{1cm}
\begin{center}
{\fontsize{20}{28}\selectfont  \sffamily \bfseries Relativistic cosmological large scale structures at one-loop
}
\end{center}

\vspace{0.5cm}

\begin{center}
{\fontsize{13}{30}\selectfont Lina Castiblanco,${}^{\rm a}$ Radouane Gannouji,${}^{\rm a}$ Jorge Nore\~{n}a,${}^{\rm a}$ and Cl\'ement Stahl${}^{\rm a}$
}
\end{center}

\begin{center}
\vskip 8pt
\textsl{${}^{\rm a}\;$Instituto de F\'{\i}sica, Pontificia Universidad Cat\'{o}lica de Valpara\'{\i}so, Casilla 4950, Valpara\'{\i}so, Chile}\\ 
\end{center}

\vspace{1.2cm}
\hrule \vspace{0.3cm}
\noindent {\sffamily \bfseries Abstract} \\[0.1cm]
The large scale structure bispectrum in the squeezed limit couples large with small scales. Since relativity is important at large scales and non-linear loop corrections are important at small scales, the proper calculation of the observed bispectrum in this limit requires a non-linear relativistic calculation. We compute the matter bispectrum in general relativity in the weak field approximation. The calculation is as involved as existing second-order results. We find several differences with the Newtonian calculation such as the non-cancellation of IR divergences, the need to renormalize the background, and the fact that initial conditions must be set at second order in perturbation theory. For the bispectrum, we find relativistic corrections to be as large as the newtonian result in the squeezed limit. In that limit relativistic one-loop contributions, which we compute for the first time, can be as large as tree level results and have the same $1/k^2$ dependence as a primordial local non-Gaussianity signal where $k$ is the momentum approaching zero. Moreover, we find the time dependence of the relativistic corrections to the bispectrum to be the same as that of a primordial non-Gaussianity signal.
\vskip 10pt
\hrule

\vspace{0.6cm}
\end{titlepage}

\tableofcontents

\newpage
\section{Introduction}

There are powerful soft theorems that fix the squeezed limit of the three-point correlation function produced in single-field inflation~\cite{Maldacena:2002vr, Creminelli:2004yq, Creminelli:2011rh, Creminelli:2012ed, Hinterbichler:2012nm}. An observation of a deviation from this behavior, called consistency relations, would be proof of the presence of one or more light fields active during inflation (with masses $m \lesssim H$, where $H$ is the Hubble parameter defined after equation \eqref{eq:metric}). This limit is also sensitive to the spin of those fields~\cite{Arkani-Hamed:2015bza}. Physically, these relations stem from the fact that a long-wavelength gravitational potential is locally unobservable if the evolution of the universe can be described by a single degree of freedom. The assumptions behind these theorems also hold in the late universe~\cite{Peloso:2013zw, Kehagias:2013yd, Creminelli:2013mca}, and thus fix the squeezed limit of galaxy correlation functions. This makes these results robust against possible astrophysical effects that may contaminate the primordial signal. Other configurations of the primordial three-point function are not so robust, as they are degenerate with the non-linear gravitational evolution of matter and galaxies~\cite{Baldauf:2016sjb}.

The Planck satellite has already put bounds on these correlation functions~\cite{Ade:2015ava}, where the ``local non-Gaussianity'' parameter has been constrained to be $f_{\rm NL}^{loc} = 2.5 \pm 5.7$ at the $68\%$ confidence level. However, these bounds are still far from the physically interesting region \hbox{$f_{\rm NL}^{loc} \sim 1$~\cite{dePutter:2016trg}}, which could be probed by galaxy surveys in the near future (see e.g.~\cite{Tellarini:2016sgp, Castorina:2018zfk} for recent analyses). Interesting constraints have already been obtained using the so-called scale-dependent bias~\cite{Dalal:2007cu, Matarrese:2008nc, Slosar:2008hx}, which measures the effect of a long-wavelength gravitational potential $\phi \sim \delta/k^2$ on the number of halos formed. Though easier to compute and measure, this observable has limited information. For example, it is not sensitive to the spin of the field inducing the deviation from the consistency relation, since it is intrinsically an averaged quantity. The galaxy bispectrum is harder to measure and compute but contains many more modes and looses no angular information.

Our purpose is to take steps towards predicting the observed galaxy bispectrum if a single light scalar field was active during inflation. In this case there can be no physical correlation between long and short modes induced by astrophysical dynamics. However, there can be ``projection effects'', meaning geometrical effects introduced for example by the change in physical volume due to perturbations, or the deviation of the path of a photon from a galaxy to the telescope. Though these effects are irrelevant for current constraints on primordial non-Gaussianity~\cite{Yoo:2012se}, if future constraints achieve $\sigma_{f_{\rm{NL}}} \sim 1$, these projection effects are expected to be degenerate with the primordial signal, as has been computed for the scale-dependent bias~\cite{Yoo:2009au,Challinor:2011bk,Bonvin:2011bg,Jeong:2011as}.

For the bispectrum, these effects will come from non-linear geometrical effects and photon propagation, and in a given gauge will be present in the dynamical equations. One may naively think that these relativistic non-linear projection effects are very suppressed since non-linearities are important at small scales while relativistic projection effects are important at large scales where the universe is linear. However, the bispectrum couples scales, and one expects most of the signal in the squeezed limit will come from the coupling of a large scale with small non-linear scales. It is thus crucial to perform a calculation which is both non-linear and relativistic. 

To develop a physical intuition of our results, we will now make a rough estimate of how large these effects can be. At linear order, the typical relativistic (projection) effect goes as $\sim \phi$, where $\phi$ is the gravitational potential. At non-linear order, one of the relativistic corrections can go as $\sim \phi \delta_l$ where $\delta_l$ is the linear density contrast. We thus write schematically
$$
\delta_{obs} \sim \delta_l + \alpha \phi + F_2 \delta_l^2 + F_2^R \phi \delta_l+\dots\,,
$$
where we have ignored several terms of the same order as the one we considered for the sake of brevity (though we include them in the full calculation). Here, $\phi$ is of order $10^{-5}$ at all scales, while $\delta_l \sim \nabla^2 \phi / H^2$ can be larger than $1$ at small enough scales. Let us now write down the bispectrum induced in the squeezed limit ($q \ll k$) and ignoring the Dirac delta from momentum conservation
$$
\langle \delta^L_{obs}({\bf q}) \delta^s_{obs}({\bf k}) \delta^s_{obs}({\bf k})\rangle \sim 2F_2 P_\delta(q) P_\delta(k) + 2 \alpha \frac{H^2}{q^2}  F_2 P_\delta(q) P_\delta(k) + \frac{H^2}{q^2} F_2^R P_\delta(k) P_\delta(q)\,.
$$
$F_2$ and $F_2^R$ are kernels that depend only of space. Their definition will be given later (equation \eqref{eq:expansion}).
Here we have ignored a term proportional to $H^2/k^2$ which is very suppressed if the short scales are deep inside the horizon, and taken the long mode to be linear. We learn from this schematic result that relativistic effects have the same dependence on the long mode momentum $q$ and the same time dependence as a primordial breaking of the consistency relation (e.g.~local non-Gaussianity), and are thus expected to be degenerate with them. Moreover, the non-linear relativistic correction to the short wavelength modes is as important as the linear relativistic correction to the long wavelength mode. This is true even at loop order. To see it, let us go one order higher in our schematic expansion
$$
\delta_{obs} \supset F_3 \delta_l^3 + F_3^R \delta_l^2 \phi+\dots\,.
$$
One of the one-loop contributions to the bispectrum in the squeezed limit thus goes as
\begin{multline*}
\langle \delta^L_{obs}({\bf q}) \delta^s_{obs}({\bf k}) \delta^s_{obs}({\bf k})\rangle \supset P_\delta(q)\int_{{\bf p}} F_3 F_2 P_\delta(|{\bf k} - {\bf p}|)P_\delta(p) + \frac{H^2}{q^2}P_\delta(q) \int_{\bf{p}} F_3^R F_2 P_\delta(|{\bf k} - {\bf p}|) P_\delta(p)\\
+ \alpha\frac{H^2}{q^2}P_\delta(q) \int_{\bf p} F_3 F_2 P_\delta(|{\bf k} - {\bf q}|)P_\delta(p)\,.
\end{multline*}
We see here that, when loop corrections to the short modes are important, the non-linear relativistic corrections \emph{to the short scales} are in principle of the same order as the linear relativistic correction to the long mode, and potentially degenerate with a primordial signal. This is especially important in light of the effective theory (EFT) of large scale structure~\cite{Baumann:2010tm,Carrasco:2012cv,Cataneo:2016suz}, since the proper renormalization of these loop corrections will induce relativistic corrections to the EFT counter-terms that are important when the short modes are of order $k \sim 0.1\,\mathrm{Mpc}^{-1}$.

Several groups have already computed the galaxy bispectrum in relativistic perturbation theory~\cite{Umeh:2012pn,Umeh:2014ana,DiDio:2014lka,DiDio:2015bua,Umeh:2016nuh,Fidler:2015npa,Fidler:2016tir,Fidler:2017pnb,Koyama:2018ttg}, and conclude that these effects are degenerate with $f_{\rm NL}^{loc}\sim 1$~\cite{DiDio:2016gpd}. In particular, the impact of general relativity on the dynamics of dark matter was investigated in~\cite{Hwang:2015jja,Goldberg:2016lcq,Gallagher:2018bdl}. Second order analyses are valid when all scales involved are large, such that the tree level bispectrum is enough. However, in order to exploit all the information contained in the bispectrum, it is necessary to push this calculation to smaller scales (for the short modes). Reference~\cite{Kehagias:2015tda} attempted to go in that direction by using consistency relation arguments which, however, are only strictly valid when the long mode is outside of the horizon. Since there are very few modes outside of the horizon even for the cosmic microwave background (CMB), a more complete computation is needed. In order to compute the one-loop bispectrum, one needs to perform perturbation theory up to the fourth order. That may seem like an insurmountable task in general relativity (GR). However, we can use the fact that the metric is close to a Friedmann Lema\^itre Robertson Walker (FLRW) solution even at small scales and (root mean squared) velocities are small
$$
\phi \sim \frac{H^2}{k^2} \delta \sim 10^{-5}\,,\quad v^i \sim \frac{H}{k} \delta \sim 10^{-3}\,.
$$
In this paper we work to next-to-leading order in $H/k$, an approximation called ``weak field approximation''~\cite{Green:2010qy,Green:2011wc,Brustein:2011dy,Kopp:2013tqa}. As a first step, we compute the dynamical equations in GR in perfect matter domination. The inclusion of radiation and a cosmological constant is straightforward though lengthy, and we will address it in a future work. We will also consider the propagation of photons in the near future.

There are already numerical simulations that incorporate relativistic effects~\cite{Adamek:2014xba,Bentivegna:2015flc,Mertens:2015ttp,Adamek:2015eda,Adamek:2016zes,Daverio:2019gql}. In order to facilitate the fit to and comparison with the most advanced research program: \texttt{gevolution}~\cite{Adamek:2016zes}, we perform our calculations in {\bf Poisson gauge}. On the other hand, what we observe are galaxies and not the dark matter density field. The connection between the two, called biasing, encodes how local dynamics depends on long-wavelength perturbations. One thus expects that this biasing should be described at the trajectory of each galaxy, and is thus more easily computed by using a gauge in which constant time hyper surfaces are fixed by the clocks of observers moving with the fluid, that is in a {\bf synchronous gauge}, which in matter domination this corresponds at all orders to the {\bf comoving gauge} that has been used in previous perturbation theory literature. Our results are gauge-dependent: they do not (yet) correspond to actual observables. When, in a follow-up work, we account for the propagation of photons from galaxies to the telescope, the final result should be gauge-independent, but the distinction between a geometric effect on the photon propagation and a dynamical effect on the local dark matter density (or the number density of galaxies) depends on the gauge being used.

Our paper is organized as follows: in section \ref{sec:Einsteinandfluid} we take the weak field approximation in general relativity and write the equations of motion to arbitrarily non-linear order in the density contrast, but perturbative in the velocitiy and gravitational potential. In section \ref{sec:PTIC} we solve these equations within perturbation theory taking the density contrast to be small, but still larger than velocities and the gravitational potential. In section \ref{sec:Correfunct} we present our results for the power spectrum and bispectrum of the density contrast to one-loop obtained from this calculation. For the bispectrum, we find that one-loop relativistic corrections are important in the squeezed limit and potentially degenerate with a primordial non-Gaussianity signal of the local type. On the other hand, the relativistic corrections to the density contrast power spectrum are negligible as expected. In section \ref{section:divergences} we remark on the infrared (IR) behavior of the resulting loop integrals, which is markedly different from the Newtonian case. We also comment briefly on the ultraviolet (UV) behavior of said integrals. Finally, we conclude in section \ref{sec:conclusions} and discuss how further steps can be taken in order to compute the observed galaxy bispectrum to one-loop in general relativity. In the appendices we present several explicit derivations that would make the main text too cumbersome. 

\section{Einstein and fluid equations in the weak field approximation}
\label{sec:Einsteinandfluid}
Our starting point is the line element of a perturbed FLRW metric 
\begin{equation}
\label{eq:metric}
ds^2 = -(1+2\phi) dt^2+2 \omega_i dx^i dt + a(t)^2 \left[(1-2\psi)\delta_{ij} + h_{ij} \right]dx^i dx^j\,,
\end{equation}
where $a$ is the background scale factor\footnote{See an insightful discussion on the physical relevance of the scale factor in this context in Ref.~\cite{Finke:2019pru}}, $t$ is the cosmic time, and $x^i$ are Cartesian comoving coordinates.\footnote{As usual,  Greek letters (e.g.~$\mu, \nu, \alpha$) represent space-time indices that run from 0 to 3, while latin letters (e.g.~$i,j,k$) represent spatial indices that run from 1 to 3. Latin indices are written arbitrary up or down as, within our approximation, they differ only by powers of $a(t)$, which are always written down explicitly.} A dot denotes a derivative with respect to the cosmic time and $H\equiv \frac{\dot{a}}{a}$. By definition, $h_{ij}$ is traceless, and we separate $\omega_i$ into a longitudinal and a transverse part
\begin{equation}
\omega_i = \partial_i\omega + w_i\,,\quad \partial_i w_i = 0\,.
\label{eq:omegatransverse}
\end{equation}
The matter content of the universe is assumed to be pressureless, barotropic and irrotational. We approximate the contents of the universe as a fluid, which is no longer a valid approximation at small scales where shell crossing takes place. We perform our calculation in a matter dominated universe. We will include the effects of radiation and a cosmological constant in the future. The so-called Einstein-de Sitter (EdS) approximation is often generalized to more fluids or a different Poisson equation, by using the kernels obtained in EdS but deriving the new time dependence of the solution due to the presence of new degrees of freedom. This has been shown to be a reasonable approximation in the case of the power spectrum~\cite{Takahashi:2008yk,Fasiello:2016qpn} in the Newtonian limit,  but its validity for the bispectrum or its impact on relativistic effects has, to our knowledge, not been studied.
We thus take the stress-energy tensor to be 
\begin{equation}
    T_{\mu\nu} = \bar{\rho}(1 + \delta) u_\mu u_\nu\,,
\end{equation}
where $\bar{\rho}$ is the background density, $\delta$ is the matter density contrast and $u_\mu$ is the matter 4-velocity. In order to describe the dark matter fluid, we will follow the approach of \cite{Boubekeur:2008kn} (section~2), summarized in Appendix~\ref{app:A}. Since the fluid is barotropic and irrotational, it can be described by a single scalar potential $\varphi$. Being irrotational, the 4-velocity of the fluid is defined to be hypersurface-orthogonal, namely orthogonal to the constant-$\varphi$ hypersurfaces, and given by
\begin{equation}
\label{eq:defumu}
u_{\mu}= \frac{\partial_{\mu} \varphi}{\sqrt{-\partial_{\mu} \varphi \partial^{\mu} \varphi}},
\end{equation}
where we take the scalar potential to be $\varphi \equiv t + U$. We will work with the spatial component of the 4-velocity $u_i$. Note that it is \emph{not} necessary curl-free as it is not possible to write it as a derivative of a potential because of its normalization (even if the 3-velocity $v_i = \de_i U$ is irrotational in the usual sense, see Appendix~\ref{app:A}).\footnote{We will see that the transverse part of $u_i$, though non-zero, satisfies equations which don't contain time derivatives (see equations \eqref{eq:EE0i0}, \eqref{eq:19} and \eqref{velocitycomoving}), and are thus fixed in terms of the scalar quantities which are our dynamical degrees of freedom, as expected from the fact that the fluid velocity can be described by a scalar field.} Note that in the Newtonian limit $v_i \approx u_i$ since the normalization is close to unity.

The continuity and Euler equations that describe the evolution of the matter fluid follow from the conservation of this stress-energy tensor, and are
\begin{align}
& \nabla_{\mu} (\bar{\rho}(1+\delta) u^{\mu}) =0\,, \label{eq:NT1} \\
& u^{\mu}\nabla_{\mu} u^{\nu} =0\,. \label{eq:NT2}
\end{align}
In the presence of radiation there will be an additional pressure force.

We will work under the ``weak-field approximation''~\cite{Green:2010qy,Green:2011wc,Brustein:2011dy,Kopp:2013tqa}. The physical reason behind it is the fact that the metric is expected to be close to the unperturbed FLRW metric even at very non-linear scales. Indeed, even in very non-linear scales such as the Solar System, relativistic effects are highly suppressed, and typical peculiar velocities of galaxies and stars are much smaller than the speed of light. This approximation breaks down when the gravitational field becomes strong, such as close to the horizon of a black hole, but that is far beyond the realm of applicability of our calculation. In order to refine our approximations further, let us estimate the size of the quantities involved by using their Newtonian expressions, which are a good approximation on small scales, and we will see that they follow from the Einstein and fluid equations at short distances. Thus, we write the Poisson equation and the linear expression for the fluid velocity
\begin{equation}
\nabla^2 \phi =\frac{3}{2} a^2H^2 \delta ,\quad v^i = -\frac{\partial_i \phi}{aH} \,.  
\end{equation}
For scales of the size of the horizon $k \sim aH$, the density contrast is a small perturbation $\delta \sim \phi \sim \mathcal{O}(10^{-5})$. On scales deep inside the horizon $aH/k \sim \mathcal{O}(10^{-3})$ the density contrast can be order $\delta \sim \mathcal{O}(1)$ while peculiar velocities are typically of the order $v^i \sim \mathcal{O}(10^{-3})$. We see then that $\phi \sim \mathcal{O}(10^{-5})$ on all scales, while derivatives on small scales can be of order $k/aH \sim \mathcal{O}(10^{3})$. Since our interest is in describing these small scales, we will take spatial derivatives to be large, and we write $\epsilon \equiv (aH/\partial)^2 \ll 1$ or in Fourier space $\epsilon \equiv (aH/k)^2 \ll 1$ depending on the context. It is this same parameter that suppresses relativistic corrections with respect to Newtonian dynamics (that is, suppresses $\phi$ with respect to $\delta$). Following this logic, in this section we perform a perturbative expansion in $\epsilon$ while keeping all orders in $\delta$ which can, in principle, be $\mathcal{O}(1)$. In section \ref{sec:PTIC}, we will solve the fully non-linear equations obtained here by using standard perturbation theory; that is, we take $\delta$ to be small but still larger than velocities or the gravitational potential.

Let us note that, by assuming spatial derivatives to be large, we are excluding the possibility that $aH/k \sim 1$, which happens on scales of the size of the horizon. These scales are important since they are already being probed by galaxy surveys. However, relativistic effects on those scales have already been studied by several groups~\cite{Umeh:2012pn,Umeh:2014ana,DiDio:2014lka,DiDio:2015bua,Umeh:2016nuh,Fidler:2015npa,Fidler:2016tir,Fidler:2017pnb,Koyama:2018ttg}, and it is trivial to write an expression that encompasses all scales: one needs simply add to our calculation those second order terms which have less spatial derivatives than ours. Let us stress that for the specific goal of computing the squeezed limit bispectrum, it is crucial to have an expression for the small scale non-linearities in general relativity, as explained in the introduction, though only the leading corrections in $aH/k$ are expected to be important. 

We perform our calculation in two different gauges. We will discuss each gauge and the resulting equations separately in the following subsections. We sum up our approximation scheme in table \ref{tab:a}, where we list the different metric and matter quantities for the two gauges under consideration.
\begin{table}
\centering
\begin{tabular}{|c|c|c|}
  \hline
  variable & order in Poisson gauge & order in comoving gauge \\
  \hline
  $\partial_i/H$ & $\mathcal{O}(\epsilon^{-1/2})$ &   $\mathcal{O}(\epsilon^{-1/2})$ \\ 
 $\partial_t/H$ & $\mathcal{O}(1)$ &  $\mathcal{O}(1)$ \\ 
  $\phi$ & $\mathcal{O}(\epsilon)$ &  $\mathcal{O}(\epsilon)$ \\ 
  $\psi$ & $\mathcal{O}(\epsilon)$ &  $\mathcal{O}(\epsilon)$ \\ 
 $\chi$ & $\mathcal{O}(\epsilon^2)$ &  - \\ 
   $w_{i}$ & $\mathcal{O}(\epsilon^{3/2})$ &  $\mathcal{O}(\epsilon^{3/2})$ \\ 
    $\omega$ & - &  $\mathcal{O}(\epsilon)$ \\ 
  $h_{ij}$ &$\mathcal{O}(\epsilon^2)$ &  $\mathcal{O}(\epsilon^2)$ \\ 
      $\delta$ & $\mathcal{O}(1)$ &  $\mathcal{O}(1)$ \\ 
    $u^i$ & $\mathcal{O}(\epsilon^{1/2})$ &  $\mathcal{O}(\epsilon^{1/2})$ \\ 
        $u^i_T$ & $\mathcal{O}(\epsilon^{3/2})$ &  $\mathcal{O}(\epsilon^{3/2})$ \\ 
  \hline
\end{tabular}
\caption{Approximation scheme for the different quantities defined in section \ref{sec:Poissongauge} and \ref{sec:comgauge}. $\epsilon$ represents the relativistic order and is defined as $\epsilon \equiv (aH/\partial)^2$ or in Fourier space $\epsilon \equiv (aH/k)^2$.}
\label{tab:a}
\end{table}
\subsection{Poisson gauge}
\label{sec:Poissongauge}
For comparison with existing literature (including similar calculations performed for the CMB, see for example~\cite{Matarrese:1997ay, Bartolo:2005kv}), we use the {\bf Poisson gauge}, defined at all orders by the conditions 
\begin{equation}
\delta^{ij}\partial_j \omega_{i}=\delta^{ij} h_{ij}=\delta^{jk}\partial_k h_{ij}=0 \,,
\end{equation}
where $\delta^{ij}$ is the Kronecker delta. In the notation of equation \eqref{eq:omegatransverse}, this means that $\omega = 0$. We take the metric perturbations to be $\phi \sim \psi \sim \mathcal{O}(\epsilon)$ as discussed above. On the other hand, we take $h_{ij} = \mathcal{O}(\epsilon^2)$ and $w_i = \mathcal{O}(\epsilon^{3/2})$, since they will be sourced only at that order (see equations \eqref{eq:wiPoiss} and \eqref{eq:hPoiss} below).

Let us begin by writing the Euler and continuity equations coming from the conservation of the stress-energy tensor. In this gauge, and in the weak-field approximation, the 4-velocity field for the perturbed dark matter fluid is given by
\begin{equation}
u^{\mu}=\left( 1-\phi+\frac{a^2 u^2}{2} + \mathcal{O}\left(\epsilon^2\right), u^i\right)\,,
\end{equation}
where $u^2 = \delta_{ij} u^i u^j$, and $u^i$ is obtained from Eq.~\eqref{eq:defumu} using \eqref{eq:metric}. Note in particular that at order $\epsilon^{3/2}$, the velocity admits a transverse part, we will give an expression for it and detail how to obtain it in appendix \ref{app:Sources}.

Using this expression for the velocity, and expanding equation (\ref{eq:NT1}), we find
\begin{multline}
\label{eq:deltaPois}
\dot{\delta} +\partial_i\left[(1+\delta)u^i\right]= \\ \dot{\delta} \left(\phi-\frac{a^2 u^2}{2}\right)- \left(1+ \delta\right)\partial_t\left(\frac{a^2 u^2}{2}\right)+3(1+\delta)\dot{\phi}+2\phi_{,i}(1+\delta)u^i + \mathcal{O}\left(\epsilon^2\right),
\end{multline}
where we have used the background equation $\dot{\bar{\rho}}+3H\bar{\rho}=0$, and set $\phi - \psi \sim \mathcal{O}(a^4 H^4/k^4)$ (for proof, see Appendix \ref{app:Einstein}).
Expanding equation (\ref{eq:NT2}), and taking its divergence, one finds
\begin{multline}
\label{eq:thetaPois}
{{\partial_i\dot{ u}^i}} + 2H \partial_i u^i + \partial_i(u^{j}\partial_j u^{i}) + \frac{\nabla^2\phi}{a^2} =\\ \partial_i\left[\left(-\phi+\frac{a^2 u^2}{2} \right) \dot{u}^i +2 \left(\frac{a^2 u^2}{2}H-H \phi-\dot{\phi} \right)u^i+2(\phi_{,i} u^2-u^j \phi_{,j}u^i)\right] + \mathcal{O}\left(\epsilon^2\right)\,.
\end{multline}
In order to obtain the equation for $w^i$, let us define a transverse projector $P_{ij}$.\footnote{This is such that $P_{ij} \partial_j f = 0$ and $P^2 = P$. In Fourier space such a projector is
$$
P_{ij} = \delta_{ij} - \frac{k_i k_j}{k^2}\,.
$$} Applying it to equation \eqref{eq:NT2}, one obtains
\begin{align}
\label{eq:wiPoiss}
& \frac{1}{a^2}\dot{\omega}_i  = - P_{ik}\left[\frac{a^2}{4}(\omega_{k,j}-\omega_{j,k})u^j + \left(-\phi+\frac{a^2 u^2}{2} \right) \dot{u}^k  + 2 \left(\frac{a^2 u^2}{2}H-H \phi-\dot{\phi} \right)u^k \right]  \nonumber \\
& + \mathcal{O}\left(\epsilon^{5/2}\right),
\end{align}
As promised, the source for $w_i$ is seen to be of order $\mathcal{O}(\epsilon^{5/2})$, such that if $w_i$ vanishes in the initial conditions, then $w_i \sim \mathcal{O}(\epsilon^{3/2})$.

Finally, in order to close the system, we use Einstein equations. The detailed derivation is presented in Appendix \ref{app:Einstein}. We obtain an equation for $\phi$, see \eqref{eq:closed}
\begin{equation}
\frac{2}{a^2} \nabla^2 \phi(1-2\phi) -6 \frac{\ddot{a}}{a}+6H(3\dot{\phi}-2H \phi) +6\ddot{\phi}-4 \phi_{,i}^2=\bar{\rho}(1+\delta) (1-2\phi+2 a^2 u^2) + \mathcal{O}\left(\epsilon^2\right)\,, \label{eq:EEiii}
\end{equation}
Equations \eqref{eq:deltaPois}, \eqref{eq:thetaPois}, and \eqref{eq:EEiii} are a closed system that can now be explicitly solved. Throughout, we have kept up to subleading terms in the weak field approximation. We remark once more that these equations are valid non-perturbatively in the density contrast $\delta$, but up to first order in the potential $\phi$ and second order in the velocity $u^i$. 

It is also of interest to write an equation for the tensor modes, which can be sourced at the non-linear level. Using the transverse projector we obtain (see equation \eqref{eq:apptensor})
\begin{align}
\label{eq:hPoiss}
   &\left(\delta_k^m \delta_l^n-\frac{1}{3}\delta^{mn}\delta_{kl} \right) P^{il}P^{jk} \left[\ddot{h}_{ij}+3H\dot{h}_{ij}-\frac{\Delta h_{ij}}{a^2}+\frac{2}{a^2}(-2 \chi \phi_{,ij}+2 \phi_{,i} \phi_{,j} +4 \phi \phi_{,ij}) \right]  =0\,.
\end{align}
We see here that the source for $h_{ij}$ is of order $\mathcal{O}(\epsilon^2)$, such that if $h_{ij}$ vanishes in the initial conditions, then $h_{ij} \sim \mathcal{O}(\epsilon^2)$, which we assumed implicitly in the equations above (see Appendix \ref{app:Einstein}). The relevant equations are \eqref{eq:deltaPois}, \eqref{eq:thetaPois}, and \eqref{eq:EEiii}, which allow to solve for the density contrast $\delta$, the velocity $u^i$, and the gravitational potential $\phi$. These in turn source frame-dragging $\omega_i$ and tensor perturbations $h_{ij}$ through equations \eqref{eq:wiPoiss} and \eqref{eq:hPoiss}.

\subsection{ Comoving (synchronous) gauge}
\label{sec:comgauge}
Though we focus on the matter density contrast, what is actually observed is the number density of galaxies. The two are related since more galaxies are expected to form in deeper potential wells. The connection between them is called biasing and we will include it in a future work. For now let us note that galaxy formation is a local effect that depends only on the time measured by an observer moving with the fluid where the galaxy will form \cite{Baldauf:2011bh, Jeong:2011as, Yoo:2014vta}. For this reason we work also in {\bf comoving gauge}, defined at all orders by the conditions
\begin{equation}
\delta^{ij} h_{ij}=\delta^{jk}\partial_k h_{ij}=0\,,\quad u_i = 0\,,
\end{equation}
We will see that, similar to what happened in the Poisson gauge, the sources for $w_i$ are of order $\mathcal{O}(\epsilon^{3/2})$, while $\partial_i \omega$ is sourced already at order $\mathcal{O}(\epsilon^{1/2})$. Tensor modes are sourced at order $h_{ij}\sim \mathcal{O}(\epsilon^2)$.

Let us show that in perfect matter domination {\bf the comoving gauge is synchronous}, in the sense that $u^0 = 1$ at all orders, up to an integration constant. Indeed, following~\cite{Yoo:2014vta}, in a comoving gauge the 4-velocity is by definition
$$
u_\mu = (-N , 0)\,,\quad u^\mu = (1/N,-N^i/N)\,,
$$
where the first expression is fixed by the normalization of the four-velocity, $N = 1/\sqrt{-g^{00}}$, and $N^i = g^{0i}/(-g^{00})$ are the lapse and shift of the ADM formalism respectively. Using this, one can write the geodesic equation
$$
\frac{d u_\alpha}{d\tau} = \frac{1}{2}g_{\mu\nu,\alpha}u^\mu u^\nu = \frac{1}{N} \partial_\alpha N\,,
$$
such that $N = 1$ is a solution of both the space $\alpha = i$ and time $\alpha = 0$ components of this equation, and thus we can take $u^0 = 1$. This ensures that the proper time along a fluid trajectory coincides with coordinate time. Therefore, the four-velocity of a fluid element in this gauge is
\begin{equation}
   u^\mu=\left(1,-\frac{1}{a^2}(1+2\psi) \partial_i\omega-\frac{1}{a^2}w_i + \mathcal{O}\left(\epsilon^2\right)\right)\,,
\end{equation}
where the $u^i$ component is fixed by the condition $u_\mu u^\mu = -1$. The condition $u^0 = 1/N = 1$ is equivalent to
\begin{equation}
    \phi = - \frac{1}{2a^2}\partial_i \omega \partial_i\omega \left(1+2\psi\right)-\frac{1}{a^2}w_i\partial_{i}\omega + \mathcal{O}\left(\epsilon^3\right)\,.
\end{equation}
We will use this equation to eliminate $\phi$ in favor of $\omega$, $\psi$ and $w_i$ .

We will work with the spatial components of the 4-velocity with an upper index, given by
\begin{equation}
    u^{i} = -\frac{1}{a^2}(1+2\psi) \partial_i\omega-\frac{1}{a^2}w_i\,.\label{velocitycomoving}
\end{equation}
With this definition, the continuity equation \eqref{eq:NT2} can be written as
\begin{equation}
\label{eq:10}
\dot{\delta}+\partial_{i}\left[(1+\delta)u^{i}\right] = 3(1+\delta)\dot{\psi}+3(1+\delta)u^{i}\partial_{i}\psi + \mathcal{O}\left(\epsilon^2\right)\,.
\end{equation}
To close the system, we again need the Einstein equations. Their derivation, and the specific combinations we use, is once more relegated to Appendix \ref{app:Einstein}. Note that in this case, we need an equation for $\psi$ and an equation for $\omega$ (or equivalently $u^i$). Using \eqref{velocitycomoving} we replace $\omega$ in favor of $u^i$ in \eqref{eq:20}, then we get
\begin{align}
\label{eq:11}
    \partial_i\dot{u}^i + 2H \partial_iu^i + \frac{3}{2}H^2\delta =& 3\ddot{\psi}+6H\dot{\psi}-\partial_j(u^{i}\partial_{i}u^{i})+3\dot{u}^i\partial_i\psi+6Hu^i\partial_i\psi+2\partial_iu^i\dot{\psi}\nonumber \\
   &+6\partial_i\dot{\psi}u^i +2u^{i}\partial_ju^{j}\partial_i\psi +3u^{i}u^{j}\partial_i\partial_j\psi+\mathcal{O}(\epsilon^3)\,,
\end{align}
which is the relativistic extension to the Euler equation in this gauge. and from  \eqref{eq:17} we obtain an equation for $\psi$
\begin{align}
    \nabla^2\psi=& \frac{3}{2}a^2H^2\delta-Ha^2\partial_iu^i +\frac{a^2}{2}\left[\left(\partial_ju^i\right)^2+ (\partial_iu^i)^2\right]+3Ha^2\dot{\psi} +Ha^2u^i\partial_i\psi
    \nonumber \\ &-\frac{1}{2}\left[3\left(\partial_i\psi\right)^2+8\psi\nabla^2\psi\right] + a^2\partial_iu^i\dot{\psi} +a^2\partial_i\psi u^j\partial_ju^i + \mathcal{O}\left(\epsilon^2\right)\,. \label{eq:12}
\end{align}
As before, we can also write an equation for $w_i$ (obtained from the Einstein equations, \eqref{eq:19})
\begin{equation}
\label{eq:wiSync}
\frac{1}{a^2}\nabla^2w_i=4\partial_i\dot{\psi}-\frac{2}{a^2}\partial_i \omega \nabla^2\psi - \frac{2}{a^2}\partial_j\omega \partial_i \partial_j \psi\,,
\end{equation}
where we see that $w_i$ is sourced at order $\mathcal{O}(\epsilon^{3/2})$ as promised (its Laplacian is sourced at order $\mathcal{O}(\epsilon^{1/2})$, and inverting it gives the order stated). Similarly, for the tensor modes see equation \eqref{tensormodess}
\begin{equation}
\label{eq:hSync}
\nabla^2h_{ij} = \psi\partial_i\partial_j\psi -\partial_j\omega\partial_i\dot{\psi} -\partial_i\omega\partial_j\dot{\psi}+\frac{1}{a^6}\partial_k\omega\partial_l\partial_i\omega\left(\partial_l\omega\partial_k\partial_j\omega-\partial_k\omega\partial_l\partial_j\omega\right)\,.
\end{equation}
As in the Poisson gauge, the source for $h_{ij}$ is of order $\mathcal{O}(\epsilon^2)$, such that if $h_{ij}$ vanishes in the initial conditions, then $h_{ij} \sim \mathcal{O}(\epsilon^2)$, which we assumed implicitly in the equations above (see Appendix \ref{app:Einstein}). The relevant equations are \eqref{eq:10}, \eqref{eq:11}, and \eqref{eq:12}, which allow to solve for the density contrast $\delta$, the velocity $\theta$, and the gravitational potential $\phi$. These in turn source the transverse component of the frame-dragging term $w_i$ and tensor perturbations $h_{ij}$ through equations \eqref{eq:wiSync} and \eqref{eq:hSync}.
\section{Perturbation theory and initial conditions}
\label{sec:PTIC}
In the previous section, we made no assumptions about the size of perturbations apart from the weak field approximation. The equations thus obtained are non-linear in $\delta$. In both the Poisson gauge and the comoving gauge, the leading order equations reduce to the familiar Newtonian fluid equations as they should. It then simplifies the algebra to separate all our fields into a Newtonian piece, denoted with a subscript $N$, that satisfies the Newtonian fluid equations 
\begin{align}
& \dot{\delta}_{N}+\partial_i\left[ (1+\delta_{N}) v^{i}_N \right]=0\,,  \label{eq:hierac1}\\
& \partial_i \dot{v}^i_N + 2H \partial_i v^i_N + \partial_i(v_N^{j} \partial_j v_N^{i}) + \frac{3}{2}H^2\delta_N=0\,, \label{eq:hierac2}
\end{align}
and a small relativistic correction denoted by a subscript $R$:
\begin{align}
& \label{eq:decd} \delta = \delta_{N}+\delta_{R}\,, \\
& \label{eq:decv} u^i = v^{i}_N + u^{i}_{R}+u^{i}_{T}\,.
\end{align}
Here we used the fact that $u^i \approx v^i$ in the Newtonian limit. The relativistic velocity, following the discussion of section \ref{sec:Einsteinandfluid} was further decomposed on a longitudinal and transverse part (subscript $R$ and $T$ respectively). The velocity divergence field is then defined as usual as $\theta_R \equiv \partial_i u^{i}_{R}$.
We can then use equations \eqref{eq:deltaPois}-\eqref{eq:EEiii}, in the Poisson gauge or \eqref{eq:10}-\eqref{eq:12} in the comoving gauge, and write in Fourier space \footnote{We work with the Fourier convention $$f(\bm{x}) = \int\frac{d^3 k}{(2\pi)^3}\, e^{i\bm{k}\cdot\bm{x}}f(\bm{k}) \equiv \int_{\bm{k}}e^{i\bm{k}\cdot\bm{x}}f(\bm{k})\,.$$
Here we have also defined a short hand notation for integrals.}
\begin{align}
&\dot{\delta}_{R}+\theta_R= - \int_{\bm{ k}_1,\bm{k}_2}\!\!(2\pi)^3\delta_D(\bm{k} - \bm{k}_{12})\Big[\alpha(\bm{k}_1,\bm{k}_2)(\theta_R(\bm{k}_1) \delta_N(\bm{k}_2) + \theta_N(\bm{k}_1) \delta_R(\bm{k}_2)) \\
& +i\bm{k}_{2} \cdot \textbf{u}_T(\textbf{k}_1)\delta_N(\bm{k}_2)\Big] +\mathcal{S}_{\delta}[\psi_N, \delta_N, \theta_N]\,, \label{eq:correc1} \\
 &\dot{\theta}_R+2 H \theta_R+ \frac{3}{2}H^2\delta_R= -\int_{\bm{ k}_1,\bm{k}_2}\!\!(2\pi)^3\delta_D(\bm{k} - \bm{k}_{12})\Big[ 2\beta(\bm{k}_1,\bm{k}_2)\theta_N(\bm{k}_1)\theta_R(\bm{k}_2)\\
 & +i\bm{k}_{2} \cdot \textbf{u}_T(\textbf{k}_1)\theta_N(\bm{k}_2)\left(1+2\frac{\textbf{k}_1 \cdot \textbf{k}_2}{\textbf{k}_2^2}\right)\Big] +  \mathcal{S}_{\theta}[\psi_N, \delta_N, \theta_N]\,, \label{eq:correc2}
\end{align} 
where, as usual, $ \alpha(\bm{k}_1 , \bm{k}_2) = (\bm{k}_{12} \cdot \bm{k}_1)/\bm{k}_1^2$, $ \beta(\bm{k}_1 , \bm{k}_2) = \bm{k}_{12}^2 (\bm{k}_1 \cdot \bm{k}_2)/2\bm{k}_1^2 \bm{k}_2^2$, $\bm{k}_{12} = \bm{k}_1 + \bm{k}_2$, and $\delta_D$ is the Dirac delta. The sources $\mathcal{S}$ to equations \eqref{eq:correc1} and \eqref{eq:correc2} are explicitly relativistic, and therefore depend only on the fields satisfying the Newtonian equations. Their explicit form in each gauge can be straightforwardly obtained from equations \eqref{eq:deltaPois}-\eqref{eq:EEiii} or \eqref{eq:10}-\eqref{eq:12} and is given in appendix \ref{app:Sources}, together with an expression for the transverse velocity $u^i_T$ which is sourced at order $\epsilon^{3/2}$, following the discussion of section \ref{sec:Einsteinandfluid}.
Due to the non-linear terms that couple relativistic corrections with the Newtonian solution, we need further approximations in order to solve these equations. To do that, {\bf we perform perturbation theory in the usual sense,} taking $\delta \ll 1$ and $\theta \ll 1$ in equations \eqref{eq:hierac1}-\eqref{eq:hierac2}, \eqref{eq:correc1}-\eqref{eq:correc2}. Note that we still keep $aH/k \ll 1$, which was used in deriving these equations. 
As in standard Newtonian perturbation theory, we combine these equations in order to obtain a second-order differential equation for $\delta_R$
\begin{equation}
\label{eq:efdelta}
\ddot{\delta}_R+2H\dot{\delta}_R-\frac{3}{2}H^2\delta_R=S\,,
\end{equation} 
where $S$ is defined to be
\begin{multline}
\label{eq:sources}
    S \equiv\dot{\mathcal{S}}_{\delta}+2H\mathcal{S}_{\delta}-\mathcal{S}_{\theta}  + \int_{\bm{k}_1, \bm{k}_2}\!\!(2\pi)^3\delta(\bm{k} - \bm{k}_{12})\bigg\{2\beta(\bm{k}_1,\bm{k}_2)\theta_N(\bm{k}_1)\theta_R(\bm{k}_2) \\
    - \frac{\partial_t}{a^2}\Big[a^2\left(\alpha(\bm{k}_1, \bm{k}_2)\theta_R(\bm{k}_1) \delta_N(\bm{k}_2) + \alpha(\bm{k}_1, \bm{k}_2)\theta_N(\bm{k}_1) \delta_R(\bm{k}_2)\right)\Big]\\+\frac{\partial_{t}}{a^2}\left(a^2\bm{k}_{2} \cdot \textbf{u}_T(\textbf{k}_1)\delta_N(\bm{k}_2)\right)+i\bm{k}_{2} \cdot \textbf{u}_T(\textbf{k}_1)\theta_N(\bm{k}_2)\left(1+2\frac{\textbf{k}_1 \cdot \textbf{k}_2}{\textbf{k}_2^2}\right)\bigg\}\,.
\end{multline}
We can now solve (\ref{eq:efdelta}) order by order in terms of its Green's function, with $t_\ast$ the initial time:
\begin{equation}
\delta_R(t,\bm{k})=c_+(\bm{k})D_+(t)+c_-(\bm{k})D_-(t)+\int_{t_\ast}^t\frac{D_+(t')D_-(t)-D_+(t)D_-(t')}{W(t')}S(t',\bm{k})\,.
\label{eq:green}
\end{equation}
Here, $c_+(\bm{k})$ and $c_-(\bm{k})$ are fixed by the initial conditions, and $D_+(t)$ and $D_-(t)$ are the growing and the decaying mode solutions to equation \eqref{eq:efdelta}. In appendices \ref{sec:icpoisson} and \ref{sec:icsync}, we discuss the determination of the initial conditions. The Wronskian is defined as $W(t) \equiv D_+(t)\dot{D}_-(t)-\dot{D}_+(t)D_-(t)$ and for the case of matter domination $D_+(t)=a(t) \propto t^{2/3}$ and $D_-(t)=a^{-3/2}$. This solution can be used to compute $\theta_R$ to the same order, using equation \eqref{eq:correc1}. We thus write
\begin{multline}
    \delta(\bm{k},t) = \sum_{n = 1}^\infty a^n(t)\int_{\bm{k}_1...\bm{k}_n}\!\!\!\!\!\!\!\! (2\pi)^3\delta_D(\bm{k}-\bm{k}_{1...n})\Big[ F_n(\bm{k}_1,\dots,\bm{k}_n) \\ + a^2(t)H^2(t) F_n^R(\bm{k}_1,\dots,\bm{k}_n)\Big]\delta_l(\bm{k}_1)\dots\delta_l(\bm{k}_n)\,,
    \label{eq:expansion}
\end{multline}
\begin{multline}
    \theta(\bm{k},t) = -H(t)\sum_{n = 1}^\infty a^n(t)\int_{\bm{k}_1...\bm{k}_n}\!\!\!\!\!\!\!\! (2\pi)^3\delta_D(\bm{k}-\bm{k}_{1...n})\Big[ G_n(\bm{k}_1,\dots,\bm{k}_n) \\ + a^2(t)H^2(t) G_n^R(\bm{k}_1,\dots,\bm{k}_n)\Big]\delta_l(\bm{k}_1)\dots\delta_l(\bm{k}_n)\,,
\end{multline}
where $\bm{k}_{1...n}\equiv \sum_{i=1}^n \bm{k}_i$, $F_n$ and $G_n$ are the usual Newtonian kernels from perturbation theory \cite{Bernardeau:2001qr}, the explicit form of the relativistic kernels $F_n^R$ and $G_n^R$ is given in appendix \ref{sec:kernels}, and $\delta_l$ is the linear Newtonian density perturbation linearly extrapolated to redshift zero in both gauges, that is
\begin{equation}
\label{eq:Poisson}
    \delta_l(\bm{k}) \equiv \frac{2}{3 H_0^2} \nabla^2 \psi_o\,.
\end{equation}
Here, $\psi_o$ is the curvature perturbation generated during inflation, assumed to be a Gaussian random field (and therefore $\delta_l$ is also a Gaussian random field). The time dependence in equation \eqref{eq:expansion} is obtained from equation \eqref{eq:green}. It can be easily deduced from the fact that a universe dominated by a single fluid component, each power of $k$ has to be accompanied by a corresponding power of $aH$. Since relativistic corrections contain two powers of derivatives less, they should be multiplied by $a^2H^2$.

We choose our initial conditions to be adiabatic, such that the universe is well described by a single clock throughout its history. For this we require that our solution matches the perturbative fully relativistic solution at early times, which was computed to second order in perturbations in~\cite{Matarrese:1997ay, Bartolo:2005kv} and~\cite{Boubekeur:2008kn} for the Poisson and comoving gauges respectively. The matching need only be done up to second order in $\delta_l$ within our approximations, as is explicitly shown in appendix \ref{app:Sources}. 
A simple way to see it is that the growing mode goes like $D_+ \sim a \sim a^{-2} H^{-2}$ which is the well-known behavior of the Newtonian solution to linear order in $\delta_l$. At second order in perturbations, this is the behavior of the relativistic correction proportional to $F_2^R$. Thus, the relativistic correction to the second order kernel (and therefore the tree level bispectrum) is sensitive to initial conditions. Corrections to higher order kernels, $F_n^R$ with $n>2$, have a different time dependence, and are thus not expected to be sensitive to the initial conditions (up to higher orders in $aH/k$), given the scaling behavior of a universe dominated by a single fluid component. The fact that the tree level density bispectrum is sensitive to initial conditions was noticed first in~\cite{2010JCAP...05..004F}, where the contribution from an early period of matter domination is considered. More complete calculations were later performed by~\cite{Huang:2013qua, Tram:2016cpy}, and in order to incorporate these effects into our results one simply needs to fix our second order kernel to be the one obtained from these codes since, at second order in perturbation theory, the source $S$ vanishes in both gauges.

The transverse velocity $\textbf{u}_T$ is sourced at third order in perturbation theory, and in order to take it into account, we define the following kernels:
   \begin{equation}
    \textbf{u}_T(\bm{k},t) = H\sum_{n = 1}^\infty a^{n+2}(t)H^2(t)\int_{\bm{k}_1...\bm{k}_n}\!\!\!\!\!\!\!\! (2\pi)^3\delta_D(\bm{k}-\bm{k}_{1...n}) \textbf{G}^{T}_n(\bm{k}_1,\dots,\bm{k}_n) \delta_l(\bm{k}_1)\dots\delta_l(\bm{k}_n)\,.
\end{equation}    
We give an expression for them in appendix \ref{sec:SPTkernpoiss} and \ref{sec:SPTkernsync}.

The calculation of the kernels of perturbation theory is essentially analogous in both gauges, however two differences deserve a special mention. In Poisson gauge, already at first order in perturbation theory a relativistic correction is present (see equation \eqref{eq:relcorrorderone}) while in comoving gauge $F_1^R(\bm{k})=0$. The gravitational potential receives relativistic corrections in the comoving gauge (equation \eqref{eq:12} ). For convenience, we therefore introduce also relativistic kernels for $\psi$:
\begin{equation}
     \psi(\bm{k},t) = a^2(t)H^2(t)\sum_{n = 1}^\infty a^{n}(t)\int_{\bm{k}_1...\bm{k}_n}\!\!\!\!\!\!\!\! (2\pi)^3\delta_D(\bm{k}-\bm{k}_{1...n}) F^{\psi}_n(\bm{k}_1,\dots,\bm{k}_n) \delta_l(\bm{k}_1)\dots\delta_l(\bm{k}_n)\,,
\end{equation}
Expressions for the kernels $F_n^{\psi}$ are given in appendix \ref{sec:SPTkernsync} for the comoving gauge, while in Poisson gauge, the gravitational potential does not receive any relativistic correction and its kernels trivially follow from the Newtonian kernels $F_n$.

\section{Correlation functions} \label{sec:Correfunct}
We use the kernels derived in section \ref{sec:PTIC}, to compute correlation functions. Note that these do not correspond to actual observables since they do not include several projection effects such as the distortion of the physical volume or the propagation of light. As such, all Redshift Space Distortion Effects have been neglegled. However, the computation of these correlation functions will give us an insight into the importance of non-linear relativistic effects and additional subtleties such as their UV and IR behavior, to be discussed in detail in section \ref{section:divergences}. We study two and three point correlation functions of the density contrast $\delta$. In terms of them, the power spectrum and bispectrum are defined in the following way:
\begin{align}
    & \langle \delta(\bm{k}_1,t) \delta(\bm{k}_2,t)\rangle= (2\pi)^3 \delta_D(\bm{k}_1+\bm{k}_2) P(k_1,t)\,, \\
    & \langle \delta(\bm{k}_1,t) \delta(\bm{k}_2,t)\delta(\bm{k}_3,t)\rangle=(2\pi)^3 \delta_D(\bm{k}_1+\bm{k}_2+\bm{k}_3) B(k_1, k_2,k_3,t)\,.
\end{align}
Due to translational and rotational invariance $P$ is a function of the magnitude of one external momentum, $P(k)$, and $B$ is a function of the three magnitudes of the external momenta $B(k_1,k_2,k_3)$. 

In this section we present the results for the one-loop power spectrum and bispectrum. In the plots we present here, we have not performed the renormalization necessary to remove the dependence on the UV cutoff~\cite{Carrasco:2012cv}, since this would require fitting the resulting counterterms to a simulation, such as the one of~\cite{Adamek:2016zes}. However, we briefly comment how this would work in section \ref{section:divergences} for the case of the power spectrum, where we find that renormalization induces a relativistic correction to the noise and a term proportional to $\delta_l$ (which could come for example from a relativistic correction to the speed of sound). As discussed in section \ref{section:divergences}, IR divergences do not automatically cancel as in the Newtonian case, and one is forced to choose an IR cutoff (which encodes the fact that averages are necessarily taken to be over a finite region of the universe). For the plots of this section we take it to be $H_0$. All the quantities are plotted at redshift $z=0$.
\subsection{One-loop power spectrum}
\label{sec:oneloopPS}
\begin{figure}[t]
\centering
\begin{tabular}{cc}
\includegraphics[width=0.5\textwidth]{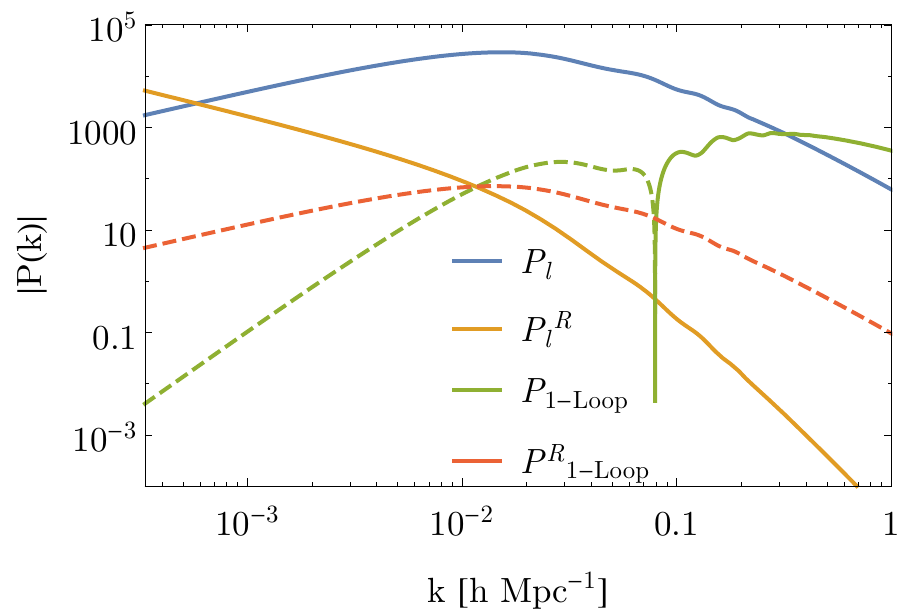} &
\includegraphics[width=0.5\textwidth]{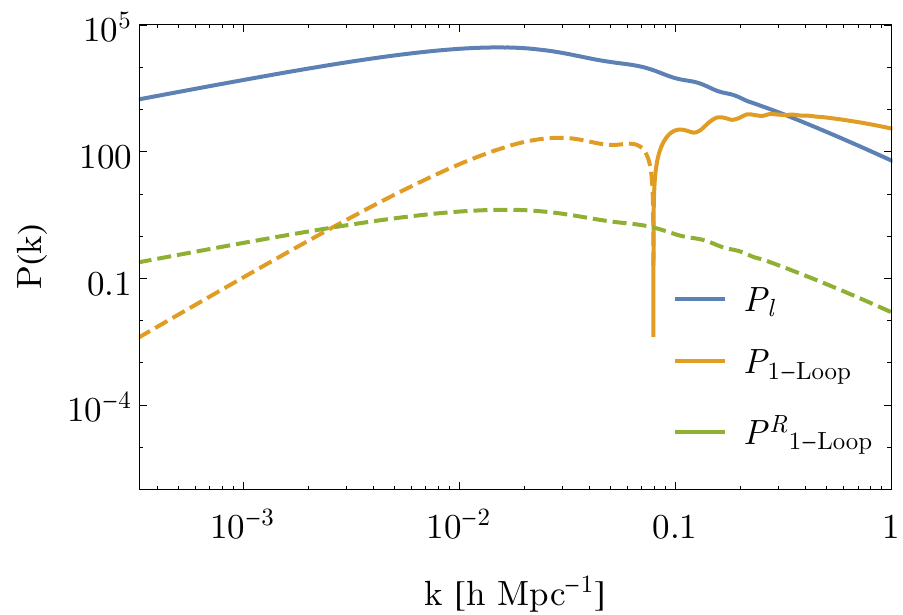}
\end{tabular}
\caption{Tree level and one-loop corrections to the linear power spectrum in Poisson gauge (left panel) and comoving gauge (right panel). As expected the relativistic corrections are negligible at small scales (large $k$) because Newtonian physics is a good approximation to the dynamics, while at large scales (small $k$), the universe is in the linear regime and relativistic corrections are subdominant. All quantities are at redshift $z=0$. For all plots, dashed lines represent a negative contribution.} \label{fig:1}
\end{figure}

The tree level relativistic correction is not present in the comoving gauge where $F_1^R = 0.$ The Newtonian one-loop correction to the power spectrum reads:
\begin{equation}
   P_{\text{1-loop}}(\bm{k},t)=a^4(t) \left(P_{13}(\bm{k})+P_{22}(\bm{k}) \right), \label{eq:p1loop} 
\end{equation}
and the leading relativistic corrections to the one-loop power spectrum are:
\begin{equation}
P_{\text{1-loop}}^R(\bm{k},t)=H_0^2a^3(t) \left(P_{13}^R(\bm{k})+P_{22}^R(\bm{k}) \right). \label{eq:p1Rloop}
\end{equation}
The two Newtonian contributions are:
\begin{align}
& P_{13}(\bm{k}) = 6P_{L}(k)\int_{\bm{q}} P_{L}(q)F_3(\bm{q},-\bm{q},\bm{k})\,, \label{eq:p13} \\
& P_{22}(\bm{k}) = 2 \int_{\bm{q}}  F_2^2(\bm{q},\bm{k}-\bm{q}) P_L(q) P_L(|\bm{k}-\bm{q}|)\,, \label{eq:p22}
\end{align}
while the relativistic corrections read:
\begin{align}
& P_{13}^R(\bm{k}) = 6P_{L}(k)\int_{\bm{q}} P_{L}(q)\left[F_3^R(\bm{q},-\bm{q},\bm{k}) + F_1^R(k) F_3(\bm{q},-\bm{q},\bm{k})\right], \label{eq:p13R} \\
& P_{22}^R(\bm{k}) = 4 \int_{\bm{q}}  F_2(\bm{q},\bm{k}-\bm{q}) F_2^R(\bm{q},\bm{k}-\bm{q})P_L(q) P_L(|\bm{k}-\bm{q}|)\,. \label{eq:p22R}
\end{align}
We numerically integrate the momentum dependant part of \eqref{eq:p1loop} and \eqref{eq:p1Rloop} and compare it to the standard numerical results for perturbation theory. Our results are presented in figures \ref{fig:1} in both gauges.
For the numerical integration of the power spectrum, we used three different codes, one based on C using the Cuba library~\cite{Hahn:2004fe}, along with one written in Mathematica. We also compared our results to the recent proposal based on the FFTLog formalism~\cite{Simonovic:2017mhp}, our results also agree with this method. As expected, the one-loop relativistic corrections to the power spectrum are negligible since relativity is important on large scales while one-loop corrections are important on small scales.
\subsection{One-loop bispectrum}

The Newtonian tree level bispectrum is:
\begin{equation}
B_{211}(k_1,k_2,k_3,t)=a^4(t)\left[ F_2(\bm{k}_1,\bm{k}_2) P_L(k_1) P_L(k_2)+\text{2 cyclic permutations}   \right]\,.
\end{equation}
The relativistic corrections to the tree level bispectrum are:
\begin{align}
B_{211}^R(k_1,k_2,k_3,t)=& a^3(t)H_0^2\left[  2 F_2^R(\bm{k}_1,\bm{k}_2) P_L(k_1) P_L(k_2)+2F_2(\bm{k}_1,\bm{k}_2)F_1^R(k_1) P_L(k_1) P_L(k_2) \right. \nonumber \\
& \left.  +2F_2(\bm{k}_1,\bm{k}_2)F_1^R(k_2) P_L(k_1) P_L(k_2)+\text{2 cyclic permutations} \right]\,.
\end{align}

\begin{figure}[t]
\centering
\begin{tabular}{cc}
\includegraphics[width=0.5\textwidth]{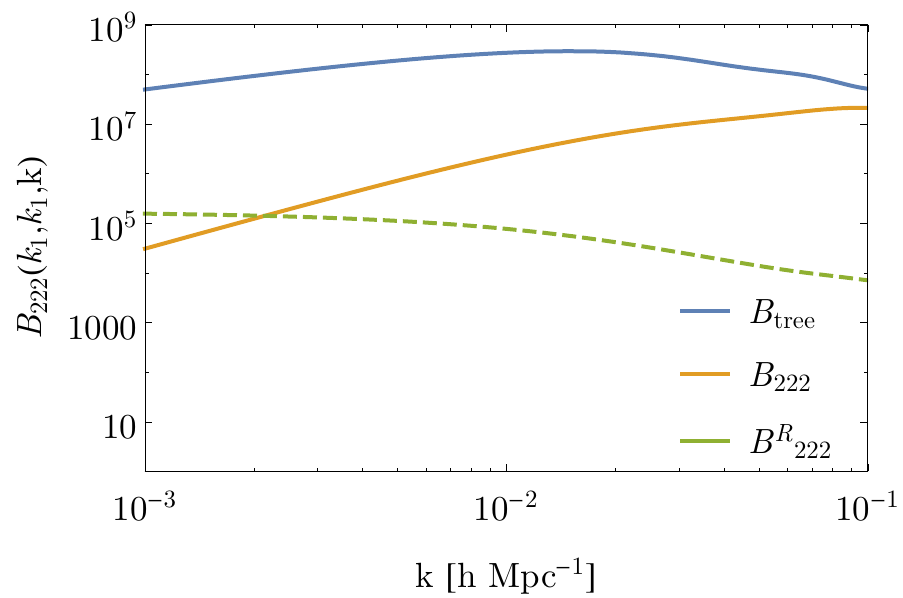} &
\includegraphics[width=0.5\textwidth]{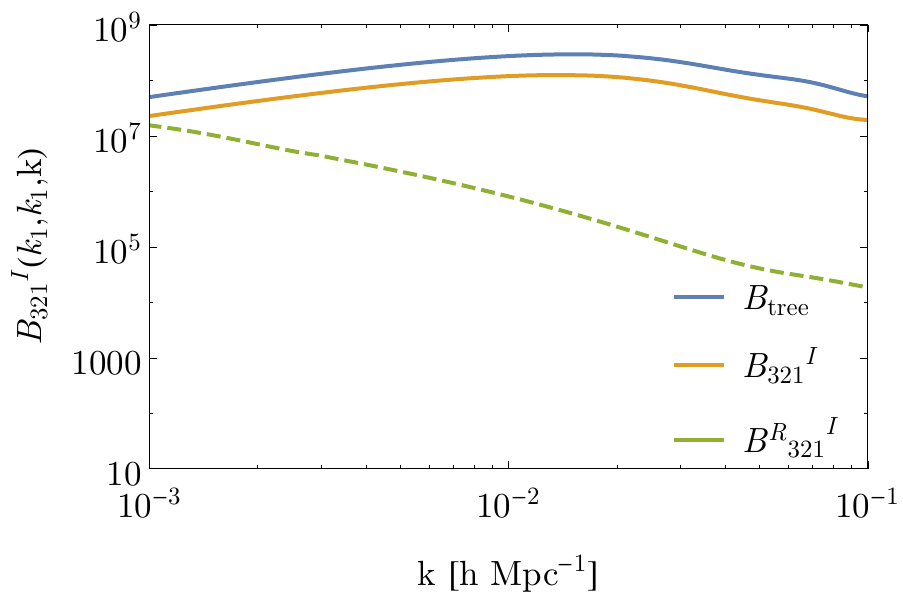} \\
\includegraphics[width=0.5\textwidth]{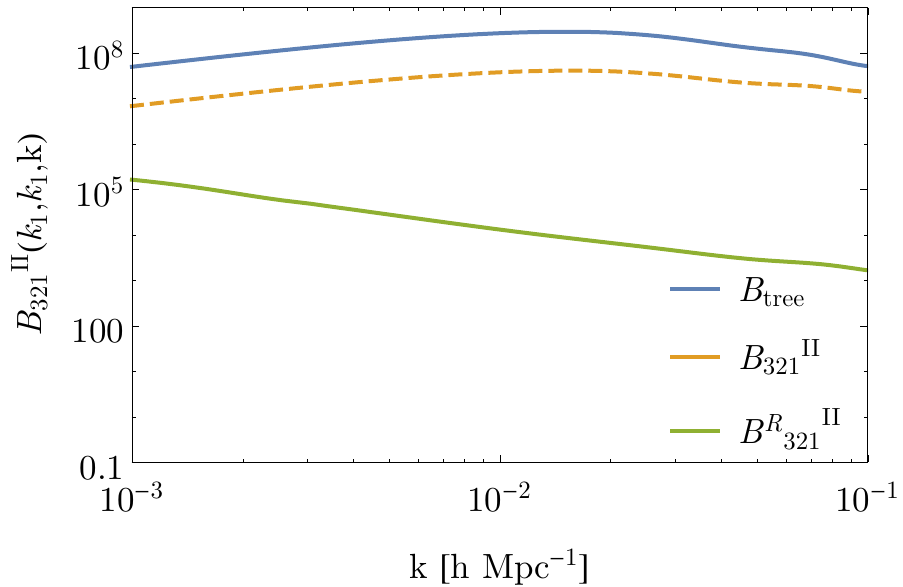} &
\includegraphics[width=0.5\textwidth]{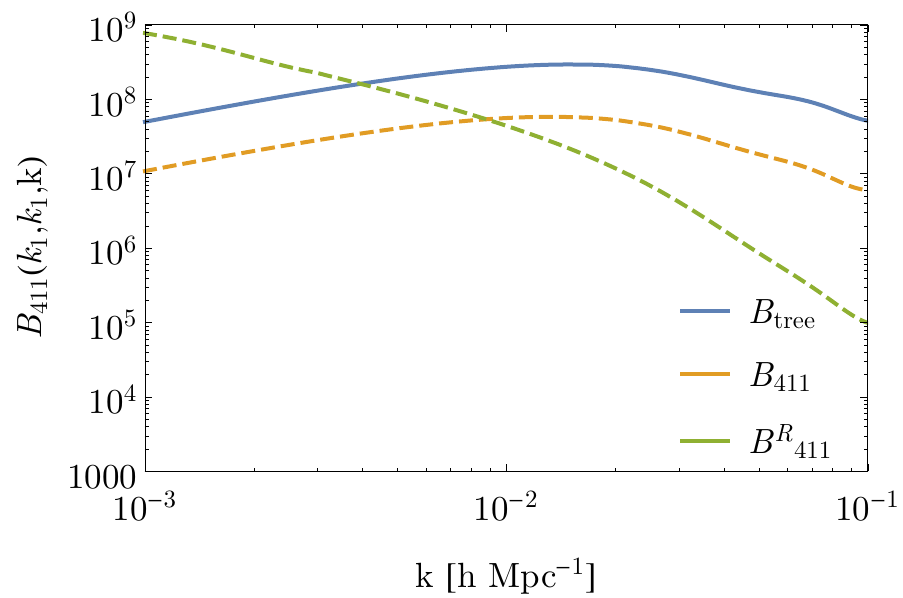} \\
\end{tabular}
\caption{In Poisson gauge $k_1=0.1$ h $\text{Mpc}^{-1}$, the 4 relativistic corrections to the 1-loop bispectrum $B(k_1,k_1,k)$ as a function of $k$ compared to its Newtonian counter part and to the tree level Newtonian result. All quantities are at redshift $z=0$.}  \label{fig:squeezedPoisson}
\end{figure}

\begin{figure}[t]
\centering
\begin{tabular}{cc}
\includegraphics[width=0.5\textwidth]{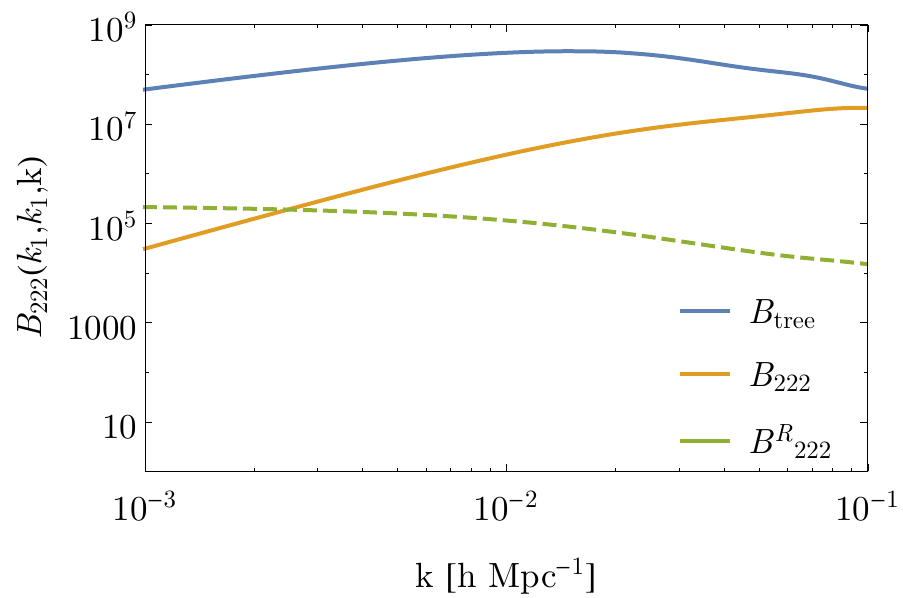} &
\includegraphics[width=0.5\textwidth]{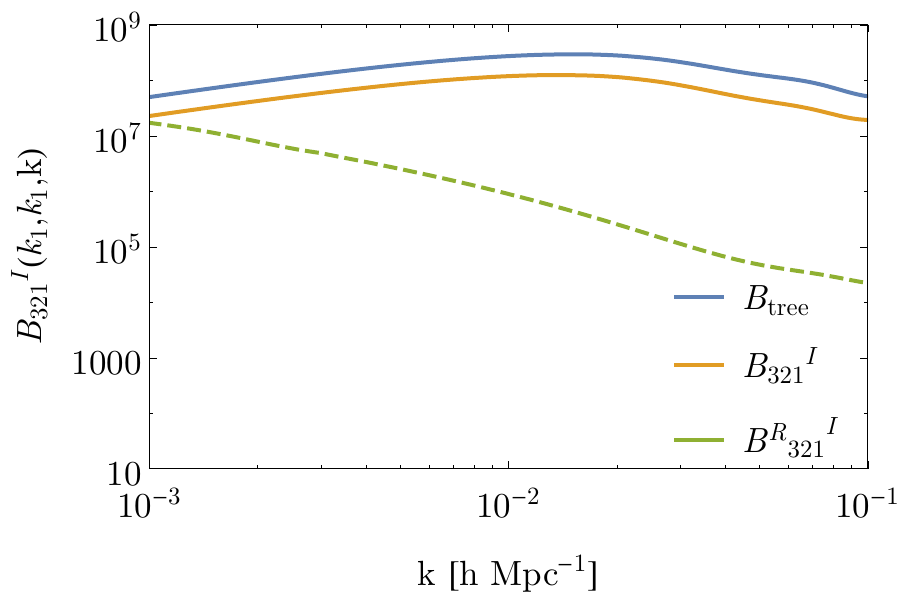} \\
\includegraphics[width=0.5\textwidth]{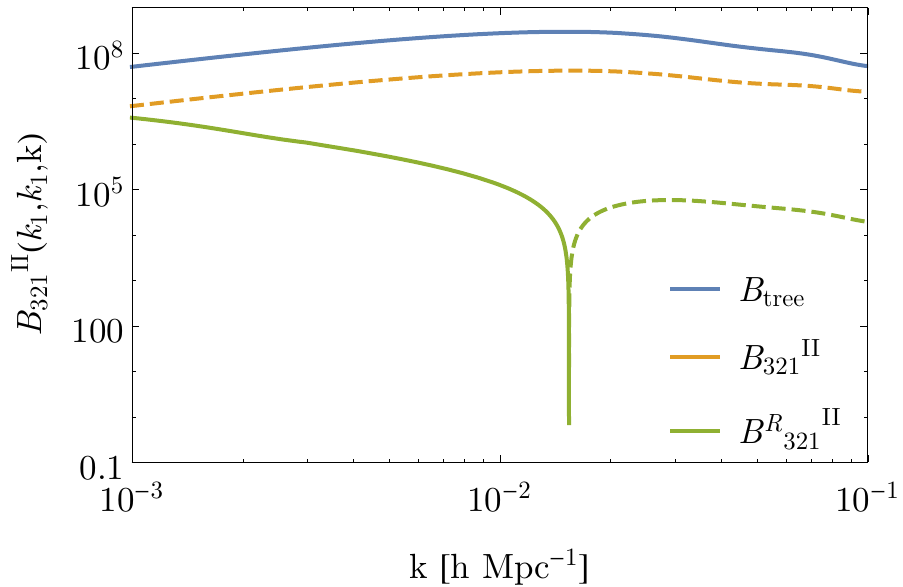} &
\includegraphics[width=0.5\textwidth]{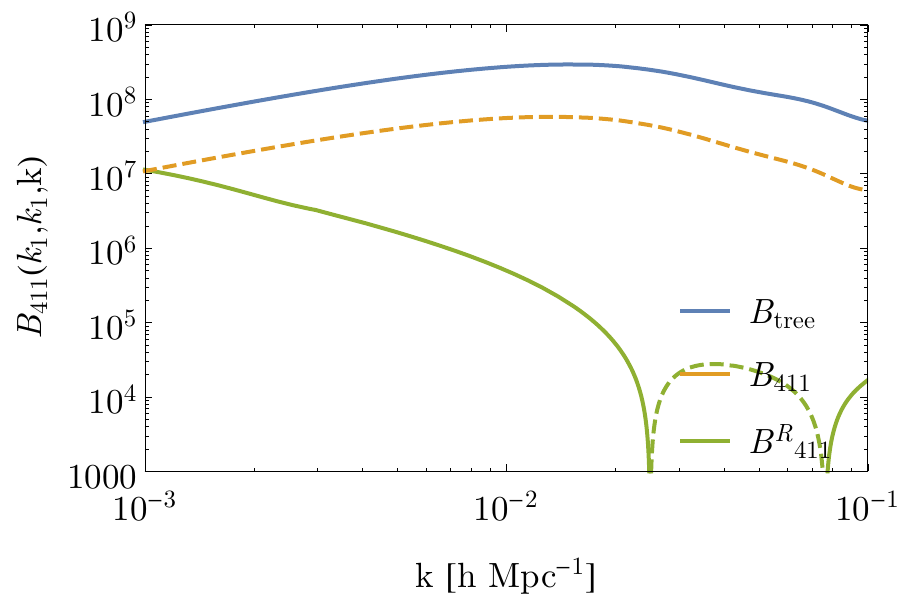} \\
\end{tabular}
\caption{In comoving gauge $k_1=0.1$ h $\text{Mpc}^{-1}$, the 4 relativistic corrections to the 1-loop bispectrum $B(k_1,k_1,k)$ as a function of $k$ compared to its Newtonian counter part and to the tree level Newtonian result. All quantities are at redshift $z=0$.}  \label{fig:squeezedsynchonous}
\end{figure}

The 1-loop bispectrum is composed of 4 diagramatic pieces:
\begin{align}
B_{\text{1-loop}}(k_1,k_2,k_3,t)=a^6(t)\left[B_{222}(k_1,k_2,k_3)+B_{321}^I(k_1,k_2,k_3)+B_{321}^{II}(k_1,k_2,k_3)+B_{411}(k_1,k_2,k_3) \right],
\end{align}
where
\begin{align}
& B_{222}(k_1,k_2,k_3) = 8 \int_{\bm{q}} F_2 (\bm{q},\bm{k}_1-\bm{q}) F_2(\bm{k}_1-\bm{q},\bm{k}_2+\bm{q}) F_2(\bm{k}_2+\bm{q},-\bm{q})\nonumber \\
& \times P_{L}(q) P_{L}(|\bm{k}_1-\bm{q}|) P_{L}(|\bm{k}_2+\bm{q}|)\,. \\
& B_{321}^I(k_1,k_2,k_3) = 6 P_{L}(k_1) \int_{\bm{q}} F_3(\bm{q},\bm{k}_2-\bm{q},\bm{k}_1) F_2(\bm{q},\bm{k}_2-\bm{q}) P_{L}(q) P_{L}(|\bm{k}_2-\bm{q}|) + 5\;\text{permutations}\,. \\
& B_{321}^{II}(k_1,k_2,k_3) = F_2(\bm{k}_1,\bm{k}_2) P_{L}(k_1) P_{13}(\bm{k}_2) + 5\;\text{permutations}\,. \\
& B_{411}(k_1,k_2,k_3) = 12 P_{L}(k_1) P_{L}(k_2) \int_{\bm{q}} F_4(\bm{q},-\bm{q},-\bm{k}_1,-\bm{k}_2) P_{L}(q) + \text{2 cyclic permutations}\,.
\end{align}
The relativistic corrections have a specific scaling with time:
\begin{align}
B_{\text{1-loop}}^R(k_1,k_2,k_3,t)=H_0^2a^5(t)&\left(  B_{222}^R(k_1,k_2,k_3)+B_{321}^{I,R}(k_1,k_2,k_3) \right. \\
& \left. +B_{321}^{II,R}(k_1,k_2,k_3)+B_{411}^R(k_1,k_2,k_3) \right)\,.
\end{align}
The expression for each diagramatic piece follows from the non-relativistic ones where one of the kernels is relativistic:
\begin{align}
& B_{222}^R(k_1,k_2,k_3) = 8 \int_{\bm{q}} \Big[ F_2^R (\bm{q},\bm{k}_1-\bm{q}) F_2(\bm{k}_1-\bm{q},\bm{k}_2+\bm{q}) F_2(\bm{k}_2+\bm{q},-\bm{q})\nonumber \\
& +F_2 (\bm{q},\bm{k}_1-\bm{q}) F_2^R(\bm{k}_1-\bm{q},\bm{k}_2+\bm{q}) F_2(\bm{k}_2+\bm{q},-\bm{q})  \nonumber \\
& + F_2 (\bm{q},\bm{k}_1-\bm{q}) F_2(\bm{k}_1-\bm{q},\bm{k}_2+\bm{q}) F_2^R(\bm{k}_2+\bm{q},-\bm{q})\Big] P_{L}(q) P_{L}(|\bm{k}_1-\bm{q}|) P_{L}(|\bm{k}_2+\bm{q}|)\,, \\
& B_{321}^{I,R}(k_1,k_2,k_3) = 6 P_{L}(k_1)  \int_{\bm{q}}\left[ F_3^R(\bm{q},\bm{k}_2-\bm{q},\bm{k}_1) F_2(\bm{q},\bm{k}_2-\bm{q})+F_3(\bm{q},\bm{k}_2-\bm{q},\bm{k}_1) F_2^R(\bm{q},\bm{k}_2-\bm{q}) \right] \nonumber \\
& \times P_{L}(q) P_{L}(|\bm{k}_2-\bm{q}|) +6 P_{L}(k_1) F_1^R(k_1)  \nonumber \\
& \times \int_{\bm{q}} F_3(\bm{q},\bm{k}_2-\bm{q},\bm{k}_1) F_2(\bm{q},\bm{k}_2-\bm{q}) P_{L}(q) P_{L}(|\bm{k}_2-\bm{q}|) + 5\;\text{permutations}\,, \\
& B_{321}^{II,R}(k_1,k_2,k_3) =6 F_2(\bm{k}_1,\bm{k}_2) P_{L}(k_1) P_{L}(k)\int_{\bm{q}} P_{L}(q)F_3^R(\bm{q},-\bm{q},\bm{k}) + F_2^R(\bm{k}_1,\bm{k}_2) P_{L}(k_1) P_{13}(\bm{k}_2) \nonumber \\ 
& +F_2(\bm{k}_1,\bm{k}_2) F_1^R(k_1) P_{L}(k_1) P_{13}(\bm{k}_2) + 5\;\text{permutations}\,, \\
& B_{411}^R(k_1,k_2,k_3)   = 12 P_{L}(k_1) P_{L}(k_2) \left[ \int_{\bm{q}} F_4^{R}(\bm{q},-\bm{q},-\bm{k}_1,-\bm{k}_2) P_{L}(q) +F_1^R(k_1) \int_{\bm{q}} F_4(\bm{q},-\bm{q},-\bm{k}_1,-\bm{k}_2) P_{L}(q) \right. \nonumber \\ 
& \left. +F_1^R(k_2)\int_{\bm{q}} F_4(\bm{q},-\bm{q},-\bm{k}_1,-\bm{k}_2) P_{L}(q)+\text{2 cyclic permutations}\right]\,.  
\end{align}
In figures \ref{fig:squeezedPoisson} and \ref{fig:squeezedsynchonous}, for the two gauges considered in this article, we present the 4 contributions to the relativistic corrections to the one-loop bispectrum in the squeezed limit, where two sides of the triangle have been fixed to be $k_1 = 0.1\,\text{Mpc}^{-1}h$. We notice that, as expected, some relativistic loop corrections to the bispectrum are as important in the squeezed limit as the Newtonian loop corrections. This is because the bispectrum in that limit couples a large relativistic scale with short non-linear scales. In Poisson gauge, there is even a break down of perturbation theory for very squeezed triangles, which could be cured if the short-scale physics is properly renormalized. Using gauge invariant quantities cures this pathology; see Ref.~\cite{Biern:2016kys} for an explicit example.

\begin{figure}[t]
\centering
\begin{tabular}{cc}
\includegraphics[width=0.5\textwidth]{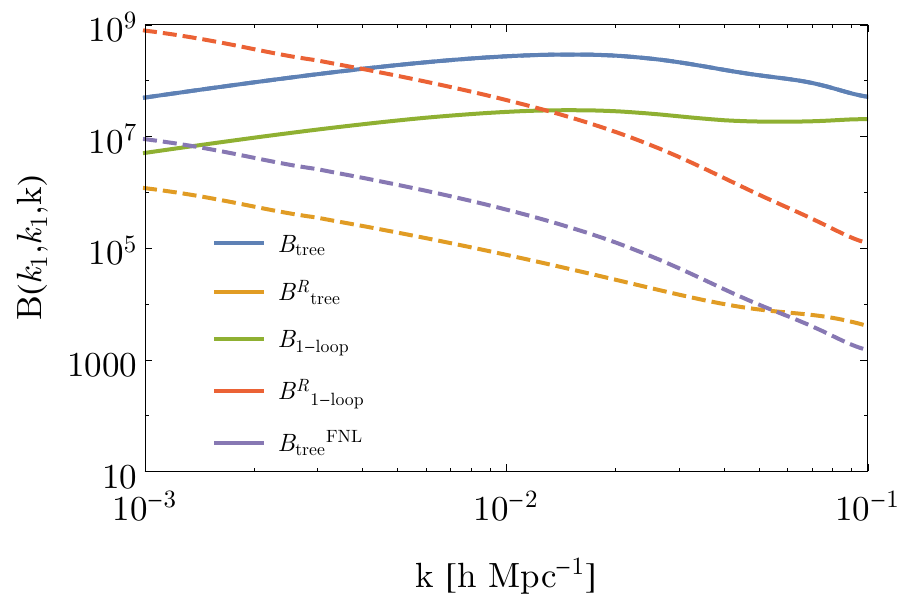} &
\includegraphics[width=0.5\textwidth]{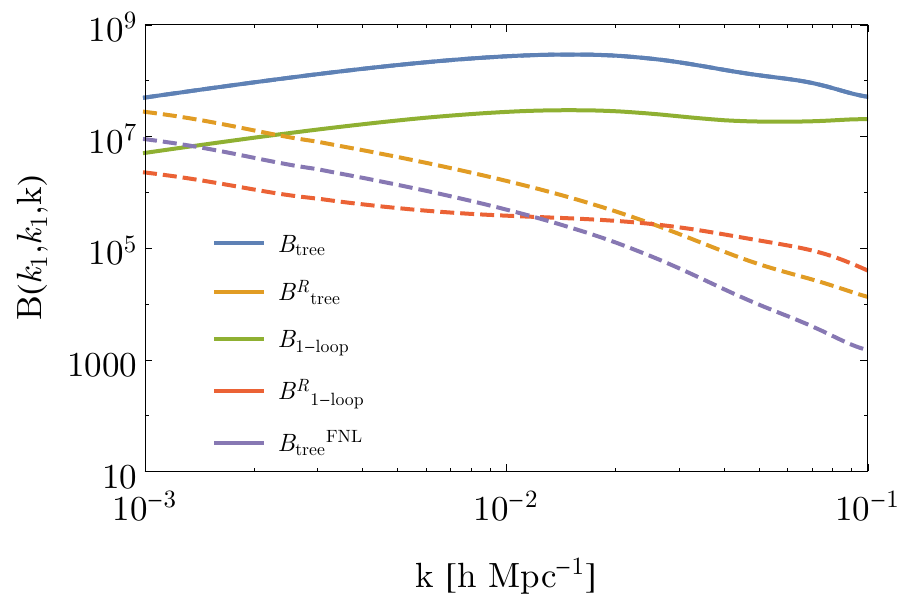}
\end{tabular}
\caption{In the squeezed configuration: $B(k_1,k_1,k)$, one-loop corrections to the linear bispectrum in Poisson (left panel) and comoving (right panel) gauge, with $k_1=0.1$ h $\text{Mpc}^{-1}$. Observe that in the Poisson gauge, $B_{411}^R$ becomes larger than the tree-level Newtonian results and the perturbation theory breaks down in the IR.}  \label{fig:SqueezedTot}
\end{figure}

In figure \ref{fig:SqueezedTot}, we present the total one-loop relativistic corrections to the 1-loop bispectrum. For comparison, we also plot the effect of a primordial non-Gaussianity of the local type on the density contrast bispectrum at tree level. In a universe which has been matter-dominated throughout its history, which we used for simplicity, it is
\begin{align}
   &  \delta(t,\bm{k})_{NG}=a^2(t)\int (2\pi)^3 \delta_D(\bm{k}-\bm{k}_{12}) F_2^{f_{\text{NL}}}(\bm{k}_1,\bm{k}_2) \delta_l(k_1) \delta_l(k_2)\,, \\
   & F_2^{f_{\text{NL}}}(\bm{k}_1,\bm{k}_2)=-\frac{3H_0^2 f_{\text{NL}}^{loc}}{2}\left(\frac{\bm{k}_1^2+\bm{k}_2^2}{\bm{k}_1^2\bm{k}_2^2}+2\frac{\bm{k}_1\cdot\bm{k}_2}{\bm{k}_1^2\bm{k}_2^2} \right)\,.
\end{align}
In the plot we set $f_{\text{NL}}^{loc} = 1$. The relativistic corrections at tree level and one-loop have the same behavior as the primordial signal as a function of the momentum being squeezed. Let us also note that the relativistic correction to the second order kernel, proportional to $F_2^R$ in eq.~\eqref{eq:expansion}, has \textit{the same time dependence} as a primordial non-Gaussianity signal. Though we don't plot it, primordial non-Gaussianity would also enter in the loop corrections, and the time dependence of this effect is the same as the one-loop relativistic corrections. We will carry out a more detailed analysis of the degeneracy of relativistic effects with the primordial signal in the future, once the full calculation (including biasing and photon propagation) has been performed. 

\begin{figure}[t]
\centering
\begin{tabular}{cc}
\includegraphics[width=0.5\textwidth]{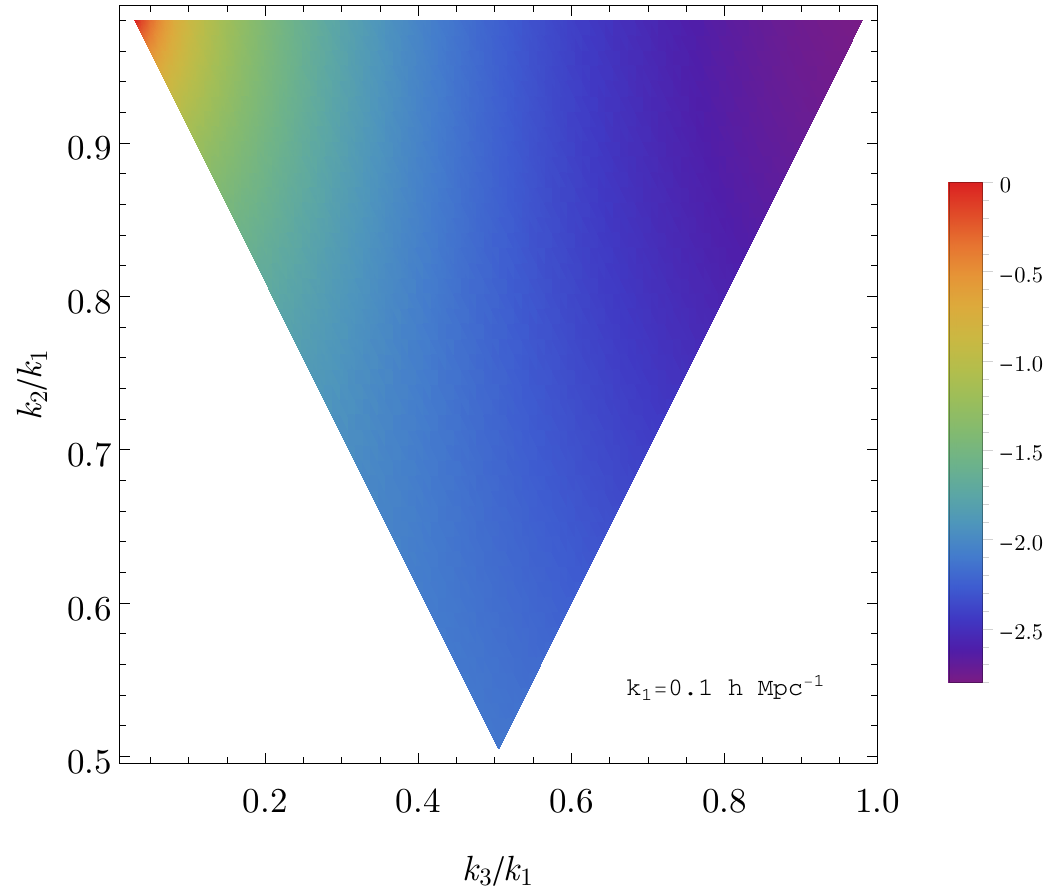} &
\includegraphics[width=0.5\textwidth]{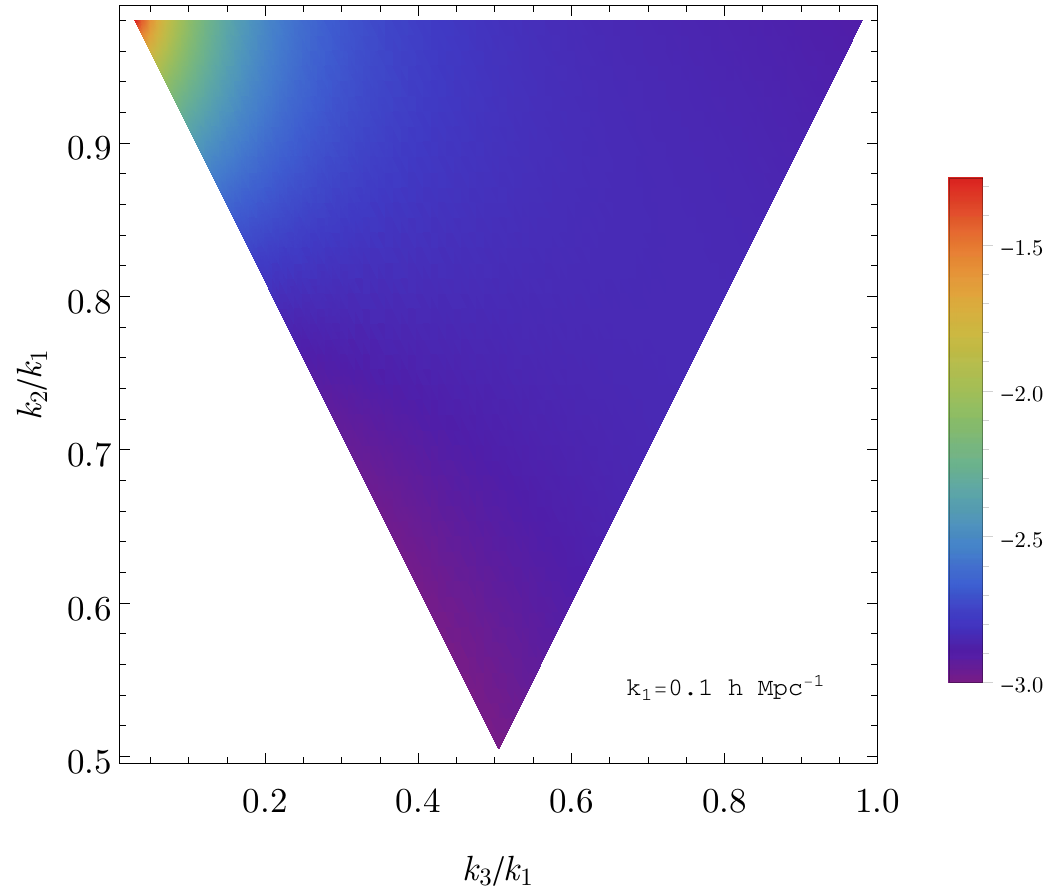}
\end{tabular}
\caption{Logarithm of the ratio of the total (tree-level plus one-loop) relativistic bispectrum $B^R(k_1,k_2,k_3)$ to its Newtonian counter part. We consider arbitrary triangle configuration in the Poisson gauge (left panel) and comoving gauge (right panel).  $k_1=0.1$ h $\text{Mpc}^{-1}$. All quantities are at redshift $z=0$.}  \label{fig:4}
\end{figure}

Finally, we plot in figure \ref{fig:4} the ratio between the relativistic and the Newtonian bispectra
\begin{equation}
\label{eq:ratioB}
  \frac{  B(k_1,k_2,k_3)^R_{211}+B(k_1,k_2,k_3)^R_{\text{1-loop}}}{ B(k_1,k_2,k_3)^N_{211}+B(k_1,k_2,k_3)^N_{\text{1-loop}}}\,.
\end{equation}
where we assumed $k_1<k_2<k_3$. In order to span all the triangle configurations, we fixed $k_1$ and plot \eqref{eq:ratioB} as a function of $k_3/k_1$ and $k_2/k_1$. As promised, the squeezed limit of the bispectrum is the one which is more affected by the relativistic corrections. For very squeezed configurations, this ratio is larger than $1$ in the Poisson gauge.


\section{IR and UV behavior of the loop integrals}
\label{section:divergences}

In this section we study the IR behavior of the loop integrals introduced in the previous section. This behavior is markedly different from the Newtonian case. In particular, the appropriate definition of perturbation theory requires the background to be renormalized, and IR ``divergences'' do not cancel automatically.
We then briefly comment on the UV behavior of the same integrals. A detailed discussion of UV renormalization and fit of the associated counterterms to simulations is beyond the scope of this paper. 

\subsection{Renormalization of the background}

In order for perturbation theory to be well defined, the average of fluctuations should be zero
\begin{equation}
\langle \delta \rangle =0\,,\quad \langle \theta \rangle =0\,,\quad \langle \psi \rangle = 0\,. \label{eq:averaget}
\end{equation}
We will see that this condition is not realized in the relativistic case, and the background needs to be appropriately redefined (``renormalized''). Let us begin by taking the average of the second order expansion for the density contrast
\begin{equation}
    \langle\delta(\bm{k})\rangle = (2\pi)^3\delta_D(\bm{k})a^2\int_{\bm q} F_2(\bm{q},- \bm{q}) P(q)\,.
\end{equation}
In the Newtonian case this expression vanishes since
\begin{equation}
    \lim_{|\bm{q}_1 + \bm{q}_2|\rightarrow 0}F_2(\bm{q}_1, \bm{q}_2) \propto (\bm{q}_1 + \bm{q}_2)^2\,.
\end{equation}
In the relativistic case this is no longer true since $F_2^R(\bm{q}, -\bm{q})$ does not vanish, as can be explicitly checked in expressions \eqref{F2RPoiss}, and \eqref{F2RSync} \footnote{As mentioned in section \ref{sec:PTIC}, these quadratic kernels are fixed by initial conditions. One can choose initial conditions such that they vanish in this limit, but such initial conditions are not adiabatic and their physical meaning is unclear.}
\begin{equation}
    \langle\delta(\bm{k})\rangle = (2\pi)^3 \delta_D(\bm{k}) A_{\delta} a^4 H^2 \int dq\, P_L(q)\,,\\
\end{equation}
where $A_\delta$ is a numerical coefficient. In Poisson gauge it reads $A_\delta = -137/56 \pi^2$, while in comoving gauge it is $A_\delta = -25/8\pi^2$. Moreover, even the average gravitational potential is different from zero in both gauges 
\begin{align}
    &\langle\phi(\bm{k})\rangle = (2\pi)^3\delta_D(\bm{k})\frac{a^4 H^2}{56 \pi^2}\int dq\,P_L(q)\quad \text{(Poisson)}\,,\\
    &\langle\psi(\bm{k})\rangle = (2\pi)^3\delta_D(\bm{k})\frac{5 a^4 H^2}{24 \pi^2}\int dq\,P_L(q)\quad \text{(comoving)}\,.
\end{align}
In order to cancel these tadpoles we need to redefine the background over which we perturb.\footnote{In actual observations the Dirac delta in front of these results will be replaced by the window function of the survey, which goes to zero for scales smaller than the largest scale observed $k > k_f$, and is a finite quantity.} The non-vanishing of the average curvature perturbation was noticed in \cite{Boubekeur:2008kn}, and first addressed in \cite{Baumann:2010tm}. We will reabsorb them explicitly in the Poisson gauge, with the discussion in comoving gauge being completely analogous.
In order to do that, let us start by rewriting the background density and pressure
\begin{align}
&\rho \rightarrow \bar{\rho}(1 + \langle\delta\rangle)\,, \\
& P \rightarrow 0+P\,,
\end{align}
where the pressure will be determined by requiring that the background metric absorbs $\langle\phi\rangle$. This induces a modified Hubble parameter through the Friedmann equation
\begin{equation}
2\Delta H - 2\langle\dot{\phi}\rangle = H(\langle \delta \rangle + 2\langle\phi\rangle) \,.
\label{deltaH}
\end{equation}
We can now write the background continuity equation, at lowest order in the new quantities, and after subtracting the unmodified background equations:
\begin{equation}
\langle\dot{\delta}\rangle + 3H\langle\delta\rangle + 3\Delta H +\frac{3HP}{\rho}=\langle\dot{\phi}\rangle + 6H \langle\phi\rangle\,.
\label{pressure}
\end{equation}
Note that if $\langle\phi\rangle$ were constant in time, it would correspond to a simple redefinition of coordinates, as it would be simply an adiabatic mode~\cite{Weinberg:2003sw}. 
Equations \eqref{deltaH} and \eqref{pressure} fix the modified Hubble parameter and pressure in terms of $\langle \delta \rangle$, and $\langle\phi\rangle$. These corrected quantities can be computed explicitly and are seen to be of the order of $10^{-5}$, and therefore negligible in all preceding calculations. On the other hand, several higher order kernels appearing in loop integrals (such as for example $F_4^R(\bm{k}_1, \bm{k_2}, \bm{q}, -\bm{q})$ inside $B_{411}$) contain factors of $F_2^R(\bm{q},-\bm{q})$, $G_2^R(\bm{q}, -\bm{q})$ or $F_2^\phi(\bm{q}, -\bm{q})$ multiplying quantities that diverge such as $\alpha(\bm{k}, \bm{q} - \bm{q})$. We set such terms to zero, understanding that they vanish exactly once the background has been appropriately renormalized.

\subsection{IR behavior}

Loop integrals can depend on the IR cutoff chosen. We focus here on the power spectrum for simplicity; the discussion of the bispectrum is completely analogous. The IR behavior is given by equation (\ref{eq:p1Rloop})
\begin{equation}
     \lim_{q\rightarrow 0}P_{22}(t,\bm{k}) = \left(\frac{1}{12\pi^2}k^2 + C_{22}^{IR}H^2a^2\right)  a^4P_L(k)\intop dq P_L(q)\,,
     \label{P22IR}
\end{equation}
\begin{equation}
    \lim_{q\rightarrow{0}}P_{13}(t,\bm{k}) = \left(-\frac{1}{6\pi^2}k^2 + C_{13}^{IR} H^2 a^2\right)a^4P_L(k)\intop dq P_L(q)\,.
    \label{P13IR}
\end{equation}
The first terms correspond to the usual IR divergence of these loops, while the second terms correspond to the relativistic corrections.
$C_{22}$ and $C_{13}$ are numerical factors which are different in Poisson and comoving gauge. 
In Poisson gauge, the coefficients read $C_{13}^{IR}=-\frac{2521}{42 \pi^2} $ and $C_{22}^{IR}=-\frac{1}{84 \pi^2} $ while in comoving gauge they are $C_{13}^{IR}=-\frac{405}{28\pi^2}$ and $C_{22}^{IR}=-\frac{95}{84\pi^2}$.

In the Newtonian case, the IR dependence of both contributions to the 1-loop power spectrum cancel each other~\cite{Scoccimarro:1995if}, after noticing that $P_{22}$ contains divergences both when $q\rightarrow 0$ or $|\bm{k} - \bm{q}| \rightarrow 0$. This cancellation is due to the equivalence principle (often called ``Galilean invariance'' in this context). Following the arguments of~\cite{Peloso:2016qdr}, the IR limit of the loop integral can be written in terms of the response of the full power spectrum to a change in the linear power spectrum, that is
\begin{equation}
    \lim_{q \rightarrow 0} (P_{13} + P_{22}) = \frac{1}{2P(q)}\int_{\bm{q}}\langle \delta(\bm{k})\delta(\bm{q})\delta(-\bm{q})\delta(-\bm{k})\rangle'\,,
\end{equation}
where the prime indicates that the Dirac delta in front of the correlation function has been ignored. Due to the equivalence principle, the integrand vanishes in the limit $q \rightarrow 0$ in the Newtonian case~\cite{Peloso:2013zw,Kehagias:2013yd}. On the other hand, this double squeezed limit is different from zero in the relativistic case~\cite{Creminelli:2013mca},\footnote{In the Newtonian case, there is no effect on short wavelength physics from a long-wavelength perturbation that can be approximated as a constant potential, since the zero of the potential can be redefined arbitrarily. The effect of a long-wavelength gradient (constant force) corresponds to an accelerated frame. In the relativistic case, the effect of such long-wavelength perturbations is still unphysical. However, this Newtonian intuition does not straightforwardly apply, the relations expressing the unphysicality of the long mode are more contrived, and give a non-zero squeezed limit for correlation functions of density perturbations.} and the IR ``divergences'' do not cancel as seen in equations \eqref{P22IR} and \eqref{P13IR}.

Due to the non-cancellation of the IR dependence, the result of loop integrals depends on the IR cutoff chosen. This can be understood physically as due to the fact that actual observations have a limited resolution in momentum space $k_f$ corresponding to the largest scale measured. All averages are taken with this resolution, meaning that all information on scales $k < k_f$ is smoothed out. One should thus convolve all integrals with a window function containing information about the shape of the survey used. Since this IR dependence is weak in our results, the details of the window function are not expected to be important and we simply work with an IR cutoff of $H_0$.\footnote{One may worry that this is not gauge invariant since we have chosen a specific coordinate system to fix the cutoff. This gauge-dependence simply corresponds to describing the same physical region in different coordinate system (for example, the region $k > H_0$ in comoving gauge will have a more complicated description in Poisson gauge). When using the window function of an actual survey, the gauge dependence will disappear.} We thus cut off any part of the region of integration where one of the arguments of the kernel, or their sum, is smaller than $H_0$.\footnote{More precisely, the contribution of those regions to the final result is a ``supersample'' redefinition of the observed power spectrum or bispectrum, see~\cite{Takada:2013bfn, Chan:2017fiv}.}

\subsection{UV behavior}

The description of matter as a fluid breaks down at small scales where highly non-linear phenomena such as shell-crossing take place. This makes the description of its evolution with perturbation theory incomplete, and small scale physics needs to be appropriately integrated out. This fact is manifest in the dependence of loop integrals on the UV cutoff. One can extend standard perturbation theory by including additional physics in the stress energy tensor that provide counterterms to renormalize this cutoff dependence, in what is called the Effective Field Theory of Large Scale Structure~\cite{Baumann:2010tm,Carrasco:2012cv} (see also~\cite{Pajer:2013jj} for additional discussion of the renormalization procedure). Let us apply the same logic to the relativistic one-loop power spectrum.

The UV dependence in the one-loop power spectrum integrals \eqref{eq:p1Rloop} for the relativistic corrections to $P_{22}$ and $P_{13}$ can be obtained by taking the large $q$ limit in the integrands
\begin{equation}
 \lim_{q\rightarrow \infty}P_{22}^{R,UV} = C_{22}^{UV}a^6 H^2 k^2\int dq \frac{P_L(q)^2}{q^2}\,,
\end{equation}
\begin{equation}
    \lim_{q\rightarrow \infty}P_{13}^{R,UV}= C_{13}^{UV} a^6 H^2 P_L(k)\int dq P_L(q)\,.
\end{equation}
In Poisson gauge, the coefficients read $C_{13}^{UV}=-\frac{10597}{98 \pi^2} $ and  $C_{22}^{UV}=\frac{233}{588 \pi^2} $ while in comoving gauge they are $C_{13}^{UV}=-\frac{405}{28\pi^2}$ and $C_{22}^{UV}=\frac{25}{84\pi^2}$.
The UV part of the $P_{22}^R$ integral has a similar behavior to its Newtonian counterpart. It is not proportional to the power spectrum of $\delta_l$, suggesting that it can also be renormalized by a noise contribution. The Newtonian noise term scales like $k^4$, while this one scales as $a^2 H^2 k^2$ as appropriate for a relativistic correction. 
A similar discussion holds for $P_{13}^R$, which has a similar behavior to its Newtonian counterpart though suppressed by a factor of $a^2H^2/k^2$. As such it can be absorbed if we add a counterterm proportional to $\delta_l$ without derivatives in the effective stress-energy tensor, and can be seen as being a relativistic correction to the speed of sound. 

The bispectrum case will require more counterterms~\cite{Baldauf:2014qfa,Angulo:2014tfa}, but a similar pattern is expected to hold, with similar behavior to the Newtonian case but with an appropriate relativistic suppression. These coefficients cannot be predicted from theory and need to be measured in  simulations, such as the ones of~\cite{Adamek:2016zes}, which is beyond the scope of this work. As such, we do not renormalize the UV behavior of loop integrals in the plots presented in section \ref{sec:Correfunct}, and leave a more systematic discussion of this for future work.

\section{Conclusions}
\label{sec:conclusions}

We have computed the one-loop power spectrum and bispectrum for the matter density contrast in general relativity. We did this by using the weak field approximation and first writing equations which are non-perturbative in the density contrast but perturbative in velocities and the gravitational potential. This was done in section \ref{sec:Einsteinandfluid}. The usual non-linear terms in the resulting equations force us to take a perturbative approach similar to the standard perturbation theory of the Large Scale Structure, as explained in section \ref{sec:PTIC}. This could be further improved by appropriately renormalizing the behavior of the loop integrals in the UV, but that is beyond the scope of this work, though we do comment on this in section \ref{section:divergences}, where we also study the IR behavior of said integrals. We plot the results for the one-loop matter power spectrum and bispectrum in general relativity in section \ref{sec:Correfunct}. As expected, we find that non-linear relativistic corrections are important in the squeezed limit of the bispectrum and are crucial for distinguishing a primordial signal from projection effects. This is due to the fact that the squeezed limit can couple a large relativistic scale with a small non-linear one. Moreover, we find the time dependence of the relativistic corrections to the bispectrum to be the same as that of a primordial non-Gaussianity signal, such that measurements at different redshifts would not be enough to break the degeneracy. 

The most important effects, going as $1/k^2$ in the squeezed limit are ``projection effects'', in the sense that they represent the correlation between a long-wavelength gravitational potential and short scale physics. These should cancel if we use coordinates which are locally Minkowski (see for example \cite{Pajer:2013ana, Dai:2015rda}). Indeed, we explicitly checked that the second and third order kernels satisfy the ``dilation'' consistency relation \cite{Creminelli:2013mca}, which encodes the fact that a long-wavelength gravitational potential has no physical effect on the small scale dynamics. We have also checked that the third order kernels satisfy the ``conformal'' consistency relation. For details on this check, see Appendix \ref{app:consistency}. However, these projection effects need to be accounted for when computing an observable quantity since not including them will give biased results when analyzing the data.

There are some qualitative differences with the Newtonian calculation:
\begin{itemize}
    \item The second-order kernel $F_2^R$ depends on the initial conditions as discussed in section \ref{sec:PTIC}. In order to compute the correct bispectrum generated by a canonical single field slow-roll model of inflation, this requires initial conditions to be fixed at second order. To our knowledge this is not currently done in relativistic simulations. If the bispectrum from these simulations is used in analyzing the data, one would incorrectly conclude that primordial non-Gaussianity has been observed.\footnote{The initial conditions in simulations are fixed using Lagrangian perturbation theory. This will include some non-linear Eulerian terms, but it does not reproduce the full second order kernel at the initial time.}
    \item As already noted in~\cite{Boubekeur:2008kn}, and discussed in section \ref{section:divergences}, we find that  at second order in perturbations $\langle \delta\rangle$ and $\langle \psi\rangle$ do not vanish a priori once relativistic corrections are taken into account. For perturbation theory to be well defined this requires an additional renormalization of the background by including a background pressure, which we find to be of order $\mathcal{O}(10^{-5})$ and not relevant for observations. This is compatible with the discussion in~\cite{Baumann:2010tm}. 
    \item IR divergences do not cancel automatically as in the Newtonian case and an IR cutoff is necessary. Contrary to the UV case, the dependence on the IR cutoff is physical in the sense that it represents the fact that we must define our averages over a finite region of the universe.  A detailed description of this is presented in section \ref{section:divergences}. 
\end{itemize}

Our calculation is necessarily gauge dependent since we have not yet computed an actual observable. In order to do that we need to consider galaxy biasing and the trajectory of a photon from the galaxy to the telescope~\cite{Yoo:2009au,Yoo:2010ni,Bonvin:2011bg,Jeong:2011as,Yoo:2014sfa,Thomas:2014aga,DiDio:2014lka,Andrianomena:2014sya,Bonvin:2014owa,Durrer:2016jzq,Giblin:2017ezj,DiDio:2018unb,Umeh:2019qyd}. Recently, \cite{DiDio:2018zmk} computed the one-loop power spectrum dipole using the weak field expansion to order $\mathcal{O}(\epsilon^{1/2})$. In their results the number counts is computed to third order in perturbation theory. In a future paper, we plan to compute the propagation effects on the one-loop bispectrum to order $\mathcal{O}(\epsilon)$. We also plan to explore the appropriate UV renormalization of loop integrals and biasing in this context. A direct generalization of our work is to include in our setup a cosmological constant $\Lambda$, that would generalize the results of \cite{Villa:2015ppa} to 4th order. Possible refinements of our approach would be to use IR-resummation techniques, such as~\cite{Senatore:2014via}, renormalization group techniques \cite{Floerchinger:2016hja} or consider viscosity \cite{Blas:2015tla} or velocity dispersion terms \cite{Erschfeld:2018zqg}. Another fruitful approach to predict LSS observable is Lagrangian perturbation theory~\cite{Matsubara:2007wj}, and relativistic version of it have been worked out, see for instance~\cite{Alles:2015vua,Li:2018hll} and references therein. The weak field approach may also simplify the calculations involved in this framework.
\section*{Acknowledgments}

It is a pleasure to thank R.~Durrer, A.~Riotto, L.~Senatore, and F.~Vernizzi for interesting discussions. L.C.~is supported by ``Beca Postgrado PUCV 2018'', R.G..~is supported by Fondecyt
project No 1171384, J.N.~is supported by Fondecyt Grant 1171466. C.S.~is supported by Fondecyt Grant 3170557. 

\appendix

\section{Description of a barotropic irrotational fluid}
\label{app:A}
In this Appendix we review the description of a barotropic irrotational fluid with a scalar field. We mostly follow what is done in reference \cite{Boubekeur:2008kn}. This fluid can be described by a single scalar degree of freedom $\varphi$ whose dynamics are governed by the action
\begin{equation}
S_f = \int d^4 x\sqrt{-g}\,P(X)\,,\quad X \equiv - g^{\mu\nu}\de_\mu \varphi \de_\nu\varphi\,,
\end{equation}
From this, we can derive the stress energy tensor
\begin{equation}
T_{\mu\nu} = 2P'(X) \de_\mu \varphi \de_\nu \varphi - P(X) g_{\mu\nu}\,,
\end{equation}
from which we can read the velocity, pressure and density of the fluid as long as $\de_\mu \varphi$ is time-like
\begin{equation}
p = P(X)\,,\quad \rho = 2X P'(X) - P(X)\,,\quad u_\mu = \frac{\de_\mu \varphi}{\sqrt{X}}\,.
\label{app:prhou}
\end{equation}
The conservation of stress-energy gives the Euler and continuity equations as usual. This describes an irrotational fluid in the sense that the 4-velocity is explicitly hypersurface orthogonal (it is orthogonal to the constant-$\varphi$ hypersurfaces). Furthermore, we can make it satisfy an equation of motion of the form $p = w\rho$ by taking
\be
P(X)= X^{\frac{1+w}{2w}}\,.
\ee
The dark matter fluid is defined in the limit $w \rightarrow 0$. This limit needs to be taken with care as discussed in \cite{Boubekeur:2008kn}, but here we will only use the conservation of stress-energy along with the expression for the 4-velocity, for which that limit is straightforward. Finally, we can take $\varphi(x) = t + U(x)$ after a redefinition of the field (taking into account that the background 4-velocity should be future-directed).

In Minkowski, hypersurface orthogonality corresponds to the usual condition that the 3-velocity be the derivative of a velocity potential. Indeed, taking as usual
$$
u_\mu = \gamma(1, v_i)\,,\quad \gamma = (1 - v^2)^{1/2}\,,
$$
one can verify that it satisfies the Frobenius condition for hypersurface orthogonality $u_{[\mu}\nabla_\nu u_{\rho]} = 0$ (see e.g. \cite{Wald:1984rg}) if and only if $v^i$ is the gradient of a velocity potential. In our case we can define the 3-velocity as being
$$
u_\mu = \frac{1}{\sqrt{X}} (1, v_i)\,,
$$
such that $v_i = \de_i U$ and irrotational in the usual sense. However, in the main body we chose to use the spatial component of the 4-velocity $u_i = v_i / \sqrt{X}$, which is not curl-free. Notice that in the non-relativistic limit, when the metric is close to Minkowski and velocities are small, $u_i \approx v_i$. Therefore the spatial component of the 4-velocity satisfies the usual Newtonian equations in that limit.

\section{Derivation of Einstein's equations}
\label{app:Einstein}
In this appendix, we fill some details of the derivation of the system of equations presented in section \ref{sec:Einsteinandfluid}. Those equation are the Euler equation (\ref{eq:NT1}), the conservation equation (\ref{eq:NT2}) and the Einstein equations:
\begin{equation}
\label{eq:EEQ}
G_{\mu \nu}= T_{\mu \nu}\,.
\end{equation}
We will present the results in the two gauges under focus in this article. To avoid clutter, we omit the order of the terms neglected in the equations of this appendix. It is understood that for any equation we keep only the leading relativistic corrections. To recover the information about the order in the weak field expansion, the reader is invited to consult Table \ref{tab:a}.
\subsection{Poisson gauge}
From the metric (\ref{eq:metric}) a direct computation gives the following relevant Christoffel symbols:
\begin{align}
& \Gamma_{\mu 0}^{\mu} = 3H + \dot{\phi} - 3\dot{\psi}\,, \\
& \Gamma_{\mu i}^{\mu}=\phi_{,i} -3 \psi_{,i}\,, \\
& \Gamma_{0i}^k=\delta_i^k (H-\dot{\psi})+\frac{1}{2a^2} (\omega_{k,i}-\omega_{i,k})\,, \\
& \Gamma_{00}^{k} = \frac{\phi_{,k}}{a^2}+\frac{1}{a^2}(2\psi \phi_{,k}+\dot{\omega_k})\,, \\
&  \Gamma_{ij}^k=-(\partial_i \psi \delta_{jk}+\partial_j \psi \delta_{ik}-\partial_k \psi \delta_{ij})\,, \\
& \Gamma_{00}^{0}=\dot{\phi}\,.
\end{align}
Calculating (\ref{eq:NT1}), we find:
\begin{equation}
\label{eq:NT11}
\dot{\delta} \left(1-\phi+\frac{a^2 u^2}{2}\right) + \left(1+ \delta\right)\left(\frac{a^2 u^2}{2}\right)_{,t}+\partial_i\left[(1+\delta)u^i\right]-3(1+\delta)\dot{\psi}+(1+\delta)(\phi_{,i}-3\psi_{,i})u^i=0\,,
\end{equation}
where the background equation has been used. Expanding equation (\ref{eq:NT2}), one finds
\begin{align}
\label{eq:NT22}
& \dot{u^i}+2Hu^i+u^j\partial_j u^i +\frac{\phi_{,i}}{a^2}+\left(-\phi+\frac{a^2 u^2}{2} \right) \dot{u^i}+\frac{\phi_{,i}}{a^2}(-2 \phi + a^2 u^2)+\frac{2}{a^2} \psi \phi_{,i}+\frac{\dot{\omega}_i}{a^2} \nonumber \\
& +2 \left(\frac{a^2 u^2}{2}H-H \phi-\dot{\psi} \right)u^i+\frac{1}{a^2}(\omega_{i,j}-\omega_{j,i})u^j+(\psi_{,i}u^2-2u^j \psi_{,j}u^i)=0\,.
\end{align}
Now to complement the generalization of the Euler and conservation equations, we calculate the Einstein equations (\ref{eq:EEQ}):
\begin{align}
&G^{00}=\frac{2}{a^2} \Delta \psi (1-2\phi+4\psi)+3H^2\left(1-4 \phi-2\frac{\dot{\psi}}{H}\right) +\frac{3}{a^2}(\partial_i \psi)^2 = \rho(1-2\phi+ a^2 u^2) \,, \label{eq:EE00} \\
& G^{0i}=\frac{1}{2a^4}\Delta \omega_i-\frac{2}{a^2}(\dot{\psi_{,i}}+H\phi_{,i})=\rho u_i\,, \label{eq:EE0i}\\
& \Sigma_i G^i_i=-\frac{\Delta}{a^2}(\phi-\psi)-\frac{6\ddot{a}}{a}-3H^2 =0 \,. \label{eq:EEii}
\end{align}
Using (\ref{eq:EEii}) in the matter dominated era, one finds that $\phi=\psi+\mathcal{O}(a^4H^4/k^4)$. Therefore, we define $\chi \equiv \phi- \psi$ and replace in equations (\ref{eq:NT11}-\ref{eq:EE0i}).
We find
\begin{align}
&  \dot{\delta} \left(1-\phi+\frac{a^2 u^2}{2}\right) + \left(1+ \delta\right)\left(\frac{a^2 u^2}{2}\right)_{,t}+\partial_i\left[(1+\delta)u^i\right]-3(1+\delta)\dot{\phi}-2\phi_{,i}(1+\delta)u^i=0\,, \label{eq:NT111} \\
& \dot{u^i}+2Hu^i+u^j\partial_j u^i +\frac{\phi_{,i}}{a^2}  +\dot{u^i}\left(-\phi+\frac{a^2 u^2}{2} \right) + \frac{\dot{\omega}_i}{a^2} \nonumber \\
&+2 \left(\frac{a^2 u^2}{2}H-H \phi-\dot{\phi} \right)u^i+\frac{1}{a^2}(\omega_{i,j}-\omega_{j,i})u^j+2(\phi_{,i}u^2-u^j \phi_{,j}u^i)=0\,, \label{eq:N222} \\
& \frac{2}{a^2} \left(\Delta \phi (1+2\phi)+\Delta \chi \right)-6H(2H \phi + \dot{\phi})+\frac{3}{a^2}(\partial_i \phi)^2 = \bar{\rho} \delta +\bar{\rho}(1+\delta)( a^2 u^2-2\phi), \label{eq:EE000} \\
& \frac{1}{2a^4}\Delta \omega_i-\frac{2}{a^2}(\dot{\phi_{,i}}+H\phi_{,i})=\rho u^i\,. \label{eq:EE0i0}
\end{align}
We stress here the choice to replace $\psi$ or $\phi$ by $\chi$ do give equivalent equations to solve. 
In order to close the system, we calculate $G^{00} + a^2 G^{ii}$ which gives:
\begin{equation}
\label{eq:closed}
\frac{2}{a^2} \Delta \phi(1-2\phi) -6 \frac{\ddot{a}}{a}+6H(3\dot{\phi}-2H \phi) +6\ddot{\phi}-4 \phi_{,i}^2=\bar{\rho}(1+\delta) (1-2\phi+2 a^2 u^2)\,.
\end{equation}
Observe that neither $\chi$ nor $h_{ij}$ appear, therefore solving (\ref{eq:closed}) will close the system and generalize the Poisson equation (\ref{eq:Poisson}) with second-order relativistic corrections in the weak field approximation.
\paragraph{} We close this appendix by presenting the equation of motion for the next-to-leading corrections which includes $\chi$ and $h_{ij}$.
\begin{equation}
    \Delta \chi=\frac{a^2 u^2 \rho}{2}+\frac{a^2}{2}\left(24 H \dot{\phi}-7 \frac{(\phi_{,i})^2}{a^2}-8 \frac{\phi \Delta \phi}{a^2}+6 \ddot{\phi} \right)\,,
\end{equation}
\begin{equation}
    \left(\delta_k^m \delta_l^n-\frac{1}{3}\delta^{mn}\delta_{kl} \right) P^{il}P^{jk} \left[\ddot{h}_{ij}+3H\dot{h}_{ij}-\frac{\Delta h_{ij}}{a^2}+\frac{4}{a^2}(\phi_{,i} \phi_{,j} +2 \phi \phi_{,ij}) \right]=0\,.
    \label{eq:apptensor}
\end{equation}
where P is a transverse projector defined in the text before equation (\ref{eq:wiPoiss}).

\subsection{Comoving gauge}

In this gauge, the required Christoffel symbols are the following:  
\begin{align}
    &\Gamma_{\mu0}^{\mu}=3H-3\dot{\psi} \,, \\
    &\Gamma_{\mu i}^{\mu}=-3\partial_i\psi\,,    \\
    &\Gamma_{0i}^{k}= (H-\dot{\psi})\delta_{ki} -\frac{1}{a^{2}}H\partial_{i}\omega\partial_{k}\omega+\frac{1}{2a^{2}}(\partial_{i}w_{k}-\partial_{k}w_{i})+\frac{1}{a^{4}}\partial_{k}\omega\partial_j\omega\partial_i\partial_j\omega\,,   \\
   & \Gamma_{00}^{k}=-\frac{1}{a^{4}}\partial_{j}\omega\partial_k\partial_j\omega+\frac{1}{a^{2}}\partial_{k}\dot{\omega}-\frac{1}{a^{4}}H\partial_{k}\omega(\partial_{j}\omega)^2+\frac{1}{a^{2}}\dot{w}^{k}+\frac{2}{a^2}\psi\partial_{k}\dot{\omega}\nonumber\\ &\qquad+\frac{1}{a^{6}}\partial_k\omega\partial_j\omega\partial_i\omega\partial_j\partial_i\omega-\frac{2}{a^4}\psi(\partial_j\omega\partial_k\partial_j\omega)\,,   \\
   &\Gamma_{ij}^{k}=\left(\partial_{k}\psi-H\partial_{k}\omega\right)\delta_{ji}-\partial_{j}\psi\delta_{ki} - \partial_{i}\psi\delta_{kj} + \frac{1}{a^2}\partial_{k}\omega\partial_{j}\partial_{i}\omega\,.
\end{align}
Equation \eqref{eq:NT1} now reads:
\begin{equation}
\dot{\delta}+\partial_{i}\left[(1+\delta)u^{i}\right]-3(1+\delta)\dot{\psi}-3(1+\delta)u^{i}\partial_{i}\psi=0\,. \label{10}
\end{equation}
Using the Einstein equations (\ref{eq:EEQ}) together with the background equations allows to close the system. The $00$ equation gives $\psi$:
\begin{align}
 & -\frac{2}{a^{2}}\nabla^{2}\psi+\frac{2}{a^{2}}H\nabla^{2}\omega+\frac{1}{2a^{4}}\left[\left(\partial_{i}\partial_{j}\omega\right)^{2}-\partial_{i}^{2}\omega\partial_{j}^{2}\omega\right]-\frac{2}{a^{2}}H\partial_{i}\psi\partial_{i}\omega+\frac{4}{a^{2}}H\psi\nabla^{2}\omega\nonumber\\
 & +\frac{1}{2a^4}\partial_i\omega\nabla^2w_i+\frac{1}{a^{4}}\partial_{i}\partial_{j}\omega\partial_{i}w_{j}+\frac{2}{a^{4}}\partial_{i}\psi(\partial_{j}\omega\partial_{i}\partial_{j}\omega)+\frac{1}{a^4}\nabla^2\psi(\partial_i\omega)^2+\frac{1}{a^4}\partial_i\partial_j\psi (\partial_i\omega \partial_j \omega)\nonumber\\
 &+\frac{2}{a^{4}}\psi\left[\left(\partial_{i}\partial_{j}\omega\right)^{2}-\partial_{i}^{2}\omega\partial_{j}^{2}\omega\right]+6H\dot{\psi}-\frac{1}{a^{2}}\left[3(\partial_{i}\psi)^{2}+8\psi\nabla^{2}\psi\right]-\frac{2}{a^{2}}\nabla^{2}\omega\dot{\psi}-\frac{2}{a^{2}}\partial_i\omega\partial_i\dot{\psi}=-\bar{\rho}\delta\,. \label{eq:17}
\end{align}
The $0i$ equation gives $w_{i}$:
\begin{equation}
\frac{1}{a^2}\nabla^2w_i=4\partial_i\dot{\psi}-\frac{2}{a^2}\partial_i \omega \nabla^2\psi - \frac{2}{a^2}\partial_j\omega \partial_i \partial_j \psi \label{eq:19}
\end{equation}
And $G^{i}_i -G^{0}_0 = \bar{\rho}(1+\delta)$ together with equation \eqref{eq:19} gives an equation for $\omega$:  
\begin{align} 
&\frac{1}{a^2}(1+2\psi)\nabla^2\dot\omega- \frac{1}{a^4}(1+4\psi)\left[(\partial_i\partial_j\omega)^2+\partial_i\omega\partial_j^2\partial_i\omega\right]-\frac{\left(\partial_i\omega\right)^2}{a^4}\nabla^2\psi-\frac{w_{i}}{a^4}\nabla^2\partial_{i}\omega -\frac{\partial_i\omega}{2a^4}\nabla^2w_{i}\nonumber\\
&-\frac{2}{a^2}\partial_{j}w_{i}\partial_i\partial_j\omega-\frac{6}{a^4}\partial_i\psi\partial_i\partial_j\omega\partial_i\partial_j\omega-\frac{2}{a^2}\partial_{i}\omega\partial_{i}\dot{\psi}-\frac{1}{a^2}\partial_i\psi\partial_i\dot{\omega}+6H\dot{\psi}+3\ddot\psi=\frac{1}{2}\bar{\rho}\delta\label{eq:20}
\end{align}
Finally in this appendix, from equation $G_{j}^{i}$ it is possible to find an equation for  tensor modes  $h^i_j$. Starting with $h^i_j$ to order $\epsilon$, the equation for this tensor is
\begin{equation}
    \nabla^2h^i_j + \mathcal{O}\left(\epsilon\right) =  T^i_j\,,
\end{equation}
where $T_j^i = -2 a^2 \rho \psi u^i u^k \delta_{kj} \sim \mathcal{O}\left(\epsilon^2\right)$, and after applying the transverse projector to terms of order $\mathcal{O}\left(\epsilon\right)$ this terms vanish, then we  conclude $h_j^i\sim \mathcal{O}\left(\epsilon^2\right)$. To next leading order the equation for the tensor modes is given by 
\begin{equation}
\nabla^2h_j^i = \psi\partial_i\partial_j\psi -\partial_j\omega\partial_i\dot{\psi} -\partial_i\omega\partial_j\dot{\psi}+\frac{1}{a^6}\partial_k\omega\partial_l\partial_i\omega\left(\partial_l\omega\partial_k\partial_j\omega-\partial_k\omega\partial_l\partial_j\omega\right). \label{tensormodess}     
\end{equation}
\section{Explicit form of the relativistic sources}
\label{app:Sources}
In this appendix, we provide explicit expressions for the relativistic sources of equations \eqref{eq:efdelta}.

\subsection{Poisson gauge}
Using, \eqref{eq:deltaPois} \eqref{eq:thetaPois} \eqref{eq:EEiii} \eqref{eq:EE0i0} in Fourier space, we find:
\begin{align}
\mathcal{S}_{\delta} =&  3\dot{\phi}_N(\bm{k})+\int_{\bm{k}_1,\bm{k}_2} (2\pi)^3 \delta_D(\bm{k}-\bm{k}_{12})\left\{\dot{\delta}_{N}(\bm{k}_1) \phi_N(\bm{k}_2)+3\delta_N(\bm{k}_1)\dot{\phi}_N(\bm{k}_2) \right.  \nonumber\\
& \left.+\frac{\bm{k}_1 \cdot \bm{k}_2}{\bm{k}_1^2 \bm{k}_2^2}\left[\frac{d}{dt}\left(\frac{a^2 \theta_N(\bm{k}_1) \theta_N(\bm{k}_2)}{2} \right)+2\theta_N(\bm{k}_1) \bm{k}_2^2 \phi_N(\bm{k}_2) \right] \right\} \nonumber \\
& +\int_{\bm{k}_1, \bm{k}_2, \bm{k}_3} (2\pi)^3 \delta_D(\bm{k}-\bm{k}_{123})\left[\frac{\bm{k}_1 \cdot \bm{k}_2}{\bm{k}_1^2 \bm{k}_2^2} \right] \times \left[ \dot{\delta}_N(\bm{k}_3)\frac{a^2 \theta_N(\bm{k}_1) \theta_N(\bm{k}_2)}{2}\right. \nonumber \\
& \left. +2\delta_N(\bm{k}_3)\theta_N(\bm{k}_1) \bm{k}_2^2 \phi_N(\bm{k}_2)+\delta_N(\bm{k}_3)\frac{d}{dt}\left(\frac{a^2 \theta_N(\bm{k}_1) \theta_N(\bm{k}_2)}{2} \right)\right]\,,
\end{align}
\begin{align}
\mathcal{S}_{\theta} =&  -3 H^2 \phi_N(\bm{k})+9H\dot{\phi}_N(\bm{k})+3\ddot{\phi}_N(\bm{k}) + \int_{\bm{k}_1,\bm{k}_2} (2\pi)^3 \delta_D(\bm{k}-\bm{k}_{12}) \bigg\{  \nonumber \\
& \alpha(\bm{k}_1,\bm{k}_2)\left[\dot{\theta}_{N}(\bm{k}_1) \phi_N(\bm{k}_2)  +2\theta_{N}(\bm{k}_1)\left(\dot{\phi}_N(\bm{k}_2)+H\phi_N(\bm{k}_2) \right) \right] +\left[\frac{\bm{k}_1 \cdot \bm{k}_2}{\bm{k}_1^2 \bm{k}_2^2}\right] \nonumber  \\
& \times \left[4\theta_N(\textbf{k}_1)\textbf{k}_2^2\left(\dot{\phi}_N(\textbf{k}_2)+H \phi_N(\textbf{k}_2) \right)-3H^2 a^2 \theta_N(\bm{k}_1) \theta_N(\bm{k}_2) +2\frac{\bm{k}_1^2\phi_N(\bm{k}_1)\bm{k}_2^2\phi_N(\bm{k}_2)}{a^2} \right] \bigg\} \nonumber \\
& +\int_{\bm{k}_1, \bm{k}_2, \bm{k}_3} (2\pi)^3 \delta_D(\bm{k}-\bm{k}_{123})\left[\frac{\bm{k}_1 \cdot \bm{k}_2}{\bm{k}_1^2 \bm{k}_2^2}\right]\bigg[\alpha(\bm{k}_3,\bm{k}_{12}) \frac{\theta_N(\bm{k}_1) \theta_N(\bm{k}_2) a^2}{2}\left(\dot{\theta}_N(\bm{k}_3)+2H\theta_N(\bm{k}_3) \right) \nonumber   \\
&  -  3H^2 a^2 \theta_N(\bm{k}_1) \theta_N(\bm{k}_2) \delta_N(\bm{k}_3)+2\alpha(\bm{k}_3,\bm{k}_{12})\theta(\bm{k}_1) \theta(\bm{k}_2) \phi(\bm{k}_3)\bm{k}_3^2 \nonumber\\
& -2\alpha(\bm{k}_3,\bm{k}_{12}) \theta(\bm{k}_1) \theta(\bm{k}_3) \phi(\bm{k}_2) \bm{k}_2^2 \bigg]\,.
\end{align}

The transverse velocity field can be obtained by pluging \eqref{eq:metric} in \eqref{eq:defumu} and noting that in matter domination: $\dot{\varphi_0}=1$. At order $\epsilon^{1/2}$, one finds: $u^i = v^i_N =\delta^{ij}\frac{\partial_j U}{a^2}$. Plugging this expression back in \eqref{eq:defumu} and going to Fourier space, it is possible to obtain an expression for the transverse velocity:
\begin{multline}
u_T^i=P_j^i \left\{\omega^j(\textbf{k})-i \int_{\bm{k}_1,\bm{k}_2} (2\pi)^3 \delta_D(\bm{k}-\bm{k}_{12}) \frac{\textbf{k}_1^j}{\textbf{k}_1^2}\theta_N(\textbf{k}_1)\Big[ 3 \phi_N(\textbf{k}_2)+a^2\frac{\dot{\theta}_N(\textbf{k}_2)+2H\theta_N(\textbf{k}_2)}{\textbf{k}_2^2} \Big] \right. \\
\left. +i\int_{\bm{k}_1,\bm{k}_2,\bm{k}_3} (2\pi)^3 \delta_D(\bm{k}-\bm{k}_{123}) \frac{\textbf{k}_1^j}{2 \textbf{k}_1^2} \left[\frac{\bm{k}_2 \cdot \bm{k}_3}{\bm{k}_2^2 \bm{k}_3^2}\right] a^2\theta(\textbf{k}_1)\theta(\textbf{k}_2)\theta(\textbf{k}_3) \right\}.
\end{multline}
We obtain the transverse part of $\omega^i$ by applying $P_i^j$ to \eqref{eq:EE0i0}.

\subsection{Comoving gauge} 
Writing equations \eqref{eq:10}, \eqref{eq:11} and \eqref{eq:12} in Fourier space the relativistic sources in comoving gauge are given by
\begin{align}
S_{\delta R}=& 3\dot\psi(t,\bm{k}) \nonumber\\ &+\int_{\bm{k}_1\bm{k}_2}\delta_D(\bm{k}-\bm{k}_{12})\left(3\delta_N(t,\bm{k_1})\dot\psi(t,\bm{k_2})+3\frac{(\bm{k_1}\cdot\bm{k_2})}{\bm{k}^2_1}\theta_N(t,\bm{k}_1)\psi(t,\bm{k}_2) \right)\nonumber\\
&+3 \int_{\bm{k}_1\bm{k}_2\bm{k}_3}\delta_D(\bm{k}-\bm{k}_{123})\frac{(\bm{k_1}\cdot\bm{k_3})}{\bm{k}^2_1}\theta_N(t,\bm{k}_1)\delta_N(t,\bm{k}_2)\psi(t,\bm{k}_3) \label{eq:36}
\end{align}
\begin{align}
 S_{\theta R} =& 3\ddot\psi+ 6H\dot\psi+\int_{\bm{k}_1\bm{k}_2}\delta_D(\bm{k}-\bm{k}_{12})\left[\frac{(\bm{k_1}\cdot\bm{k_2})}{\bm{k}^2_1}\left(3\dot\theta_N(t,\bm{k_1})\psi(t,\bm{k_2})+6H\theta_N(t,\bm{k_1})\psi(t,\bm{k_2})\right.\right.\nonumber \\
&\left.\left.+ 6 \theta_N(t,\bm{k}_1)\dot{\psi}(t,\bm{k}_2) \right) +2 \theta_N(t,\bm{k_1})\dot{\psi}(t,\bm{k_2})\right]\nonumber \\ 
& +\int_{\bm{k}_1\bm{k}_2\bm{k}_3}\delta_D(\bm{k}-\bm{k}_{123})\left[ 2\frac{(\bm{k}_1\cdot\bm{k}_3)}{\bm{k}^2_1}+3\frac{(\bm{k}_1\cdot\bm{k}_3)(\bm{k}_2\cdot\bm{k}_3)}{\bm{k}^2_1 \bm{k}^2_2}\right]\theta_N(t,\bm{k}_1)\theta_N(t,\bm{k}_2){\psi}(t,\bm{k}_3)\nonumber\\ \label{eq:37}
\end{align}
\begin{align}
\psi(t,\bm{k})=& -\frac{a^2}{\bm{k}^2}\left(\frac{3}{2}H^2\delta_N(t,\bm{k})-H \theta_N(t,\bm{k})\right)\nonumber\\
& - \frac{a^2}{4k^2} \int_{\bm{k}_1\bm{k}_2} \delta_D(\bm{k}-\bm{k}_{12})\left[\frac{(\bm{k}_1\cdot\bm{k}_2)^2}{k_1^2k^2_2}-1\right]\theta_N(t,\bm{k_2})\theta_N(t,\bm{k_1})
\end{align}
The transverse velocity in comoving gauge is found by applying the transverse projector on the velocity \eqref{velocitycomoving} and replacing $w_i$ given by equation \eqref{eq:19}, then in Fourier space the transverse velocity in comoving gauge is
\begin{align}
\bm{u}_T(t,\bm{k}) = \frac{4i\bm{k}}{\bm{k}^2}\dot{\psi}(t,\bm{k}) + 2i\int_{\bm{k}_1\bm{k}_2} \delta_D(\bm{k}-\bm{k}_{12})\left(-\frac{\bm{k}_1}{\bm{k}_1^2} + \frac{\bm{k}}{\bm{k}^2}\frac{\bm{k}.\bm{k}_1}{\bm{k}_1^2} + \frac{\bm{k}_1}{\bm{k}_1^2} \frac{\bm{k}_2^2}{\bm{k}^2} + \frac{\bm{k}_1.\bm{k}_2}{\bm{k}_1^2}\frac{\bm{k}_2}{\bm{k}^2}\right)\theta(t,\bm{k}_1)\psi(t,\bm{k}_2)
\end{align}

\section{Derivation of the kernels of the Perturbation Theory}
\label{sec:kernels}
In this appendix, we fill some steps in the determination of the kernels of SPT defined in equation \eqref{eq:expansion} for each gauge under consideration. In section \ref{sec:icpoisson} and \ref{sec:icsync}, we use dimensional analysis together with the time behavior of the solutions to determine whether the initial conditions contribute to the final solution. For instance, in both gauges, the sources $S$ given in \eqref{eq:sources} vanish at second order in perturbation theory, however, the initial conditions contribute to $F_2^R$. We stress that this fact was sometimes omitted in the literature, such as reference~\cite{Hwang:2015jja} who computes the power spectrum. However, as seen in section \ref{sec:oneloopPS}, the terms they omitted give a subdominant contribution to the power spectrum in Poisson gauge, though they are relevant for the bispectrum.
In section \ref{sec:SPTkernpoiss}, \ref{sec:SPTkernsync}, we give detailed expressions for the kernels used in this article.
\subsection{Poisson gauge}
\subsubsection{Initial conditions}
\label{sec:icpoisson}
While the weak field limit allows for compact expression, the initial conditions are taken to match the full GR calculation to leading order. For simplicity, we work in full matter domination, and use the results of appendix A of Ref.~\cite{2010JCAP...05..004F}. To match with the notation of~\cite{2010JCAP...05..004F}, only in this section, we sometimes work in conformal time defined as $dt=a(\eta) d \eta$ and then transform the result back to cosmological time. 
The equation of motion \eqref{eq:efdelta} has the following form in conformal time:
\begin{equation}
\label{eq:conformdelta}
    \delta'' + \frac{2}{\eta}\delta' - \frac{6}{\eta^2}\delta = 12 \eta^{-2}\Phi + \eta^{-2} S_N^{(2)} + \eta^{-4} S_R^{(2)} + \eta^{-4} S_R^{(3)} + \mathcal{O}(\eta^{-6})\,,
\end{equation}
where a prime denotes here a derivative with respect to conformal time $\eta$. The source $S$ have been sorted by decreasing powers of $\eta$, the superscript indicate the order in perturbation theory for each source. Their time behavior can be derived by explicitly calculating \eqref{eq:sources}. From dimensional analysis, each power of $k$ comes with a power of $\eta$, so that the ``weak field'' approximation (large $k$) is equivalent to counting powers of $\eta$, and we work to subleading order in this approximation.
The homogeneous solution to \eqref{eq:conformdelta} takes the form
\begin{equation}
    \delta(\eta,\bm{k}) = c_{-}(\bm{k}) \eta^{-3} + c_{+}(\bm{k}) \eta^2\,,
\end{equation}
\paragraph{Linear order} The relativistic adiabatic growing mode solution at linear order is equation (54) of~\cite{2010JCAP...05..004F}:
    \begin{equation}
        \delta(\eta,\bm{k}) = -\left(2 + \frac{k^2\eta^2}{6} \right)\psi(\bm{k})\,, \label{linearpoisson}
    \end{equation}
    where the gravitational potential is $\psi$ instead of the $\Phi$ in~\cite{2010JCAP...05..004F}. The solution to the linear equation is
    \begin{equation}
        \delta(\eta,\bm{k}) = c_{1}^{(1)}(\bm{k}) \eta^{-3} + c_{2}^{(1)}(\bm{k}) \eta^2 - 2\psi(\bm{k)}\,,
    \end{equation}
    such that we choose $c_{1}^{(1)}(\bm{k}) = 0$ and $c_{2}^{(1)}(\bm{k}) = -k^2 \psi(\bm{k}) / 6$ for the linear solution. We see that the second term corresponds to the Newtonian solution while the last term is the relativistic correction suppressed by two powers of $k\eta$. Going back to cosmic time $t$ by using a factor $\eta_0^2=\frac{4}{H_0^2}$ and using \eqref{eq:Poisson}, it corresponds to setting $c_{-}^{(1)}(\bm{k}) = 0$ and $c_{+}^{(1)}(\bm{k}) = \delta_0$, where $c_{+}(k)$ and $c_{-}(k)$ are defined in equation \eqref{eq:green}. This allows us to conclude that:
    \begin{align}
        & F_1(\bm{k})=1\,, \\
        & F_1^R(\bm{k})=\frac{3}{\bm{k}^2}\label{eq:relcorrorderone}\,.
    \end{align}
    \paragraph{Second order} The relativistic contributions to the sources cancel, and the resulting equation has the form of the Newtonian equation
    \begin{equation}
        \delta'' + \frac{2}{\eta}\delta' - \frac{6}{\eta^2}\delta = \eta^2 F_N^{(2)}\,.
    \end{equation}
    The solution is the sum of the homogeneous solution and a particular solution
    \begin{equation}
        \delta^{(2)}(\eta,\bm{k}) =  c_{1}^{(2)}(\bm{k}) \eta^{-3} + c_{2}^{(2)}(\bm{k}) \eta^2 + \eta^4 F_2\,,
    \end{equation}
    where $F_2$ is the usual Newtonian kernel. Within our approximation $c_{1}^{(2)}(\bm{k})$ is again suppressed by too many negative powers of $\eta$ with respect to the leading term: $c_{1}^{(2)}(\bm{k})=0$ . However $c_{2}^{(2)}(\bm{k})$ must be fixed by matching with the initial conditions. Taking the $\eta \rightarrow \infty$ approximation in the fully relativistic second-order solution (equation (59) of~\cite{2010JCAP...05..004F}), we get
    \begin{align}
        \delta^{(2)}(\eta,\bm{k}) &= \int_{\bm{k}_1,\bm{k}_2}(2 \pi)^3\delta_D(\bm{k}-\bm{k}_{12}) \Bigg[\left(\frac{\bm{k}^4}{14} + \frac{3}{28}(\bm{k}_1^2 + \bm{k}_2^2)\bm{k}^2 - \frac{5}{28}(\bm{k}_1^2 - \bm{k}_2^2)^2\right)\eta^4 \\
        &\phantom{=} + \left(\frac{59\bm{k}^2}{14} - \frac{125}{14}(\bm{k}_1^2 + \bm{k}_2^2) - \frac{9(\bm{k}_1^2 - \bm{k}_2^2)^2}{7\bm{k}^2}\right)\eta^2 + \mathcal{O}(\bm{k}^0 \eta^0)\Bigg]\frac{\psi(\bm{k}_1)\psi(\bm{k}_2)}{36}\,,
    \end{align}
    where the first line is the Newtonian result, and thus
    \begin{equation}
      c_{2}^{(2)}(\bm{k}) =  \int_{\bm{k}_1,\bm{k}_2}(2 \pi)^3\delta_D(\bm{k}-\bm{k}_{12})\left[\left(\frac{59\bm{k}^2}{14} - \frac{125}{14}(\bm{k}_1^2 + \bm{k}_2^2) - \frac{9(\bm{k}_1^2 - \bm{k}_2^2)^2}{7\bm{k}^2}\right)\eta^2 + \mathcal{O}(\bm{k}^0 \eta^0)\right]\frac{\psi(\bm{k}_1)\psi(\bm{k}_2)}{36}\,.
    \end{equation}
    Transforming back to cosmic time, using $\eta_0^2=\frac{4}{H_0^2}$ and \eqref{eq:Poisson} we find:
    \begin{align}
      F_2(\bm{k}_1,\bm{k}_2)  &=  \frac{5}{7} \alpha(\bm{k}_1,\bm{k}_2)+\frac{2}{7} \beta(\bm{k}_1,\bm{k}_2)\,, \\
 F_2^R(\bm{k}_1,\bm{k}_2) &= \frac{1}{4 \bm{k}_1^2 \bm{k}_2^2}   \left[\left(\frac{59(\bm{k}_1+\bm{k}_2)^2}{14} - \frac{125}{14}(\bm{k}_1^2 + \bm{k}_2^2) - \frac{9(\bm{k}_1^2 - \bm{k}_2^2)^2}{7\bm{k}^2}\right)\right]\,. \\   
    \end{align}
    Though we have set initial conditions for a universe which has been dominated by matter throughout its history, it is straightforward to set $F_2$ by matching with perturbative results to second order in a more realistic case, such as \cite{Tram:2016cpy}.
    \paragraph{Third order} \eqref{eq:conformdelta} is
    \begin{equation}
    \delta'' + \frac{2}{\eta}\delta' - \frac{6}{\eta^2}\delta =     \eta^4 \tilde{F}_N^{(3)} + \eta^2 \tilde{F}_R^{(3)}\,,
    \end{equation}
    where the second order solution is plugged into the second order source to obtain a third order term. The solution is again given by
    \begin{equation}
        \delta^{(3)} = c_{1}^{(3)}(\bm{k}) \eta^{-3} +c_{2}^{(3)}(\bm{k}) \eta^2 + \eta^6 F_3 + \eta^4 F^{(R)}_3\,,
    \end{equation}
   Both the term proportional to $c_{1}^{(3)}(\bm{k})$ and the one proportional to $c_{2}^{(3)}(\bm{k})$ are subdominant compared to the source terms, and can be set to zero: $c_{1}^{(3)}(\bm{k})=c_{1}^{(3)}(\bm{k})=0$. Note that this reasoning also holds for higher order in perturbation theory, and the initial condition can be neglected.

\subsubsection{SPT kernels}
\label{sec:SPTkernpoiss}
The kernels are given by:
\begin{align}
 F_1^R(\bm{k})  &= \frac{3}{\bm{k}^2}\,, \label{F1RPoiss}\\
 F_2(\bm{k}_1,\bm{k}_2)  &=  \frac{5}{7} \alpha(\bm{k}_1,\bm{k}_2)+\frac{2}{7} \beta(\bm{k}_1,\bm{k}_2)\,, \\
 F_2^R(\bm{k}_1,\bm{k}_2) &= \frac{1}{4 \bm{k}_1^2 \bm{k}_2^2}   \left[\left(\frac{59(\bm{k}_1+\bm{k}_2)^2}{14} - \frac{125}{14}(\bm{k}_1^2 + \bm{k}_2^2) - \frac{9(\bm{k}_1^2 - \bm{k}_2^2)^2}{7(\bm{k}_1+\bm{k}_2)^2}\right)\right]\,, \label{F2RPoiss}\\
 G_2(\bm{k}_1,\bm{k}_2) & =2F_2(\bm{k}_1,\bm{k}_2)-\alpha(\bm{k}_1,\bm{k}_2)=  \frac{3}{7} \alpha(\bm{k}_1,\bm{k}_2)+\frac{4}{7} \beta(\bm{k}_1,\bm{k}_2)\,, \\
 G_2^R(\bm{k}_1,\bm{k}_2) &= F_2^R(\bm{k}_1,\bm{k}_2)+ \frac{9}{2}\frac{F_2(\bm{k}_1,\bm{k}_2)}{(\bm{k}_1+\bm{k}_2)^2}- \frac{13}{2}\frac{\bm{k}_1 \cdot \bm{k}_2}{\bm{k}_1^2 \bm{k}_2^2} -\frac{3}{4\bm{k}_1^2}-\frac{3}{4\bm{k}_2^2}\,, \label{G2RPoiss}\\
\bm{G}^T_{2R}(\bm{k}_1,\bm{k}_2)&=-6\left(\bm{k}_1 -\frac{\bm{k}_1 \cdot \bm{k}_{12} \bm{k}_{12} }{\bm{k}_{12}^2} \right) \left(\frac{\bm{k}_{12}^2+\bm{k}_{2}^2}{\bm{k}_{1}^2\bm{k}_2^2\bm{k}_{12}^2} \right)\label{G2RTPoiss}\\
 F_3(\bm{k}_1,\bm{k}_2,\bm{k}_3) &= \frac{1}{18}\left[7 F_2(\bm{k}_1,\bm{k}_2) \alpha(\bm{k}_3,\bm{k}_{12})+7 G_2(\bm{k}_1,\bm{k}_2) \alpha(\bm{k}_{12},\bm{k}_3) + 4 G_2(\bm{k}_1,\bm{k}_2) \beta(\bm{k}_3,\bm{k}_{12})\right]\,, \label{eq:F3} \\
 F_3^R(\bm{k}_1,\bm{k}_2,\bm{k}_3) & =  \frac{1}{14}\left\{18\frac{F_3(\bm{k}_1,\bm{k}_2,\bm{k}_3)}{(\bm{k}_1+\bm{k}_2+\bm{k}_3)^2}+\left[52+3\alpha(\bm{k}_3,\bm{k}_{12})\right]\frac{\bm{k}_1 \cdot \bm{k}_2}{\bm{k}_1^2 \bm{k}_2^2} \right.  \nonumber \\
 & \left.  +G_2^R(\bm{k}_1,\bm{k}_2)\left[10\alpha(\bm{k}_{12},\bm{k}_3)+8\beta(\bm{k}_{12},\bm{k}_3)\right] +10F_2^R(\bm{k}_1,\bm{k}_2) \alpha(\bm{k}_3,\bm{k}_{12}) \right. \nonumber \\
 & \left. +F_2(\bm{k}_1,\bm{k}_2)\left[-\frac{30}{\bm{k}_{3}^2}-\frac{81}{\bm{k}_{12}^2}-75 \frac{\bm{k}_3 \cdot \bm{k}_{12}}{\bm{k}_3^2 \bm{k}_{12}^2}\right]+G_2(\bm{k}_1,\bm{k}_2)\left[65\frac{\bm{k}_3 \cdot \bm{k}_{12}}{\bm{k}_3^2 \bm{k}_{12}^2}+\frac{15}{\bm{k}_3^2} \right] \right\} \nonumber  \\
&+\frac{1}{7}\bm{G}^{T}_{2R}(\bm{k}_1,\bm{k}_3)\cdot\bm{k}_2\left(7+4\frac{\bm{k}_{13}\cdot\bm{k}_2}{\bm{k}_2^2}\right)
 \,,  \label{eq:F3R} \\
  G_3(\bm{k}_1,\bm{k}_2,\bm{k}_3) &= 3F_3(\bm{k}_1,\bm{k}_2,\bm{k}_3)-F_2(\bm{k}_1,\bm{k}_2)\alpha(\bm{k}_3,\bm{k}_{12})-G_2(\bm{k}_1,\bm{k}_2)\alpha(\bm{k}_{12},\bm{k}_3) \,,
\end{align}
\begin{align} 
  &G_3^R(\bm{k}_1,\bm{k}_2,\bm{k}_3) = 9 \frac{F_3(\bm{k}_1,\bm{k}_2,\bm{k}_3)}{(\bm{k}_1+\bm{k}_2+\bm{k}_3)^2}+2 F_3^R(\bm{k}_1,\bm{k}_2,\bm{k}_3)-F_2^R(\bm{k}_1,\bm{k}_2)\alpha(\bm{k}_{3},\bm{k}_{12}) \nonumber \\ & -G_2^R(\bm{k}_1,\bm{k}_2)\alpha(\bm{k}_{12},\bm{k}_3) -4\frac{\bm{k}_1 \cdot \bm{k}_2}{\bm{k}_1^2 \bm{k}_2^2} +F_2(\bm{k}_1,\bm{k}_2)\left[\frac{3}{\bm{k}_{3}^2}+\frac{6}{\bm{k}_{12}^2}-3 \frac{\bm{k}_3 \cdot \bm{k}_{12}}{\bm{k}_3^2 \bm{k}_{12}^2}\right]  \nonumber \\
 & -G_2(\bm{k}_1,\bm{k}_2)\left[8 \frac{\bm{k}_3 \cdot \bm{k}_{12}}{\bm{k}_3^2 \bm{k}_{12}^2}+\frac{3}{\bm{k}_{3}^2}\right] -\bm{k}_2\cdot\bm{G}^T_{2R}(\bm{k}_1,\bm{k}_3)  \,,
 \end{align}
 \begin{align}
 \bm{G}^T_{3R}(\bm{k}_1,\bm{k}_2,\bm{k}_3) = & -6 \frac{G_2(\bm{k}_1,\bm{k}_3)}{\bm{k}_{13}^2}\frac{(\bm{k}_2^2+\bm{k}_{123}^2)}{\bm{k}_2^2 \bm{k}_{123}^2}\left(\bm{k}_{13}+\frac{\bm{k}_{123}\cdot\bm{k}_{13} \bm{k}_{123}}{\bm{k}_{123}^2}\right) \nonumber \\
 &  -\left[\frac{F_2(\bm{k}_2,\bm{k}_3)}{\bm{k}_1^2}\left(\frac{6}{\bm{k}_{123}^2}+\frac{9}{\bm{k}_{23}^2}\right)+\frac{5 G_2(\bm{k}_2,\bm{k}_3)}{2 \bm{k}_1^2 \bm{k}_{23}^2}+\frac{\bm{k}_2\cdot\bm{k}_3}{2\bm{k}_1^2 \bm{k}_2^2 \bm{k}_3^2}\right]\left(\bm{k}_1+\frac{\bm{k}_1\cdot\bm{k}_{123}\bm{k}_{123}}{\bm{k}_{123}^2}\right)
 \end{align}
 \begin{align} 
  & F_4^R(\bm{k}_1,\bm{k}_2,\bm{k}_3,\bm{k}_4) = \frac{1}{36}\left(36\frac{F_4(\bm{k}_1,\bm{k}_2,\bm{k}_3,\bm{k}_4)}{(\bm{k}_1+\bm{k}_2+\bm{k}_3+\bm{k}_4)^2}+F_2(\bm{k}_1,\bm{k}_2)G_2(\bm{k}_3,\bm{k}_4)\left[-33\frac{\bm{k}_{12} \cdot \bm{k}_{34}}{\bm{k}_{12}^2 \bm{k}_{34}^2}-\frac{27}{\bm{k}_{12}^2}\right] \right. \nonumber \\
& \left. +F_2(\bm{k}_1,\bm{k}_2)F_2(\bm{k}_3,\bm{k}_4)\left[-18\frac{\bm{k}_{12} \cdot \bm{k}_{34}}{\bm{k}_{12}^2 \bm{k}_{34}^2}-\frac{105}{\bm{k}_{12}^2}\right]+33G_2(\bm{k}_1,\bm{k}_2)G_2(\bm{k}_3,\bm{k}_4)\frac{\bm{k}_{12} \cdot \bm{k}_{34}}{\bm{k}_{12}^2 \bm{k}_{34}^2} \nonumber \right. \\
& \left. +F_3(\bm{k}_1,\bm{k}_2,\bm{k}_3)\left[-99\frac{\bm{k}_{123} \cdot \bm{k}_{4}}{\bm{k}_{123}^2 \bm{k}_{4}^2}-\frac{180}{\bm{k}_{123}^2}-\frac{63}{\bm{k}_{4}^2}  \right] +G_3(\bm{k}_1,\bm{k}_2,\bm{k}_3)\left[105\frac{\bm{k}_{123} \cdot \bm{k}_{4}}{\bm{k}_{123}^2 \bm{k}_{4}^2}+\frac{21}{\bm{k}_{4}^2}  \right] \right. \nonumber \\
& \left. +14\alpha(\bm{k}_{12},\bm{k}_{34})\left[G^R_2(\bm{k}_1,\bm{k}_2)F_2(\bm{k}_3,\bm{k}_4)+G_2(\bm{k}_1,\bm{k}_2)F^R_2(\bm{k}_3,\bm{k}_4)\right] \right. \nonumber \\
& \left.+G^R_3(\bm{k}_1,\bm{k}_2,\bm{k}_3)\left[14 \alpha(\bm{k}_{123},\bm{k}_4)+8\beta(\bm{k}_{123},\bm{k}_4) \right] \right. \nonumber \\
& \left.+8G^R_2(\bm{k}_1,\bm{k}_2)G_2(\bm{k}_3,\bm{k}_4)\beta(\bm{k}_{12},\bm{k}_{34})+14 F_3^R(\bm{k}_1,\bm{k}_2,\bm{k}_3) \alpha(\bm{k}_4,\bm{k}_{123}) \right. \nonumber \\
& \left. +F_2(\bm{k}_3,\bm{k}_4)\left\{\frac{\bm{k}_1 \cdot \bm{k}_2}{\bm{k}_1^2 \bm{k}_2^2}\left[75+12 \alpha(\bm{k}_{34},\bm{k}_{12}) \right]+\frac{\bm{k}_1 \cdot \bm{k}_{34}}{\bm{k}_1^2 \bm{k}_{34}^2}\left[42-12 \alpha(\bm{k}_2,\bm{k}_{134}) \right] \right\} \right. \nonumber \\
& \left. +G_2(\bm{k}_3,\bm{k}_4)\left\{\frac{\bm{k}_1 \cdot \bm{k}_2}{\bm{k}_1^2 \bm{k}_2^2}\left[-7 \alpha(\bm{k}_{34},\bm{k}_{12}) \right]+\frac{\bm{k}_1 \cdot \bm{k}_{34}}{\bm{k}_1^2 \bm{k}_{34}^2}\left[33+15 \alpha(\bm{k}_2,\bm{k}_{134}) \right]\right. \right.  \nonumber \\
& \left. \left. +\frac{\bm{k}_2 \cdot \bm{k}_{34}}{\bm{k}_2^2 \bm{k}_{34}^2}\left[75+3 \alpha(\bm{k}_1,\bm{k}_{234}) \right] \right\}\right) +\frac{1}{18}\bm{G}_{3R}^T(\bm{k}_1,\bm{k}_3,\bm{k}_4)\cdot\bm{k}_2\left(9+4\frac{\bm{k}_{134}\cdot\bm{k}_2}{\bm{k}_2^2}\right)\nonumber\\
&+\frac{1}{18}\bm{G}_{2R}^{T}(\bm{k}_1,\bm{k}_3)\cdot\bm{k}_{24}\left(7F_2(\bm{k}_2,\bm{k}_4)+2G_2(\bm{k}_2,\bm{k}_4)\frac{\bm{k}_{13}\cdot\bm{k}_{24}}{\bm{k}_{24}^2}\right) \,.
\end{align}
Note that at second order in perturbation theory, $S=0$: all the sources cancel, however relativistic corrections arise from initial conditions as it was discussed in section \ref{sec:icpoisson}. The kernels we give here remain to be symetrized and are expressed recursively. We conjecture that a general recursive relation could be derived for our weak field framework. In the Newtonian case, it is given for instance in equations (43)-(44) of~\cite{Bernardeau:2001qr}. We reserve this calculation for the future.
\subsection{Comoving Gauge}

\subsubsection {Initial conditions}
\label{sec:icsync}
In the comoving gauge, the calculation is analogous to the Poisson case. We borrow the results for the primordial inflationary perturbations $\phi_{\zeta}$, $\omega_{\zeta}$,  $w_{\zeta i}$ and  $\psi_{\zeta}$ given in equations (77)-(80) of \cite{Boubekeur:2008kn}. With equation \eqref{eq:17}, it is possible to express $\delta(t,\bm{x})$ in term of these quantities as
\begin{align}
    \delta(t,\bm{x}) =  \frac{2}{3H^2a^2}&\left[\nabla^2\psi_{\zeta}-a H\nabla^2\omega_{\zeta}-\frac{1}{4}(\partial_i\partial_j\omega_{\zeta})^2+\frac{1}{4}\partial^2\omega_{\zeta}\partial^2\omega_{\zeta}+Ha\partial_i\psi_{\zeta}\partial_i\omega_{\zeta}-2Ha\psi_{\zeta}\nabla^2\omega_{\zeta}\right.\nonumber \\
    &\left.-3Ha^2\dot{\psi}_{\zeta}+\frac{3}{2}(\partial_i\psi_{\zeta})^2 +4\psi_{\zeta}\nabla^2\psi_{\zeta}+a\nabla^2\omega_{\zeta}\dot{\psi}+a\partial_i\omega_{\zeta}\partial_i\dot{\psi}_{\zeta}\right]\,.
\end{align} 
As in the Poisson gauge, the only contribution from the initial conditions to the perturbation theory kernels come from the second order
\begin{align}
\label{eq:delta_synchro}
\delta^{(2)}(t,\bm{k}) =&\frac{a^2}{2} \int_{\bm{k}_1,\bm{k_2}}(2 \pi)^3\delta_D(\bm{k}-\bm{k}_{12})\left[\left(\frac{10}{7}\bm{k}_1^2 \bm{k}_2^2+(\bm{k}_1\cdot\bm{k}_2)(\bm{k}_1^2+\bm{k}_2^2)+\frac{4}{7}(\bm{k}_1\cdot\bm{k}_2)^2\right)\right]\frac{\delta_l(\bm{k}_1)\delta_l(\bm{k}_2)}{k_1^2 k_2^2}\nonumber\\
&+a^4H^2 \frac{25}{6}\int_{\bm{k}_1,\bm{k_2}}(2 \pi)^3\delta_D(\bm{k}-\bm{k}_{12})\left[\bm{k}^2 -\frac{17}{10}\bm{k}_1\cdot\bm{k}_2-\frac{8}{5}(\bm{k}_1^2+\bm{k}_2^2)\right]\frac{\delta_l(\bm{k}_1)\delta_l(\bm{k}_2)}{k_1^2 k_2^2}\,,
\end{align}
where we used the linear solution for $\psi(\bm{k})=-\frac{5  a^2H^2 }{2 \bm{k}^2}\delta_l(\bm{k},t)$. We conclude that the initial conditions induce:
\begin{equation}
F_{2}^{R}(\bm{k}_1,\bm{k}_2)=\left(-\frac{5}{2}\frac{(\bm{k}_1^2+\bm{k}_2^2)}{\bm{k}_1^2\bm{k}_2^2}+\frac{5}{4}\frac{\bm{k}_1\cdot\bm{k}_2}{\bm{k}_1^2\bm{k}_2^2}\right)\,. \label{F2R}
\end{equation}
\subsubsection{SPT kernels}
\label{sec:SPTkernsync}
In comoving gauge the relativistic kernels are given by
\begin{align}
F_1^R(\bm{k}) = & 0\,, \\
F_{2}^{R}(\bm{k}_1,\bm{k}_2)= & -\frac{5}{2}\frac{(\bm{k}_1^2+\bm{k}_2^2)}{\bm{k}_1^2\bm{k}_2^2}+\frac{5}{4}\frac{\bm{k}_1\cdot\bm{k}_2}{\bm{k}_1^2\bm{k}_2^2}\,, \label{F2RSync} \\
F_{2}^\psi(\bm{k}_1,\bm{k}_2) = & \frac{1}{4 \bm{k}_{12}^{2}}\left[1-6F_2(\bm{k}_1,\bm{k}_2)-4G_2(\bm{k}_1,\bm{k}_2 )-\frac{(\bm{k}_1\cdot\bm{k}_2)^2}{\bm{k}_1^2\bm{k}_2^2}\right]\,,\\ 
G_{2}^{R}(\bm{k}_1,\bm{k}_2)= &  F_{2}^{R}(\bm{k}_1,\bm{k}_2)-3F_{2}^{\psi}(\bm{k}_1,\bm{k}_2)-\frac{15}{2}\frac{\bm{k}_1\cdot\bm{k}_2}{\bm{k}_1^2\bm{k}_2^2}\,, \label{G2RSync}
\end{align}
\begin{equation}
	\bm{G}^{T}_{2R}(\bm{k}_1,\bm{k}_2)=  4\frac{\bm{k}}{\bm{k}^2} F_{2}^{\psi}(\bm{k}_1,\bm{k}_2)+ \frac{5}{\bm{k}_2^2}\left(-\frac{\bm{k}_1}{\bm{k}_1^2} + \frac{\bm{k}}{\bm{k}^2}\frac{\bm{k}.\bm{k}_1}{\bm{k}_1^2} + \frac{\bm{k}_1}{\bm{k}_1^2} \frac{\bm{k}_2^2}{\bm{k}^2} + \frac{\bm{k}_1.\bm{k}_2}{\bm{k}_1^2}\frac{\bm{k}_2}{\bm{k}^2}\right)
\end{equation}
\begin{align}
F_{3}^R(\bm{k}_1,\bm{k}_2,\bm{k}_3) = & \frac{5}{7}\alpha(\bm{k_1},\bm{k}_{23})F_{2R}(\bm{k}_2,\bm{k}_3)+G_{2R}(\bm{k}_2,\bm{k}_3)\left(\frac{5}{7}\alpha(\bm{k}_{23},\bm{k}_1)+\frac{4}{7}\beta(\bm{k}_1,\bm{k}_{23})\right)\nonumber \\
& +F_{2}^{\psi}(\bm{k}_2,\bm{k}_3)\left(\frac{19}{7}+\frac{6}{7}\frac{\bm{k}_1\cdot\bm{k}_{23}}{\bm{k}_1^2}\right)+\frac{95}{14}\frac{(\bm{k}_1\cdot\bm{k}_3)}{\bm{k}^2_1 \bm{k}_3^2}\nonumber \\
&+\frac{1}{7}\bm{G}^{T}_{2R}(\bm{k}_1,\bm{k}_3)\cdot\bm{k}_2\left(7+4\frac{\bm{k}_{13}\cdot\bm{k}_2}{\bm{k}_2^2}\right)+\frac{15}{7}\frac{(\bm{k}_1\cdot\bm{k}_2)(\bm{k}_2\cdot\bm{k}_3)}{\bm{k}^2_1 \bm{k}^2_2\bm{k}_3^2}\label{F3R}
\end{align}
\begin{equation}
F_{3}^{\psi}(\bm{k}_1,\bm{k}_2,\bm{k}_3)=\frac{1}{2\bm{k}_{123}^2}\left[G_2(\bm{k}_1,\bm{k}_{3})\left(1-\frac{(\bm{k}_{13}\cdot\bm{k}_{2})^2}{\bm{k}_{13}^2\bm{k}^2_{2}}\right)-3F_3(\bm{k}_1,\bm{k}_2\bm{k}_3)-2G_3(\bm{k}_1,\bm{k}_2\bm{k}_3) \right]
\end{equation}
\begin{align}
G_{3}^R(\bm{k}_1,\bm{k}_2,\bm{k}_3) = & 2F_{3}^{R}(\bm{k}_1,\bm{k}_2,\bm{k}_3) -6F_{3}^{\psi}(\bm{k}_1,\bm{k}_2,\bm{k}_3)-\alpha(\bm{k}_1,\bm{k}_{23})F_{2}^{R}(\bm{k}_2,\bm{k}_3)\nonumber\\
&-\alpha(\bm{k}_{13},\bm{k}_2)G_{2}^{R}(\bm{k}_1,\bm{k}_3)-\frac{15}{2}\frac{\bm{k}_1\cdot\bm{k}_3}{\bm{k}_1^2\bm{k}_3^2}-\frac{15}{2}G_{2}(\bm{k}_{1},\bm{k}_3)\frac{\bm{k}_{13}\cdot\bm{k}_2}{\bm{k}_{13}^2\bm{k}_2^2}\nonumber \\
& -3F_{2}^{\psi}(\bm{k}_2,\bm{k}3)\left(1-\frac{\bm{k}_1\cdot\bm{k}_{23}}{\bm{k}_1^2}\right) -\bm{k}_2\cdot\bm{G}^T_{2R}
\end{align}
\begin{align}
\bm{G}^T_{3R}(\bm{k}_1,\bm{k}_2,\bm{k}_3) =&  \frac{8\bm{k}_{123}}{\bm{k}_{123}^2}F_3^{\psi}(\bm{k}_1,\bm{k}_2,\bm{k}_3) \nonumber \\
&+\frac{5}{\bm{k}_2^2}G_2(\bm{k}_1,\bm{k}_3)\left(-\frac{\bm{k}_{13}}{\bm{k}_{13}^2}+\frac{\bm{k}_{123}}{\bm{k}_{123}^2}\frac{\bm{k}_{123}\cdot\bm{k}_{13}}{\bm{k}_{13}^2}+\frac{\bm{k}_{13}}{\bm{k}_{13}^2}\frac{\bm{k}_2^2}{\bm{k}_{123}^2}+\frac{\bm{k}_{13}\cdot\bm{k}_2}{\bm{k}_{123}^2}\right)\nonumber\\
&-2 F_2^\psi(\bm{k}_2,\bm{k}_3)\left(-\frac{\bm{k}_1}{\bm{k}_1^2}+\frac{\bm{k}_{123}}{\bm{k}_{123}^2}\frac{\bm{k}_{123}\cdot\bm{k}_1}{\bm{k}_1^2}+\frac{\bm{k}_1}{\bm{k}_1^2}\frac{\bm{k}_{23}^2}{\bm{k}_{123}^2}+\frac{\bm{k}_1\cdot\bm{k}_{23}}{\bm{k}_1^2}\frac{\bm{k}_{23}}{\bm{k}_{123}^2}\right)
\end{align}
\begin{align}
&F_4^R(\bm{k}_1,\bm{k}_2,\bm{k}_3,\bm{k}_4)=  \frac{7}{18}\alpha\left(\bm{k}_1,\bm{k}_{234}\right)F_3^R\left(\bm{k}_2,\bm{k}_3,\bm{k}_4\right) + \frac{7}{18} \alpha\left(\bm{k}_{12},\bm{k}_{34}\right)G_2\left(\bm{k}_1,\bm{k}_2\right) F_{2R}\left(\bm{k}_3,\bm{k}_4 \right)\nonumber \\
&+\frac{7}{18}\alpha\left(\bm{k}_{134},\bm{k}_2\right)G_3^R\left(\bm{k}_1,\bm{k}_3,\bm{k}_4\right)+\frac{7}{18} \alpha\left(\bm{k}_{12},\bm{k}_{34}\right)G_{2R}\left(\bm{k}_1,\bm{k}_2\right) F_{2}\left(\bm{k}_3,\bm{k}_4 \right)\nonumber\\
&+\frac{2}{9}\beta\left(\bm{k}_{134},\bm{k}_2\right)G_3^R\left(\bm{k}_1,\bm{k}_3,\bm{k}_4\right)+\frac{2}{9}\beta\left(\bm{k}_{12},\bm{k}_{34}\right)G_2\left(\bm{k}_1,\bm{k}_2\right)G_{2R}\left(\bm{k}_3,\bm{k}_4\right)\nonumber\\
&+\frac{1}{18}F_{2}^{\psi}(\bm{k}_3,\bm{k}_4)\left[9F_2\left(\bm{k}_1,\bm{k}_2\right)+G_2\left(\bm{k}_1,\bm{k}_2\right)\left(4+6\frac{\bm{k}_{12}\cdot\bm{k}_{34}}{\bm{k}_{12}^2}\right)\right] \nonumber\\
&+ \frac{5}{36}G_2\left(\bm{k}_1,\bm{k}_4\right)\left(29\frac{\bm{k}_{14}\cdot\bm{k}_3}{\bm{k}_{14}^2\bm{k}_3^2}+12\frac{(\bm{k}_{14}\cdot\bm{k}_2)(\bm{k}_2\cdot\bm{k}_3)}{\bm{k}^2_{14}\bm{k}_3^2\bm{k}_2^2}\right)+\frac{105}{36}\frac{\bm{k}_1\cdot\bm{k}_3}{\bm{k}_1^2\bm{k}_3^2}F_2\left(\bm{k}_2,\bm{k}_4\right)\nonumber\\
& -\frac{1}{18}F_{2}^{\psi}(\bm{k}_3,\bm{k}_4)\left(\frac{(\bm{k}_1\cdot\bm{k}_{34})}{\bm{k}_1^2}+6\frac{(\bm{k}_1\cdot\bm{k}_2)(\bm{k}_2\cdot\bm{k}_{34})}{\bm{k}_1^2\bm{k}_2^2}\right)+\frac{1}{18}\bm{G}_{3R}^T(\bm{k}_1,\bm{k}_3,\bm{k}_4)\cdot\bm{k}_2\left(9+4\frac{\bm{k}_{134}\cdot\bm{k}_2}{\bm{k}_2^2}\right)\nonumber\\
&+\frac{1}{18}\bm{G}_{2R}^{T}(\bm{k}_1,\bm{k}_3)\cdot\bm{k}_{24}\left(7F_2(\bm{k}_2,\bm{k}_4)+2G_2(\bm{k}_2,\bm{k}_4)\frac{\bm{k}_{13}\cdot\bm{k}_{24}}{\bm{k}_{24}^2}\right)
\end{align}

\subsection{Gauge transformation: From comoving to Poisson.}  
In this subsection, we check our results with the existing literature up to second order. Following the notations of \cite{Matarrese:1997ay}, we perform a gauge transformation between Poisson and comoving gauge.

\par First our derivation of $F_1^R$ in equation \eqref{F1RPoiss} agrees with known results, for instance with equation (6.3) of \cite{Matarrese:1997ay}: $\delta_P^{(1)}=\delta_S^{(1)}-2 \psi$. From this equality, we also deduce: $\alpha_{(1)}=\frac{\eta}{3} \psi$.
\par Second, with equation\footnote{Note that we define unlike \cite{Matarrese:1997ay}, $\delta= \delta^{(1)}+ \delta^{(2)}$.} (3.29) of \cite{Matarrese:1997ay}:
\begin{equation}
\delta_P^{(2)}=\delta_S^{(2)} + \frac{\rho'}{\rho} \alpha_{(2)}+\alpha_{(1)} \frac{\delta \rho'}{\rho},
\end{equation}
 $\alpha_{(2)}$ given by equation (78) and (86) of \cite{Boubekeur:2008kn} and our expression in the weak field limit of $\delta_S^{(2)}$ \eqref{eq:delta_synchro}, we obtain $\delta_P^{(2)}$. The expression we obtain agrees with equation (59) of \cite{2010JCAP...05..004F} which has been used to construct $F_2^R$ in \eqref{F2RPoiss}.
\par The third and forth order relativistic kernels are new.    
   
\section{Consistency Relation}
\label{app:consistency}
In order to check the validity of our solutions,  we used the consistency relations from Ref. \cite{Creminelli:2013mca}. The consistency relation is the relation between a n-point function of the density perturbation $\delta$ and a $(n+1)$-point function in the squeezed limit. This relation comes from the fact that the effect of a long wavelength mode on short scales is equivalent to a coordinate transformation at scales where the curvature perturbation in comoving gauge $\zeta$ is conserved (as is the case in matter domination even deep inside the horizon). 
\subsection{Poisson Gauge}
In this section we want to prove that the relativistic kernels from section \ref{sec:SPTkernpoiss} satisfy the relativistic consistency relation for large scale structure in a matter dominated  universe. Using equation (60) in \cite{Creminelli:2013mca}, equations \eqref{linearpoisson} and \eqref{eq:Poisson} the consistency relation at equal times is given by  
\begin{align}
&\langle \delta_{\bm{q}}(\eta)\delta_{\bm{k}_1}(\eta_1)\cdots\delta(\eta_n)_{\bm{k}_n}\rangle'_{\bm{q}\rightarrow{0}}=-\frac{3H_0^2a(\eta)}{2q^2}P_L(q)\left[3n-5+ \frac{1}{3}\sum_a \left(5 \bm{k}_a\cdot\bm{\partial}_{k_a} + \eta_a\partial_{\eta_a}\right)+\frac{1}{6}\sum_a\bm{q}\cdot\bm{k}_a\eta_a^2\right. \nonumber\\
&\left.\left.+\frac{5}{6}q^iD_i-2\sum_a\left(1-\frac{1}{6}\eta_a\partial_{\eta_a}\right)\bm{q}\cdot\bm{\partial}_{k_a}+\sum_a\frac{8\bm{q}\cdot\bm{k}_a}{k_a^2\left(2+k_a^2\eta_a^2/6\right)}\right]\langle\delta_{\bm{k}_1}(\eta_1)\cdots\delta(\eta_n)_{\bm{k}_n}\rangle'\right|_{\eta_1=\cdots=\eta_n=\eta}\label{CRP},
\end{align}
where the primes mean that momentum conserving delta was removed. And
\begin{equation}
q^iD_i\equiv\sum_{a=1}^n\left[6\bm{q}\cdot\bm{\partial}_{k_a}-\bm{q}\cdot\bm{k}_a\bm{\partial}^2_{k_a}+2\bm{k}_a\cdot\bm{\partial}_{k_a}(\bm{q}\cdot\bm{\partial}_{k_a})\right].
\end{equation}

For $n=2$ the left hand side of equation \eqref{CRP} gives 
\begin{align}
\left\langle \delta_{\bm{q}}(\eta)\delta_{\bm{k}_1}(\eta_1)\delta_{\bm{k}_2}(\eta_2)\right\rangle'_{q\rightarrow{0}} & = 
2H_0^2\left[F_2^R(-\bm{q},\bm{k}_1 + \bm{q})P(|\bm{k}_1 + \bm{q}|) + F_2^R(-\bm{q},-\bm{k}_1)P(k_1)\right]a^{3}(\eta)P_L(q) \nonumber\\
&= -\frac{a^{3}(\eta)H_0^2}{q^2}\left(6 +3\frac{\bm{q}\cdot\bm{k}_1}{k_1^2}\right)P_L(q)P_L(k_1), 
\end{align}
where the last  line was obtained by taking the limit for $q\rightarrow{0}$ of equation \eqref{F2RPoiss}. For the 2-point function appearing on the right  hand side we obtain
\begin{equation}
\langle\delta_{\bm{k}_1}(\eta_1)\delta_{\bm{k}_2}(\eta_2)\rangle = \left[ a(\eta_1)a(\eta_2)+2 a(\eta_1)F_1^R(k_1)+F_1^R(k_1)F_1^R(k_2)\right] P_L(k_1),
\end{equation}
therefore, the first line in the l.h.s of  the squared bracket is  
\begin{align}
\left.\left[1+ \frac{1}{3}\left(5 \bm{k}_1\cdot\bm{\partial}_{k_1} + \eta\partial_{\eta}\right)\right]\langle\delta_{\bm{k}_1}(\eta_1)\delta_{\bm{k}_2}(\eta_2)\rangle \right|_{\eta_1=\eta_2=\eta}= 4a^2(\eta)P_L(k_1)
\end{align}
Note that we only used the Newtonian power spectrum since the operator acting on it is already suppressed by $\epsilon^2 = a^2H^2/k^2$ with respect to the Newtonian expression, and we have used the fact that $P_{L}(k)\sim k$ since we assumed matter domination throughout.\footnote{If more realistic initial conditions, including a period of radiation domination, are used, both the power spectrum and the second order relativistic kernel need to be modified as discussed in the text. We expect everything to be consistent in that case.} The first term in the second line corresponds to the Newtonian consistency relation \cite{Peloso:2013zw, Kehagias:2013yd}
\begin{equation}
\frac{1}{6}\sum_a\bm{q}\cdot\bm{k}_a\eta_a^2\langle\delta_{\bm{k}_1}(\eta_1)\delta_{\bm{k}_2}(\eta_2)\rangle = \frac{1}{6}\bm{q}\cdot\left(\bm{k}_1\eta_1^2+\bm{k}_2\eta_2^2\right) \langle\delta_{\bm{k}_1}(\eta_1)\delta_{\bm{k}_2}(\eta_2)\rangle.
\end{equation}
However, since $\bm{k}_1+\bm{k}_2 = \bm{q}$, we neglect this term as it is suppressed by too many powers of $q$ in the squeezed limit. It is also easy to see that in the weak field approximation $k\eta \ll 1$, the last term in equation~\eqref{CRP} is also negligible. The remaining terms give

\begin{align}
&\left.\left[\frac{5}{6}q^iD_i-2\sum_a\left(1-\frac{1}{6}\eta_a\partial_{\eta_a}\right)\bm{q}\cdot\bm{\partial}_{k_a}\right]\langle\delta_{\bm{k}_1}(\eta_1)\delta_{\bm{k}_2}(\eta_2)\rangle'\right|_{\eta_1=\eta_2=\eta} = 2 \frac{\bm{q}\cdot\bm{k}_1}{k_1^2} a^{2}(\eta) P_L(k_1) 
\end{align}

Putting everything together, the left hand side gives:
\begin{equation}
\langle\delta_{\bm{q}}\delta_{\bm{k}_1}(\eta_1)\delta_{\bm{k}_2}(\eta_2)\rangle'_{\bm{q}\rightarrow{0}} = -\frac{a^3(\eta)H_0^2}{q^2}\left[6 +3\frac{\bm{q}\cdot\bm{k}_1}{\bm{k}_1^2} \right]P_L(q)P_L(k_1)
\end{equation}
which is in agreement with right hand side.

Now For n=3 we have on the left hand side
\begin{align}
\left\langle\delta_{\bm{q}}(\eta)\delta_{\bm{k}_1}(\eta_1)\delta_{\bm{k}_2}(\eta_2)\delta_{\bm{k}_3}(\eta_3)\right\rangle'_{q\rightarrow{0}} =&  \langle \delta^{(1)}_{\bm{q}}(\eta)\delta^{(3)}_{\bm{k}_1}(\eta_1)\delta^{(1)}_{\bm{k}_2}(\eta_2)\delta^{(1)}_{\bm{k}_3}(\eta_3)\rangle + 2 \text{ perm} \nonumber \\
&\left\langle\delta^{(1)}_{\bm{q}}(\eta)\delta^{(2)}_{\bm{k}_1}(\eta_1)\delta^{(2)}_{\bm{k}_2}(\eta_2)\delta^{(1)}_{\bm{k}_3}(\eta_3)\right\rangle + 2\text{ perm} 
\end{align}
note that in order to check the consistency relation in  this case it is enough to compare only a specific combination of the momentum, we will compare the terms on both sides which are proportional to $P_{L}(k_2)P_{L}(k_3)$. On the l.h.s we have:   
\begin{align}
	 \left\langle\delta_{\bm{q}}(\eta)\delta_{\bm{k}_1}(\eta_1)\delta_{\bm{k}_2}(\eta_2)\delta_{\bm{k}_3}(\eta_3)\right\rangle'_{q\rightarrow{0}} \supset &\left[ 6 \left(F_1^R(\bm{k}_2)+F_1^R(\bm{k}_3)\right)F_3(-\bm{q},-\bm{k}_2,-\bm{k}_3) +6 F_3^R(-\bm{q},-\bm{k}_2,-\bm{k}_3)\right. \nonumber \\
	 &+4F_2^R(-\bm{q},\bm{q}+\bm{k}_2)F_2(-\bm{q}-\bm{k}_2,-\bm{k}_3)\left(1+\frac{\bm{q}\cdot\bm{k}_2}{\bm{k}_2^2}\right) + \bm{k}_2\leftrightarrow\bm{k}_3 \nonumber \\   
	&\left.+4F_2(-\bm{q},\bm{q}+\bm{k}_2)F^R_2(-\bm{q}-\bm{k}_2,-\bm{k}_3)\left(1+\frac{\bm{q}\cdot\bm{k}_2}{\bm{k}_2^2}\right)+ \bm{k}_2\leftrightarrow\bm{k}_3\right]  \nonumber \\   
	&\left.a(\eta) a^2(\eta_1)a(\eta_2)a(\eta_3) H_0^2P_L(q)P_L(k_2)P_L(k_3)\right|_{\eta_{1,2,3}=\eta}\nonumber \\
	=& -30\frac{a^5(\eta) H_0^2}{q^2} F_2(\bm{k}_2,\bm{k}_3)P_L(q)P_L(k_2)P_L(k_3) + \mathcal{O}(q^{-1}). \label{D.10}
\end{align} 
For the r.h.s we have
\begin{align}
\langle\delta_{\bm{k}_1}(\eta_1)\delta_{\bm{k}_2}(\eta_2)\delta_{\bm{k}_3}(\eta_3)\rangle &\supset \langle\delta^{(2)}_{\bm{k}_1}(\tau_1)\delta^{(1)}_{\bm{k}_2}(\eta_2)\delta^{(1)}_{\bm{k}_3}(\eta_3)\rangle \nonumber \\ 
&= 2 a^2(\eta_1)a(\eta_2)a(\eta_3) F_2(\bm{k}_2,\bm{k}_3)P_L(k_2)P_L(k_3)
\end{align}
therefore, the part corresponding to the dilation transformation gives:
\begin{align}
\left.\left[4+\frac{1}{3}\left(5\sum_a \bm{k}_a\cdot\bm{\partial}_a + \eta\partial_{\eta_a}\right)\right]\langle\delta^{(2)}_{\bm{k}_1}(\eta_1)\delta^{(1)}_{\bm{k}_2}(\eta_2)\delta^{(1)}_{\bm{k}_3}(\eta_3)\rangle\right|_{\eta_{1,2,3}=\eta}= 20a^4(\eta)F_2(k_2,k_3)P_L(k_2)P_L(k_3).
\end{align}
We thus have
\begin{align}
\left.\left\langle\delta_{\bm{q}}(\eta)\delta_{\bm{k}_1}(\eta_1)\delta_{\bm{k}_2}(\eta_2)\delta_{\bm{k}_3}(\eta_3)\right\rangle'_{q\rightarrow{0}}\right|_{\eta_{1,2,3}=\eta}  & \supset  -30\frac{ a^5(\eta)H^2_0}{q^2} F_2(\bm{k}_2,\bm{k}_3) P_L(q)P_L(k_2)P_L(k_3) 
\end{align}
which is in agreement with \eqref{D.10} up to order $\mathcal{O}(q^{-2})$.

\subsection{Comoving Gauge}
We now check the consistency condition for the comoving gauge. It is given by
%
\begin{align}
\left\langle \delta_{\bm{q}}\delta_{\bm{k}_1}(\eta_1)\cdots\delta(\eta_n)_{\bm{k}_n}\right\rangle'_{q\rightarrow{0}}=-\frac{5H_0^2}{2q^2}a(\eta)P_L(q)&\left[ 3(n-1)+\sum_a \bm{k}_a\cdot\bm{\partial}_{k_a}\right.\nonumber \\
& \left.+\frac{1}{2}q^iD_i-\frac{1}{5}\sum_a \bm{q}\cdot\bm{k}_a\eta^2_a\right]\left\langle\delta_{\bm{k}_1}(\eta_1)\cdots\delta(\eta_n)_{\bm{k}_n}\right\rangle'. \label{CRS}
\end{align}
We start by checking that this relation is satisfied for $n=2$. For the l.h.s at equal times we obtain
\begin{align}
\left\langle \delta_{\bm{q}}(\eta)\delta_{\bm{k}_1}(\eta)\delta(\eta)_{\bm{k}_n}\right\rangle'_{q\rightarrow{0}} & =2H_0^2\left[F_2^R(-\bm{q},\bm{k}_1 + \bm{q})P(|\bm{k}_1 + \bm{q}|) + F_2^R(-\bm{q},-\bm{k}_1)P(k_1)\right]a^3(\eta)P_L(q) \nonumber\\
&=2H_0^2 \left[F_2^R(-\bm{q},\bm{k}_1)\left(1+\frac{\bm{q}\cdot\bm{k}_1}{k_1^2}\right)+F_2^R(-\bm{q},-\bm{k}_1)\right]a^3(\eta)P_L(q)P(k_1) \nonumber\\
&= -\frac{a^3(\eta)H_0^2}{q^2}\left(10 +5\frac{\bm{q}\cdot\bm{k}_1}{k_1^2}\right)P_L(q)P_L(k_1), \label{CRF2RS}
\end{align}
where the last line was obtained by using equation \eqref{F2RSync} and taking the corresponding limit up to order $\mathcal{O}(q^{-1})$.

Now for the r.h.s we obtain 
\begin{align}
\left[3+\sum_a \bm{k}_a\cdot\bm{\partial}_{k_a}+\frac{1}{2} q^iD_i\right]\langle\delta_{\bm{k}_1}(\eta)\delta_{\bm{k}_n}(\eta)\rangle'&=\left[3+\bm{k_1}\cdot\bm{\partial}_{k_1}+\frac{1}{2}\bm{q}\cdot\bm{k_1}\left(\frac{4}{k_1}\partial_{k_1}+\partial^2_{k_1}\right)\right]a^{2}(\eta)P(k_1)\nonumber \\
& = \left(4 + 2\frac{\bm{q}\cdot\bm{k}_1}{k_1^2} \right)a(\eta)^{2}P(k_1),
\end{align}
we have neglected last term in the consistency relation \eqref{CRS} since for equal times it is order $\mathcal{O}(q )$, and too suppressed by $q$ to be captured by this relation. 
%
%
Therefore, for the r.h.s we get 
\begin{equation}
\langle\delta_{\bm{q}}(\eta)\delta_{\bm{k}_1}(\eta)\delta_{\bm{k}_n}(\eta)\rangle' = -\frac{a^3(\eta)H_0^2}{q^2}\left(10+5\frac{\bm{q}\cdot\bm{k}_1}{k_1^2}\right)  P_L(k_1)P_L(q),
\end{equation}
which is in agreement with equation \eqref{CRF2RS}.

For $n=3$ on the left hand side we use the same combination as in Poisson gauge \eqref{D.10}. Taking the limit $q\rightarrow0$ on the left hand side and computing the dilation part on the right hand side we obtain the  same result given by 
\begin{align}
\left\langle\delta_{\bm{q}}(\eta)\delta_{\bm{k}_1}(\eta)\delta_{\bm{k}_2}(\eta)\delta_{\bm{k}_3}(\eta)\right\rangle'_{q\rightarrow{0}} \supset  -40\frac{a^5(\eta)H_0^2}{q^2} F_2(\bm{k}_2,\bm{k}_3)  P_L(q)P_L(k_2)P_L(k_3) + \mathcal{O}(\bm{q}^{-1}).\label{D.19}
\end{align}

\bibliographystyle{ieeetr}
\bibliography{bibPTwithRG}

\begin{thebibliography}{10}

\bibitem{Maldacena:2002vr}
J.~M. Maldacena, ``{Non-Gaussian features of primordial fluctuations in single
  field inflationary models},'' {\em JHEP}, vol.~05, p.~013, 2003.

\bibitem{Creminelli:2004yq}
P.~Creminelli and M.~Zaldarriaga, ``{Single field consistency relation for the
  3-point function},'' {\em JCAP}, vol.~0410, p.~006, 2004.

\bibitem{Creminelli:2011rh}
P.~Creminelli, G.~D'Amico, M.~Musso, and J.~Nore\~na, ``{The (not so) squeezed
  limit of the primordial 3-point function},'' {\em JCAP}, vol.~1111, p.~038,
  2011.

\bibitem{Creminelli:2012ed}
P.~Creminelli, J.~Nore\~na, and M.~Simonović, ``{Conformal consistency
  relations for single-field inflation},'' {\em JCAP}, vol.~1207, p.~052, 2012.

\bibitem{Hinterbichler:2012nm}
K.~Hinterbichler, L.~Hui, and J.~Khoury, ``{Conformal Symmetries of Adiabatic
  Modes in Cosmology},'' {\em JCAP}, vol.~1208, p.~017, 2012.

\bibitem{Arkani-Hamed:2015bza}
N.~Arkani-Hamed and J.~Maldacena, ``{Cosmological Collider Physics},'' 2015.

\bibitem{Peloso:2013zw}
M.~Peloso and M.~Pietroni, ``{Galilean invariance and the consistency relation
  for the nonlinear squeezed bispectrum of large scale structure},'' {\em
  JCAP}, vol.~1305, p.~031, 2013.

\bibitem{Kehagias:2013yd}
A.~Kehagias and A.~Riotto, ``{Symmetries and Consistency Relations in the Large
  Scale Structure of the Universe},'' {\em Nucl. Phys.}, vol.~B873,
  pp.~514--529, 2013.

\bibitem{Creminelli:2013mca}
P.~Creminelli, J.~Nore\~na, M.~Simonović, and F.~Vernizzi, ``{Single-Field
  Consistency Relations of Large Scale Structure},'' {\em JCAP}, vol.~1312,
  p.~025, 2013.

\bibitem{Baldauf:2016sjb}
T.~Baldauf, M.~Mirbabayi, M.~Simonović, and M.~Zaldarriaga, ``{LSS constraints
  with controlled theoretical uncertainties},'' 2016.

\bibitem{Ade:2015ava}
P.~A.~R. Ade {\em et~al.}, ``{Planck 2015 results. XVII. Constraints on
  primordial non-Gaussianity},'' {\em Astron. Astrophys.}, vol.~594, p.~A17,
  2016.

\bibitem{dePutter:2016trg}
R.~de~Putter, J.~Gleyzes, and O.~Doré, ``{Next non-Gaussianity frontier: What
  can a measurement with {$\sigma(f_{\text{NL}})\lesssim 1$} tell us about
  multifield inflation?},'' {\em Phys. Rev.}, vol.~D95, no.~12, p.~123507,
  2017.

\bibitem{Tellarini:2016sgp}
M.~Tellarini, A.~J. Ross, G.~Tasinato, and D.~Wands, ``{Galaxy bispectrum,
  primordial non-Gaussianity and redshift space distortions},'' {\em JCAP},
  vol.~1606, no.~06, p.~014, 2016.

\bibitem{Castorina:2018zfk}
E.~Castorina, Y.~Feng, U.~Seljak, and F.~Villaescusa-Navarro, ``{Primordial
  non-Gaussianities and zero bias tracers of the Large Scale Structure},'' {\em
  Phys. Rev. Lett.}, vol.~121, no.~10, p.~101301, 2018.

\bibitem{Dalal:2007cu}
N.~Dalal, O.~Dore, D.~Huterer, and A.~Shirokov, ``{The imprints of primordial
  non-gaussianities on large-scale structure: scale dependent bias and
  abundance of virialized objects},'' {\em Phys. Rev.}, vol.~D77, p.~123514,
  2008.

\bibitem{Matarrese:2008nc}
S.~Matarrese and L.~Verde, ``{The effect of primordial non-Gaussianity on halo
  bias},'' {\em Astrophys. J.}, vol.~677, pp.~L77--L80, 2008.

\bibitem{Slosar:2008hx}
A.~Slosar, C.~Hirata, U.~Seljak, S.~Ho, and N.~Padmanabhan, ``{Constraints on
  local primordial non-Gaussianity from large scale structure},'' {\em JCAP},
  vol.~0808, p.~031, 2008.

\bibitem{Yoo:2012se}
J.~Yoo, N.~Hamaus, U.~Seljak, and M.~Zaldarriaga, ``{Going beyond the Kaiser
  redshift-space distortion formula: a full general relativistic account of the
  effects and their detectability in galaxy clustering},'' {\em Phys. Rev.},
  vol.~D86, p.~063514, 2012.

\bibitem{Yoo:2009au}
J.~Yoo, A.~L. Fitzpatrick, and M.~Zaldarriaga, ``{A New Perspective on Galaxy
  Clustering as a Cosmological Probe: General Relativistic Effects},'' {\em
  Phys. Rev.}, vol.~D80, p.~083514, 2009.

\bibitem{Challinor:2011bk}
A.~Challinor and A.~Lewis, ``{The linear power spectrum of observed source
  number counts},'' {\em Phys. Rev.}, vol.~D84, p.~043516, 2011.

\bibitem{Bonvin:2011bg}
C.~Bonvin and R.~Durrer, ``{What galaxy surveys really measure},'' {\em Phys.
  Rev.}, vol.~D84, p.~063505, 2011.

\bibitem{Jeong:2011as}
D.~Jeong, F.~Schmidt, and C.~M. Hirata, ``{Large-scale clustering of galaxies
  in general relativity},'' {\em Phys. Rev.}, vol.~D85, p.~023504, 2012.

\bibitem{Baumann:2010tm}
D.~Baumann, A.~Nicolis, L.~Senatore, and M.~Zaldarriaga, ``{Cosmological
  Non-Linearities as an Effective Fluid},'' {\em JCAP}, vol.~1207, p.~051,
  2012.

\bibitem{Carrasco:2012cv}
J.~J.~M. Carrasco, M.~P. Hertzberg, and L.~Senatore, ``{The Effective Field
  Theory of Cosmological Large Scale Structures},'' {\em JHEP}, vol.~09,
  p.~082, 2012.

\bibitem{Cataneo:2016suz}
M.~Cataneo, S.~Foreman, and L.~Senatore, ``{Efficient exploration of cosmology
  dependence in the EFT of LSS},'' {\em JCAP}, vol.~1704, no.~04, p.~026, 2017.

\bibitem{Umeh:2012pn}
O.~Umeh, C.~Clarkson, and R.~Maartens, ``{Nonlinear relativistic corrections to
  cosmological distances, redshift and gravitational lensing magnification: I.
  Key results},'' {\em Class. Quant. Grav.}, vol.~31, p.~202001, 2014.

\bibitem{Umeh:2014ana}
O.~Umeh, C.~Clarkson, and R.~Maartens, ``{Nonlinear relativistic corrections to
  cosmological distances, redshift and gravitational lensing magnification. II
  - Derivation},'' {\em Class. Quant. Grav.}, vol.~31, p.~205001, 2014.

\bibitem{DiDio:2014lka}
E.~Di~Dio, R.~Durrer, G.~Marozzi, and F.~Montanari, ``{Galaxy number counts to
  second order and their bispectrum},'' {\em JCAP}, vol.~1412, p.~017, 2014.
\newblock [Erratum: JCAP1506,no.06,E01(2015)].

\bibitem{DiDio:2015bua}
E.~Di~Dio, R.~Durrer, G.~Marozzi, and F.~Montanari, ``{The bispectrum of
  relativistic galaxy number counts},'' {\em JCAP}, vol.~1601, p.~016, 2016.

\bibitem{Umeh:2016nuh}
O.~Umeh, S.~Jolicoeur, R.~Maartens, and C.~Clarkson, ``{A general relativistic
  signature in the galaxy bispectrum: the local effects of observing on the
  lightcone},'' {\em JCAP}, vol.~1703, no.~03, p.~034, 2017.

\bibitem{Fidler:2015npa}
C.~Fidler, C.~Rampf, T.~Tram, R.~Crittenden, K.~Koyama, and D.~Wands,
  ``{General relativistic corrections to $N$-body simulations and the
  Zel'dovich approximation},'' {\em Phys. Rev.}, vol.~D92, no.~12, p.~123517,
  2015.

\bibitem{Fidler:2016tir}
C.~Fidler, T.~Tram, C.~Rampf, R.~Crittenden, K.~Koyama, and D.~Wands,
  ``{Relativistic Interpretation of Newtonian Simulations for Cosmic Structure
  Formation},'' {\em JCAP}, vol.~1609, no.~09, p.~031, 2016.

\bibitem{Fidler:2017pnb}
C.~Fidler, T.~Tram, C.~Rampf, R.~Crittenden, K.~Koyama, and D.~Wands,
  ``{General relativistic weak-field limit and Newtonian N-body simulations},''
  {\em JCAP}, vol.~1712, no.~12, p.~022, 2017.

\bibitem{Koyama:2018ttg}
K.~Koyama, O.~Umeh, R.~Maartens, and D.~Bertacca, ``{The observed galaxy
  bispectrum from single-field inflation in the squeezed limit},'' {\em JCAP},
  vol.~1807, no.~07, p.~050, 2018.

\bibitem{DiDio:2016gpd}
E.~Di~Dio, H.~Perrier, R.~Durrer, G.~Marozzi, A.~Moradinezhad~Dizgah,
  J.~Nore\~na, and A.~Riotto, ``{Non-Gaussianities due to Relativistic
  Corrections to the Observed Galaxy Bispectrum},'' {\em JCAP}, vol.~1703,
  no.~03, p.~006, 2017.

\bibitem{Hwang:2015jja}
J.-c. Hwang, D.~Jeong, and H.~Noh, ``{Cosmological non-linear density and
  velocity power spectra including non-linear vector and tensor modes},'' {\em
  Mon. Not. Roy. Astron. Soc.}, vol.~459, no.~1, pp.~1124--1136, 2016.

\bibitem{Goldberg:2016lcq}
S.~R. Goldberg, T.~Clifton, and K.~A. Malik, ``{Cosmology on all scales: a
  two-parameter perturbation expansion},'' {\em Phys. Rev.}, vol.~D95, no.~4,
  p.~043503, 2017.

\bibitem{Gallagher:2018bdl}
C.~S. Gallagher and T.~Clifton, ``{Relativistic Euler equations in cosmologies
  with nonlinear structures},'' {\em Phys. Rev.}, vol.~D98, no.~10, p.~103516,
  2018.

\bibitem{Kehagias:2015tda}
A.~Kehagias, A.~Moradinezhad~Dizgah, J.~Nore\~na, H.~Perrier, and A.~Riotto,
  ``{A Consistency Relation for the Observed Galaxy Bispectrum and the Local
  non-Gaussianity from Relativistic Corrections},'' {\em JCAP}, vol.~1508,
  no.~08, p.~018, 2015.

\bibitem{Green:2010qy}
S.~R. Green and R.~M. Wald, ``{A new framework for analyzing the effects of
  small scale inhomogeneities in cosmology},'' {\em Phys. Rev.}, vol.~D83,
  p.~084020, 2011.

\bibitem{Green:2011wc}
S.~R. Green and R.~M. Wald, ``{Newtonian and Relativistic Cosmologies},'' {\em
  Phys. Rev.}, vol.~D85, p.~063512, 2012.

\bibitem{Brustein:2011dy}
R.~Brustein and A.~Riotto, ``{Evolution Equation for Non-linear Cosmological
  Perturbations},'' {\em JCAP}, vol.~1111, p.~006, 2011.

\bibitem{Kopp:2013tqa}
M.~Kopp, C.~Uhlemann, and T.~Haugg, ``{Newton to Einstein — dust to dust},''
  {\em JCAP}, vol.~1403, p.~018, 2014.

\bibitem{Adamek:2014xba}
J.~Adamek, R.~Durrer, and M.~Kunz, ``{N-body methods for relativistic
  cosmology},'' {\em Class. Quant. Grav.}, vol.~31, no.~23, p.~234006, 2014.

\bibitem{Bentivegna:2015flc}
E.~Bentivegna and M.~Bruni, ``{Effects of nonlinear inhomogeneity on the cosmic
  expansion with numerical relativity},'' {\em Phys. Rev. Lett.}, vol.~116,
  no.~25, p.~251302, 2016.

\bibitem{Mertens:2015ttp}
J.~B. Mertens, J.~T. Giblin, and G.~D. Starkman, ``{Integration of
  inhomogeneous cosmological spacetimes in the BSSN formalism},'' {\em Phys.
  Rev.}, vol.~D93, no.~12, p.~124059, 2016.

\bibitem{Adamek:2015eda}
J.~Adamek, D.~Daverio, R.~Durrer, and M.~Kunz, ``{General relativity and cosmic
  structure formation},'' {\em Nature Phys.}, vol.~12, pp.~346--349, 2016.

\bibitem{Adamek:2016zes}
J.~Adamek, D.~Daverio, R.~Durrer, and M.~Kunz, ``{gevolution: a cosmological
  N-body code based on General Relativity},'' {\em JCAP}, vol.~1607, no.~07,
  p.~053, 2016.

\bibitem{Daverio:2019gql}
D.~Daverio, Y.~Dirian, and E.~Mitsou, ``{General Relativistic Cosmological
  N-body Simulations I: time integration},'' 2019.

\bibitem{Finke:2019pru}
A.~Finke, ``{The perturbed FLRW metric on all scales: Newtonian limit and
  top-hat collapse},'' 2019.

\bibitem{Takahashi:2008yk}
R.~Takahashi, ``{Third Order Density Perturbation and One-loop Power Spectrum
  in a Dark Energy Dominated Universe},'' {\em Prog. Theor. Phys.}, vol.~120,
  pp.~549--559, 2008.

\bibitem{Fasiello:2016qpn}
M.~Fasiello and Z.~Vlah, ``{Nonlinear fields in generalized cosmologies},''
  {\em Phys. Rev.}, vol.~D94, no.~6, p.~063516, 2016.

\bibitem{Boubekeur:2008kn}
L.~Boubekeur, P.~Creminelli, J.~Nore\~na, and F.~Vernizzi, ``{Action approach
  to cosmological perturbations: the 2nd order metric in matter dominance},''
  {\em JCAP}, vol.~0808, p.~028, 2008.

\bibitem{Matarrese:1997ay}
S.~Matarrese, S.~Mollerach, and M.~Bruni, ``{Second order perturbations of the
  Einstein-de Sitter universe},'' {\em Phys. Rev.}, vol.~D58, p.~043504, 1998.

\bibitem{Bartolo:2005kv}
N.~Bartolo, S.~Matarrese, and A.~Riotto, ``{The full second-order radiation
  transfer function for large-scale cmb anisotropies},'' {\em JCAP}, vol.~0605,
  p.~010, 2006.

\bibitem{Baldauf:2011bh}
T.~Baldauf, U.~Seljak, L.~Senatore, and M.~Zaldarriaga, ``{Galaxy Bias and
  non-Linear Structure Formation in General Relativity},'' {\em JCAP},
  vol.~1110, p.~031, 2011.

\bibitem{Yoo:2014vta}
J.~Yoo, ``{Proper-time hypersurface of nonrelativistic matter flows: Galaxy
  bias in general relativity},'' {\em Phys. Rev.}, vol.~D90, no.~12, p.~123507,
  2014.

\bibitem{Bernardeau:2001qr}
F.~Bernardeau, S.~Colombi, E.~Gaztanaga, and R.~Scoccimarro, ``{Large scale
  structure of the universe and cosmological perturbation theory},'' {\em Phys.
  Rept.}, vol.~367, pp.~1--248, 2002.

\bibitem{2010JCAP...05..004F}
A.~L. {Fitzpatrick}, L.~{Senatore}, and M.~{Zaldarriaga}, ``{Contributions to
  the dark matter 3-Point function from the radiation era},'' {\em JCAP},
  vol.~5, p.~004, May 2010.

\bibitem{Huang:2013qua}
Z.~Huang and F.~Vernizzi, ``{The full CMB temperature bispectrum from
  single-field inflation},'' {\em Phys. Rev.}, vol.~D89, no.~2, p.~021302,
  2014.

\bibitem{Tram:2016cpy}
T.~Tram, C.~Fidler, R.~Crittenden, K.~Koyama, G.~W. Pettinari, and D.~Wands,
  ``{The Intrinsic Matter Bispectrum in $\Lambda$CDM},'' {\em JCAP}, vol.~1605,
  no.~05, p.~058, 2016.

\bibitem{Hahn:2004fe}
T.~Hahn, ``{CUBA: A Library for multidimensional numerical integration},'' {\em
  Comput. Phys. Commun.}, vol.~168, pp.~78--95, 2005.

\bibitem{Simonovic:2017mhp}
M.~Simonović, T.~Baldauf, M.~Zaldarriaga, J.~J. Carrasco, and J.~A. Kollmeier,
  ``{Cosmological perturbation theory using the FFTLog: formalism and
  connection to QFT loop integrals},'' {\em JCAP}, vol.~1804, no.~04, p.~030,
  2018.

\bibitem{Biern:2016kys}
S.~G. Biern and J.~Yoo, ``{Gauge-Invariance and Infrared Divergences in the
  Luminosity Distance},'' {\em JCAP}, vol.~1704, no.~04, p.~045, 2017.

\bibitem{Weinberg:2003sw}
S.~Weinberg, ``{Adiabatic modes in cosmology},'' {\em Phys. Rev.}, vol.~D67,
  p.~123504, 2003.

\bibitem{Scoccimarro:1995if}
R.~Scoccimarro and J.~Frieman, ``{Loop corrections in nonlinear cosmological
  perturbation theory},'' {\em Astrophys. J. Suppl.}, vol.~105, p.~37, 1996.

\bibitem{Peloso:2016qdr}
M.~Peloso and M.~Pietroni, ``{Galilean invariant resummation schemes of
  cosmological perturbations},'' {\em JCAP}, vol.~1701, no.~01, p.~056, 2017.

\bibitem{Takada:2013bfn}
M.~Takada and W.~Hu, ``{Power Spectrum Super-Sample Covariance},'' {\em Phys.
  Rev.}, vol.~D87, no.~12, p.~123504, 2013.

\bibitem{Chan:2017fiv}
K.~C. Chan, A.~Moradinezhad~Dizgah, and J.~Nore\~na, ``{Bispectrum Supersample
  Covariance},'' {\em Phys. Rev.}, vol.~D97, no.~4, p.~043532, 2018.

\bibitem{Pajer:2013jj}
E.~Pajer and M.~Zaldarriaga, ``{On the Renormalization of the Effective Field
  Theory of Large Scale Structures},'' {\em JCAP}, vol.~1308, p.~037, 2013.

\bibitem{Baldauf:2014qfa}
T.~Baldauf, L.~Mercolli, M.~Mirbabayi, and E.~Pajer, ``{The Bispectrum in the
  Effective Field Theory of Large Scale Structure},'' {\em JCAP}, vol.~1505,
  no.~05, p.~007, 2015.

\bibitem{Angulo:2014tfa}
R.~E. Angulo, S.~Foreman, M.~Schmittfull, and L.~Senatore, ``{The One-Loop
  Matter Bispectrum in the Effective Field Theory of Large Scale Structures},''
  {\em JCAP}, vol.~1510, no.~10, p.~039, 2015.

\bibitem{Pajer:2013ana}
E.~Pajer, F.~Schmidt, and M.~Zaldarriaga, ``{The Observed Squeezed Limit of
  Cosmological Three-Point Functions},'' {\em Phys. Rev.}, vol.~D88, no.~8,
  p.~083502, 2013.

\bibitem{Dai:2015rda}
L.~Dai, E.~Pajer, and F.~Schmidt, ``{Conformal Fermi Coordinates},'' {\em
  JCAP}, vol.~1511, no.~11, p.~043, 2015.

\bibitem{Yoo:2010ni}
J.~Yoo, ``{General Relativistic Description of the Observed Galaxy Power
  Spectrum: Do We Understand What We Measure?},'' {\em Phys. Rev.}, vol.~D82,
  p.~083508, 2010.

\bibitem{Yoo:2014sfa}
J.~Yoo and M.~Zaldarriaga, ``{Beyond the Linear-Order Relativistic Effect in
  Galaxy Clustering: Second-Order Gauge-Invariant Formalism},'' {\em Phys.
  Rev.}, vol.~D90, no.~2, p.~023513, 2014.

\bibitem{Thomas:2014aga}
D.~B. Thomas, M.~Bruni, and D.~Wands, ``{Relativistic weak lensing from a fully
  non-linear cosmological density field},'' {\em JCAP}, vol.~1509, no.~09,
  p.~021, 2015.

\bibitem{Andrianomena:2014sya}
S.~Andrianomena, C.~Clarkson, P.~Patel, O.~Umeh, and J.-P. Uzan, ``{Non-linear
  relativistic contributions to the cosmological weak-lensing convergence},''
  {\em JCAP}, vol.~1406, p.~023, 2014.

\bibitem{Bonvin:2014owa}
C.~Bonvin, ``{Isolating relativistic effects in large-scale structure},'' {\em
  Class. Quant. Grav.}, vol.~31, no.~23, p.~234002, 2014.

\bibitem{Durrer:2016jzq}
R.~Durrer and V.~Tansella, ``{Vector perturbations of galaxy number counts},''
  {\em JCAP}, vol.~1607, no.~07, p.~037, 2016.

\bibitem{Giblin:2017ezj}
J.~T. Giblin, J.~B. Mertens, G.~D. Starkman, and A.~R. Zentner, ``{General
  Relativistic Corrections to the Weak Lensing Convergence Power Spectrum},''
  {\em Phys. Rev.}, vol.~D96, no.~10, p.~103530, 2017.

\bibitem{DiDio:2018unb}
E.~Di~Dio, R.~Durrer, R.~Maartens, F.~Montanari, and O.~Umeh, ``{The Full-Sky
  Angular Bispectrum in Redshift Space},'' 2018.

\bibitem{Umeh:2019qyd}
O.~Umeh, K.~Koyama, R.~Maartens, F.~Schmidt, and C.~Clarkson, ``{General
  relativistic effects in the galaxy bias at second order},'' 2019.

\bibitem{DiDio:2018zmk}
E.~Di~Dio and U.~Seljak, ``{The relativistic dipole and gravitational redshift
  on LSS},'' 2018.

\bibitem{Villa:2015ppa}
E.~Villa and C.~Rampf, ``{Relativistic perturbations in $\Lambda$CDM: Eulerian
  \& Lagrangian approaches},'' {\em JCAP}, vol.~1601, no.~01, p.~030, 2016.
\newblock [Erratum: JCAP1805,no.05,E01(2018)].

\bibitem{Senatore:2014via}
L.~Senatore and M.~Zaldarriaga, ``{The IR-resummed Effective Field Theory of
  Large Scale Structures},'' {\em JCAP}, vol.~1502, no.~02, p.~013, 2015.

\bibitem{Floerchinger:2016hja}
S.~Floerchinger, M.~Garny, N.~Tetradis, and U.~A. Wiedemann,
  ``{Renormalization-group flow of the effective action of cosmological
  large-scale structures},'' {\em JCAP}, vol.~1701, no.~01, p.~048, 2017.

\bibitem{Blas:2015tla}
D.~Blas, S.~Floerchinger, M.~Garny, N.~Tetradis, and U.~A. Wiedemann, ``{Large
  scale structure from viscous dark matter},'' {\em JCAP}, vol.~1511, p.~049,
  2015.

\bibitem{Erschfeld:2018zqg}
A.~Erschfeld and S.~Floerchinger, ``{Evolution of dark matter velocity
  dispersion},'' 2018.

\bibitem{Matsubara:2007wj}
T.~Matsubara, ``{Resumming Cosmological Perturbations via the Lagrangian
  Picture: One-loop Results in Real Space and in Redshift Space},'' {\em Phys.
  Rev.}, vol.~D77, p.~063530, 2008.

\bibitem{Alles:2015vua}
A.~Alles, T.~Buchert, F.~Al~Roumi, and A.~Wiegand, ``{Lagrangian theory of
  structure formation in relativistic cosmology. III. Gravitoelectric
  perturbation and solution schemes at any order},'' {\em Phys. Rev.},
  vol.~D92, no.~2, p.~023512, 2015.

\bibitem{Li:2018hll}
Y.-Z. Li, P.~Mourier, T.~Buchert, and D.~L. Wiltshire, ``{Lagrangian theory of
  structure formation in relativistic cosmology. V. Irrotational fluids},''
  {\em Phys. Rev.}, vol.~D98, no.~4, p.~043507, 2018.

\bibitem{Wald:1984rg}
R.~M. Wald, {\em {General Relativity}}.
\newblock Chicago, USA: Chicago Univ. Pr., 1984.

\end{thebibliography}

\end{document}